%% file: monsaraz.tex
\documentclass[12pt]{report}
\usepackage{epsf,amsfonts}
\def\epsfsize#1#2{1.0#1}
\epsfverbosetrue
\def\goth{\mathfrak}          
\def\double{\mathbb}         
\def\ccal{\cal}           

\def\cc{{\double C}}     
\def\nn{{\double N}}       
\def\rr{{\double R}}     
\def\zz{{\double Z}}

\def\aa{{\cal A}}
\def\ccc{{\cal C}}
\def\dd{{\cal D}} 
\def\gg{{\goth g}}        
\def\hh{{\cal H}}
\def\hhh{{{\double H}}}   
\def\ff{{\cal F}}
\def\mm{{{\ccal M}}}

\def\aa{{\cal A}}
\def\dd{{\cal D}} 
\def\hh{{\cal H}}
\def\ff{{\cal F}}
\def\lll{{\cal L}}
\def\sss{{\cal S}}
\def\jj{{\cal J}}
\def\t{{\rm tr}} 
\def\re{{\rm Re}\,}
\def\tt{\,{\rm tr}_\omega\,}
\def\ddd{{\,\hbox{$\partial\!\!\!/$}}}
\def\ddee{{\,\hbox{${\rm D}\!\!\!\!/\,$}}} 
\def\aaa{{\,\hbox{$A\!\!\!/\,$}}} 
\def\dee{\hbox{\rm D}}
\def\de{\hbox{\rm d}} 

\def\pa{{\partial}}
\def\Box{\,\hbox{$\sqcap\!\!\!\!\sqcup$}\,}
\def\semi{\hbox{$\odot\,\!\!\!\!\!s\,$}}

\def\lb{\left[} 
\def\rb{\right]}

\def\ul{\underline}
\def\ot{\otimes}
\def\op{\oplus}

\def\bb{\begin{eqnarray}}
\def\ee{\end{eqnarray}}
\def\eee{\nonumber\end{eqnarray}}
\def\pp{\pmatrix}
\def\qq{\quad}

\begin{document}

\hsize 17truecm
\vsize 24truecm
\font\twelve=cmbx10 at 13pt
\font\eightrm=cmr8
\baselineskip 18pt

\input{monsa1}

\input{monsa2}

\input{monsa3}
\input{monsa4}

\input{monsa5}

\input{monsa6}

\input{monsa7}

 \end{document}

%% file: monsa1
\begin{titlepage}

\centerline{\twelve CENTRE DE PHYSIQUE TH\'EORIQUE}
\centerline{\twelve CNRS - Luminy, Case 907}
\centerline{\twelve 13288 Marseille Cedex 9}
\vskip 3truecm

\centerline{\twelve Geometries and Forces}

\bigskip

\begin{center}
\bf Thomas SCH\"UCKER 
\footnote{\, schucker@cpt.univ-mrs.fr, also at 
Universit\'e de Provence} \\

\end{center}

\vskip 2truecm
\leftskip=1cm
\rightskip=1cm
\centerline{\bf Abstract} 

\bigskip

The present status of Connes' noncommutative view
at the four forces is reviewed. 

\vskip 4 truecm
PACS-92: 11.15 Gauge field theories, 04.20 General
relativity\\ 
\indent
MSC-91: 81T13 Yang-Mills and other gauge theories,\\
\indent $\ \qq$ $\ \qq$ $\ \qq$
83D05 Relativistic gravitation theories others than
Einstein's 
 
\vskip 1truecm

\noindent july 1997
\vskip 1truecm
\noindent CPT-97/P.3518\\
\noindent hep-th/9712095
 
\vskip1truecm

 \end{titlepage}

\tableofcontents \vfil\eject

\chapter{Two dreis\"atze for Maxwell}

The dream is older than Democrite: describe the
universe as a game of Lego: a few `elementary'
particles are held together by a few `elementary'
forces. The universe is complicated, the dream naive.
Still, it has seen impressive successes explaining e.g.
our solar system, chemical elements, light, the
hydrogen atom, nuclear reactions. To avoid
misunderstanding, the successes were on purely
scientific level, often followed by human failure. The
aim of these notes is to review today's version of the
game and Connes' attempt to understand its rules 
as geometry.

\section{A qualitative vocabulary}

Today we believe that there are four forces: gravity,
electromagnetism, weak and strong forces.

Gravity describes the falling apple, the motion of
earth around the sun, the dynamics inside a galaxy
and maybe even the dynamics of galaxies. But the last
item is the cosmological part of theology. In any case 
all items are macroscopic phenomena and we do not
know of any microscopic manifestation of gravity.
What is more, we have so far no consistent quantum
theory of gravity.

\begin{figure}[hbt]
\hspace{6cm}
\def\epsfsize#1#2{0.8#1}
\epsfbox{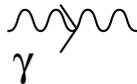}
\caption{A photon from $\gamma$ decay}
\label{gamma}
\end{figure}

\begin{figure}[hbt]
\hspace{5cm}
\def\epsfsize#1#2{0.8#1}
\epsfbox{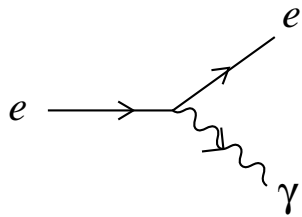}
\caption{Bremsstrahlung}
\label{bremsstrahlung}
\end{figure}

\begin{figure}[hbt]
\hspace{5cm}
\def\epsfsize#1#2{0.8#1}
\epsfbox{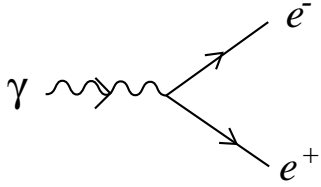}
\caption{Pair production}
\label{pair}
\end{figure}

Electromagnetism describes, on the macroscopic side,
e.g. electric generators and motors, light, radio
transmission. On quantum level, it is responsible for
$\gamma$ decay, bremsstrahlung, and pair
creation. The first refers to an unstable nucleus, a
bound state of protons and neutrons. The protons and
neutrons rearrange by emitting a photon with
enormous energy, figure \ref{gamma}. This photon is
a killer and you better hide behind a solid screen of
lead. Bremsstrahlung says that a high energy
electron for instance from a $\beta$ decay may slow
down by emitting a high energy photon, figure
\ref{bremsstrahlung}. To protect yourself against
these electrons typically a layer of cheap plexiglass is
sufficient. Bremsstrahlung makes radioprotection
expensive, before getting stuck in the plexiglass, the
electron emits a photon that goes through plexiglass
like through butter and you better buy lead. Pair
creation is a process where a photon traveling with
sufficient energy changes into an electron and a
positron, figure \ref{pair}. With it, quantum
electrodynamics teaches us two important lessons:
even an `elementary' particle, here the photon, may
be unstable, it may change identity or said differently
it may {\it decay}. This makes quantum field theory so
complicated. Fortunately the decays are not arbitrary.
They are governed by precise laws, e.g. conservation
laws for which group theory will play a fundamental
role. An important task for physicists is to compute
life times and branching ratios from these laws and to
confront the numbers with experiment. What is a
branching ratio in our example of a decaying photon?
If the photon has enough energy it may decay into
any pair of a charged particle and antiparticle. The
branching ratios are the corresponding probabilities.
The word interaction is often used instead of force to
underline that now a force not only changes the state
of motion of the concerned particles but also their
identity, their state in an {\it internal space}. We owe
the second lesson to Dirac who generalized
Schr\"odinger's equation to high energies, that
means to special relativity or Minkowskian geometry.
This generalization forces the introduction of
antimatter. To every particle there must exist an
antiparticle, with same mass and spin but with
opposite charges. For instance, the antiparticle of the
electron is the positron. Electromagnetism is {\it the}
show off theory of physics. It is successful both on
macroscopic and quantum level, it operates with clean
mathematics and has many applications to every day
life. It should be used to set the scale of success in our
Lego game.


\begin{figure}[hbt]
\hspace{5cm}
\def\epsfsize#1#2{0.8#1}
\epsfbox{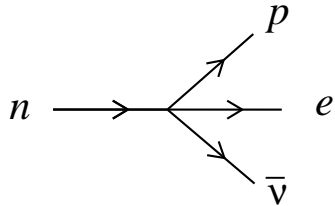}
\caption{A $\beta$ decay}
\label{beta}
\end{figure}

Weak interactions describe the $\beta$ decay and they
are popular since Chernobyl. Take an iodine-131
isotope. It is a bound state of 53 protons and 78
neutrons. One of the neutrons changes identity. It
decays to a proton, an electron and an anti-neutrino,
figure \ref{beta}. The proton is heavy and lazy. It
stays in the nucleus which becomes a Xenon-131. The
neutrino has zero mass and  zero
electric charge and is therefore harmless for man. It
can pass through the entire earth without losing
energy. The damage is done by the electron that
deposits its energy in the immediate vicinity of its
point of decay. This is e.g. the thyroid of babies
where iodine likes to accumulate 
\cite{insight}. Let us be macabre and note an
academic property of the killer electron: its {\it
chirality}. The electron goes at almost the speed of
light and it has spin 1/2. Quantum mechanics tells us,
that in this situation, there are only two possibilities,
the spin is parallel to its velocity, the electron has
chirality {\it left} or the spin is anti-parallel to its
velocity, the electron has chirality {\it right}. Here
comes the surprising observation, the electron from
$\beta$ decay is always left-handed. The spin is a
vector describing the axis of rotation of the electron
around itself. Therefore the spin is an {\it axial}
vector, a vector that changes sign under {\it parity},
space reflection. Weak interactions break parity
maximally, you never observe a right-handed
electron or neutrino coming out of a $\beta$ decay.
The (electric) charge of a particle indicates to what
extent it is subject to the electric force. Likewise
there is the weak charge called {\it (weak) isospin}.
The left-handed electron and the left-handed
neutrino have non-vanishing isospin. The
right-handed electron has zero isospin. A
right-handed neutrino has never been observed. If
the neutrino has no right-handed part then it must be
massless, in agreement with observation.

We do not know of any macroscopic manifestation of
weak forces. We will try to understand this with the
help of spontaneous breaking of gauge symmetry that
will be in the center of our discussion. Let us
anticipate a little. Maxwell tells us that the
electromagnetic force between two charged particles
results from another particle being exchanged
between them. This particle, the photon or
generically the gauge boson, has spin 1 and is
massless. The gauge symmetry implies that the gauge
boson is massless, which in turn implies that the force
is long range, falls off like the inverse square of the
distance. Weak interactions are also mediated by
gauge bosons, the $W$ or weak boson. In fact, the
neutron in the example from Chernobyl first decays
into a $W$ and the proton. Then the $W$ decays into a
neutrino and an electron like in pair creation from
the photon. The $W$ has spin 1 as the photon, but
must be very massive to render the weak interactions
short range in accordance with experiment.

\begin{figure}[hbt]
\hspace{4.5cm}
\def\epsfsize#1#2{0.8#1}
\epsfbox{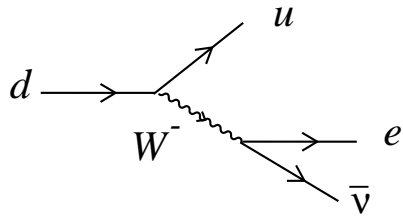}
\caption{Same $\beta$ decay with better resolution }
\label{beta2}
\end{figure}

The strong force was invented to bind protons and
neutrons inside the nucleus. Protons have electric
charge and according to Coulomb's law they repel
each other with an electric force that increases as the
inverse square of their distance. The size of the
nucleus being only $10^{-15}$ meters, we must invoke
a strong force to explain the stability of the nucleus.
Once accepted, the strong force also explains $\alpha$
decay, that is the emission of a helium nucleus, two
protons and two neutrons, from a heavy nucleus like
plutonium-239. Moreover, the strong force explains
fusion and fission and thereby the energy production
in the sun, in diverse nuclear bombs and in nuclear
`facilities'. Again we have to face the question, why do
we see no macroscopic manifestation. The gauge
bosons of the strong force are called gluons and they
are massless. Nevertheless the strong force is short
range because of {\it confinement}. Confinement has
so far resisted every attempt of proof. All we have is
clue from perturbation theory indicating that the
strong force decreases with energy, {\it asymptotic
freedom}. Extrapolating to low energies we do assume
an extremely strong static force law that confines all
particles with nonvanishing strong
charge, called {\it colour}. The idea then is that the
proton and the neutron are colourless bound states of
three coloured quarks. Quarks are supposed
elementary. We have the
$up$ quark with electric charge 2/3 (in units of the
absolute value of the electron charge) and the $down$
quark of charge $-1/3$. The quarks carry also the
strong charge, colour. On the other hand the proton, a
$uud$ bound state, and the neutron, $udd$, are
colourless and therefore they can be isolated. The
force tying protons and neutrons to a nucleus are
imagined of van der Waals' type and consequently
short range. When in our Chernobyl example one of
the neutrons inside the iodine nucleus suffers $\beta$
decay to a proton and a
$W$, it is in fact one of the two $down$ quarks in this
neutron that decays to an $up$ quark and a $W^-$,
figure \ref{beta2}.  Just as gravity and
electromagnetism, the strong force preserves parity,
it is {\it vectorial}.

Let us recapitulate our elementary particles and start
with the gauge bosons. They mediate the
non-gravitational forces and have spin 1. There is the
photon $\gamma$. Its mass, electric charge and
colour all vanish, but not its isospin. There is the weak
boson $W^{\pm}$. It is very massive, it has unit
electric charge, non-zero isospin, no colour. A second
weak boson, the $Z^0$ was discovered in the
seventies. Its quantum numbers are as for the $W$
except for zero electric charge. Finally there are
eight gluons, no mass, no charge, no isospin, but
non-zero colour. Gravity is also mediated by a boson,
the graviton. It is {\it not} a gauge boson, it has spin 2.
It is massless and has no charge, no isospin, no colour.
Let us anticipate that we shall need another boson, the
Brout-Englert-Higgs scalar, with spin 0, no charge, no
colour, but with isospin, and massive. We need it to
give masses to bosons and fermions via spontaneous
symmetry break down. We need it, but we have not
seen it and it is the last missing particle.

All other elementary particles are fermions, they
have spin 1/2. They fall into two classes, leptons and
quarks. Leptons, from the greek word mild, do not
participate in strong interaction, they are colourless.
There is the (electronic) neutrino $\nu_e$, purely
left-handed and therefore massless, no charge but
isospin. The electron $e$ is massive, charge $-1$. Its
left-handed part $e_L$ has isospin, its right-handed
part $e_R$ has isospin 0. Confinement suggests that
quarks (and gluons) will never be observed alone and
today we only have indirect measurements of their
quantum numbers. All quarks are thought to be
massive, the $u$ has charge 2/3, the $d$ has
$-1/3$. Their left-handed parts have isospin, the
right-handed parts do not. 

We are now ready for another mystery of particle
physics. More elementary fermions have been
observed that are boring copies of the above ones. Let
us call the $(u,\ d,\ \nu_e,\ e)$ first generation. Then
we have two more generations $(c,\ s,\ \nu_\mu,\
\mu)$ and $(t,\ b,\ \nu_\tau,\ \tau)$. The names of the
quarks, {\it charm, strange, top, bottom}, recount well
how the physicists felt about their discovery,
estranged, charmed, blas\'e. Nature has simply copied
the quantum numbers of the first generation, except
for the masses, that remain a puzzle. The $top$ is
extremely heavy and consequently was the last to be
seen, only two years ago. Experiments also indicate --
of course only indirectly -- that the $top$ is the last,
there should not be a fourth generation. Here is
today's periodic table of elementary fermions:
\bb \pp{u\cr d}_L,\ \pp{c\cr s}_L,\ \pp{t\cr b}_L,\ 
\pp{\nu_e\cr e}_L,\ \pp{\nu_\mu\cr\mu}_L,\ 
\pp{\nu_\tau\cr\tau}_L\eee
\bb\matrix{u_R,\cr d_R,}\qq \matrix{c_R,\cr s_R,}\qq
\matrix{t_R,\cr b_R,}\qq  e_R,\qq \mu_R,\qq 
\tau_R\eee
The
parentheses indicate {\it isospin doublets}, i.e.
particles that can be produced pairwise from a
decaying $W$.

\section{The gauge dreisatz}

In this section we want to be a little more quantitative
about weak and strong charges. Behind the decay
laws, there are conservation laws. Behind
conservation laws, there is group theory by Emmy
Noether's theorem. This is well known from
electromagnetism. Maxwell's equations are invariant
under the group $U(1)$. This invariance explains the
experimentally well established charge conservation.
For instance, the electricly neutral photon can only
decay into an electricly neutral pair. Charge
conservation also implies that the $W$ has unit
charge. Moreover, the electromagnetic gauge group is
Abelian. This implies that the photon has zero electric
charge, that it does not itself feel the force which it
mediates. On the other hand, we expect the weak and
strong gauge groups to be non-Abelian. Maxwell's
equations are also Lorentz invariant, if we suppose
that the conserved electric charge is Lorentz
invariant, i.e. does not depend on velocity. Then the
photon must have spin 1, and likewise for weak and
strong bosons. However, the gravitational `charge' is
the mass or more precisely energy, which is not a
Lorentz scalar. Energy is a component of a
four-vector in Minkowskian geometry. Therefore the
graviton has spin 2.
To cut a long story short here is today's credo for
playing Lego:
\begin{itemize}\item
{ Elementary particles are orthonormal basis
vectors of a unitary group representation.} The group
$G$  falls from heaven, most of the time.
\item
 { The
charge parameterizes the choice of the
representation}.
\item
{  Composite particles are obtained from
tensor products.} 
\end{itemize}
Wigner proposed the credo. His starting point, the
Poincar\'e group or its spin cover, does not fall from
heaven. It comes from Minkowskian geometry. The
Poincar\'e group is non-compact and its unitary
representations are infinite dimensional. They are
characterized by a continuous variable, the mass, and
a discrete one, the spin. The spin parameterizes the
finite dimensional part of the representation under
the compact subgroup $SU(2)$, the cover of the
rotation group in three dimensional Euclidean space.
We denote by $\ul {2j+1}$ the $2j+1$ dimensional
irreducible representation of
$SU(2)$. It has spin $j$, $j=0,\
\frac{1}{2},\ 1,...$ A composite particle consisting of
two spin $ \frac{1}{2}$ particles, 
\bb \ul 2\ot\ul 2 = \ul 1 \op \ul 3,\ee can have spin 0,
the antisymmetric part of the tensor product: the two
spin
$ \frac{1}{2}$ are anti-parallel, or it can have spin 1,
the symmetric part of the tensor product: the two spin
$ \frac{1}{2}$ are parallel. This is the Clebsch-Gordan
decomposition and you must know that physicists
working at CERN carry a pocket size table with 
 two hundred Clebsch-Gordon coefficients
\cite{pocket}.

Motivated from charge conservation and Emmy
Noether, let us have $G=U(1)$ fall from heaven. Its
irreducible, unitary representations are all one
dimensional, $\hh=\cc\owns\psi$ with
$\rho(\exp {i\theta})\psi=\exp {i(q/e)\theta}\,\psi$.
$q$ is the electric charge, it is additive under tensor
products. Indeed,
\bb &&(\rho_1\ot\rho_2)(\exp
i\theta)(\psi_1\ot\psi_2) =(\rho_1(\exp
i\theta)\psi_1)\ot (\rho_2(\exp i\theta)\psi_2)\cr
&&\qq = (\exp {i(q_1/e)\theta}\,\psi_1)\ot
(\exp {i(q_2/e)\theta}\,\psi_2)=
\exp {i((q_1+q_2)/e)\theta}\,\psi.\ee
 Heisenberg found that one dimensional
representations are boring and tried $G=SU(2)$ which
he called (strong) isospin in order to distinguish it
from the spin
$SU(2)$. Instead of spin up and spin down, he puts the
proton and the neutron -- or in today's picture the
$up$ and $down$ quarks -- in the $\ul 2$. Gell-Mann
was more successful with $G=SU(3)$ and discovered
the first three quarks $(u,d,s)$ sitting in the
fundamental representation,
$\hh=\cc^3$, $\rho(g)=g$. Indeed, this hypothesis
allowed him to classify the baryons and mesons of his
time as bound states of three quarks or of
quark-antiquark. Heisenberg's $SU(2)$ of strong
isospin and Gell-Mann's
$SU(3)$ of flavour should not be confused with the
{\it gauged} $SU(2)$ of weak isospin and the {\it
gauged} $SU(3)$ of coulour. The latter will play a
fundamental role and generate the forces. At the same
time they will allow to derive the non-gauged ones.
Consequently, the non-gauged ones play a secondary
role today and we mentioned them for historical
reasons. They lead to the discovery of quarks and to
the establishment of the credo. A newcomer should be
warned however: the confusion is still present today
and not only in the terminology isospin. 

At the root of this confusion is the gauge miracle. The
ungauged $U(1)$ of electric charge conservation can
be gauged and its gauging produces
electromagnetism. 

Here is the story in short. The ungauged $U(1)$ of
electric charge conservation does {\it not} fall from
heaven, it is given to us free of charge by quantum
mechanics, via the conservation of probability. The
representation space of quantum mechanics is
$\lll^2(\rr^3,\cc)$. Its elements, complex valued,
square integrable functions on our Euclidean space
$\rr^3$, are the wave functions, $\psi(\vec x)$. A
natural group of unitaries acts in this Hilbert space.
Its elements are translations $U_{\vec\xi}$, rotations
$U_R$ and phase transformations $U_{\exp i\theta}$,
\bb(U_{\vec\xi}\psi)(\vec x)&:=&\psi(\vec
x-\vec\xi),\\ (U_R\psi)(\vec x)&:=&\psi(R^{-1}\vec
x),\\ (U_{\exp
i\theta}\psi)(\vec x)&:=&\exp
(i(q/e)\theta)\,\psi(\vec x).\ee
 Their generators are momentum, angular momentum
and electric charge,
\bb \,\frac{\pa}{\pa\vec x}\, ,\qq
\vec x\,\wedge\,\frac{\pa}{\pa\vec x}\, ,\qq i1.\ee 
The
associated conserved quantities have the same names.
The free Schr\"odinger equation follows via the
Euler-Lagrange variational principle from the action 
\bb \int\de t\int\de\vec x \ \bar\psi(t,\vec x)\,
(i\hbar\,\frac{\pa}{\pa
t}\,+{\textstyle\frac{\hbar
^2}{2m}}\,\frac{\pa^2}{\pa\vec x^2}\, )\psi(t,\vec x)
.\ee   Schr\"odinger's version of quantum
mechanics treats space and time differently, the
position $\vec x_{\rm op}$ is an observable, a
Hermitian operator, $(\vec x_{\rm
op}\psi)(t,\vec x)=\vec x\psi (t,\vec x),$ and as such it
has an uncertainty, time $t$ is just a parameter.
$\hbar= 1.055\cdot 10^{-34}\ {\rm m^2\ kg/s}$ is
Planck's constant, most of the time we adopt units
such that
$\hbar=1$.
$m$ denotes the mass of the free `matter'.
Schr\"odinger's action is obviously invariant under
phase transformations, in agreement with the
postulate that only $|\psi|^2$ has physical
significance: it is the probability density of location.
However, the choice of phase must be rigid, constant
over the entire universe. One might object that a
physicists somewhere in Andromeda should be able to
do his quantum mechanical calculations with a phase
convention that should not be tied to a phase
convention used by a colleague on Earth. This leads us
to consider spacetime dependent phase
transformations $\exp i\theta(t, \vec x)$. They form
the infinite dimensional gauge group or gauged
$U(1)$. Its elements are functions from spacetime into
$U(1)$ with pointwise multiplication. How can we
render Schr\"odinger's action gauge invariant? The
trick goes by the name of minimal coupling. Postulate
the existence of a connection or gauge field
$A_\mu$, $\mu=0,1,2,3$ with the affine
transformation law 
\bb \rho_V(\exp { i\theta})A_\mu=A_\mu+i
{\textstyle\frac{\hbar}{e}}\exp { i\theta}\,
\frac{\pa}{\pa x^\mu}\,\exp{ -i\theta}=
A_\mu+{\textstyle\frac{\hbar}{e}}
\,\frac{\pa}{\pa x^\mu}\,\theta
\ee 
where
we have put $x^0=ct$. The subscript $V$ stands for
vector because the gauge field is a vector field, it has
spin 1. Now replace all derivatives
 $\frac{\pa}{\pa x^\mu}$ in the free action by
covariant derivatives $\frac{\pa}{\pa
x^\mu}+i{\textstyle\frac{q}{\hbar}}
A_\mu$  and you get a gauge invariant
action. Physically the free matter particle, we started
from, is now coupled to an electromagnetic field,
`radiation', whose vector potential is $A_\mu$. In a
second stroke, we want to make the gauge field
dynamical. We look for a kinetic term, i.e. a term
involving derivatives of the gauge field, and that is
gauge invariant. In lowest order, two derivatives, the
answer is unique, it is Maxwell's action with the
$1/r^2$ fall off in its static force field. Genesis is
rewritten, {\it Let there be light} is to be replaced by
{\it Let there be gauge} and we can  
 summarize the gauge miracle in form of a
{\it dreisatz} or {\it regra de tr\^es:}
\begin{itemize}\item
{ \qq quantum mechanics \ + \ gauge invariance \ \ =
\ \ Maxwell.  }
\end{itemize}
 Note that this dreisatz works for
non-relativistic quantum mechanics, Schr\"odinger,
or relativistic quantum mechanics, Klein-Gordon and
Dirac. 

We anticipate that the gauge miracle also works for
weak and strong interactions. {\it Let there be
non-Abelian gauge.} However, the groups
$G=SU(2),\ SU(3)$, and their representations fall from
heaven and we cherish the dream to
derive them from first principles just as the $U(1)$ was
derived from quantum mechanics.  This is precisely
what Connes proposes. {\it Let there be
noncommutative geometry.}
We end this section with a warning: there is a
semantical ambiguity, should a particle be an entire
representation space or only a vector therein? In
Wigner's point of view, it is the same particle that can
have different energies and spin orientations. An
applied force can change the energy or spin
orientation of a particle without changing its identity.
Weak interactions force us to treat the different spin
orientations or chiralities of the electron as different
particles with different charges and at the same time
forces us to allow for interactions that change the spin
orientation in spacetime {\it and} the isospin
orientation in an internal space. The latter is the
change of identity in pair production.

\section{The Minkowski dreisatz}

 Our next dreisatz is of even more geometric nature:

\begin{itemize}\item
{ \qq Coulomb \ + \ Minkowskian geometry \ \ = \ \
Maxwell.}
\end{itemize}
This dreisatz does not do justice to the
historical development, Maxwell's theory existed when
Einstein discovered special relativity and it came as
surprise that Maxwell's theory was already Lorentz
invariant. 

We start with Coulomb's static force law,
\bb F=\,\frac{1}{4\pi
\epsilon_0}\,\,\frac{qQ}{r^2}\,\ee 
describing the
force between two electric charges $q$ and $Q$ at rest
at a distance $r$. The proportionality constant $
\epsilon_0=8.8544\cdot 10^{-12}{\rm
s^2\,C^2/(m^3\,kg)}$ will be referred to as the inverse
square of the {\it coupling constant}. In the following
we will measure electric charge not in Coulomb C but
in units of minus the electron charge,
$e=1.6021\cdot 10^{-19}$ C. Note that this
normalization can be changed at will, only the ratio
$e^2/\epsilon_0$ is physical. Often we also use units
of electric charge such that $\epsilon_0=1$. Then $e$
is the coupling constant. Now we perform a Lorentz
boost
\bb c\bar t=\,\frac{ct+vx/c}{\sqrt{1-v^2/c^2}}\,,\qq
 \bar x=\,\frac{x+vt}{\sqrt{1-v^2/c^2}}\,,\qq
\bar y= y,\qq \bar z=z,\ee
with the speed of light $c=2.9979\cdot 10^8$
m/s, and the magnetic field pops up. The force
involving two time derivatives has a complicated
transformation law under the Lorentz group and we
take advantage of the fact that Coulomb's force derives
from a potential,
\bb \vec E= \vec F/q=-\,\frac{\pa}{\pa \vec x}\, V,\qq
V=\,\frac{1}{4\pi
\epsilon_0}\,\frac{Q}{r}\label{coupot}\ee
In terms of the potential we can write Coulomb's law
as the following differential equation 
\bb \,\frac{\pa^2}{\pa \vec x^2}\, V=-\rho/
\epsilon_0,\label{coudiff}\ee
where $\rho$ is the charge density.
The Coulomb potential $V$, equation (\ref{coupot}) is
the elementary solution or Green function of the
Laplacian for a pointlike source
$\rho(\vec x)=\frac{Q}{\epsilon_0}\delta(\vec x)$.
The charge being a conserved quantity we must
suppose that it is Lorentz invariant. Then the charge
density transforms as the zero component $j^0=c\rho$
of a four vector $j^\mu$ whose spatial components are
the current density $\vec j$. Consequently, the lhs of
the differential equation (\ref{coudiff}) must be a four
vector. Therefore we introduce the (four) vector
potential
$A^\mu$ with $A^0=V/c$. The force a `test'
charge or matter particle $q$ feels in the
electromagnetic field is obtained from the static force
law,  
\bb m\,\frac{\de^2\vec x}{\de t^2}\,=q\vec E,\ee
by a Lorentz boost:
\bb m\,\frac{\de^2 x^\mu}{\de \tau^2}\,=q
{F^\mu}_\nu \,\frac{\de x^\nu}{\de
\tau}\,.\label{test}\ee 
We use Einstein's summation convention, summing
over repeated indices is always understood. We denote
by
$\tau$ the Lorentz invariant proper time defined only
on the trajectory
$x^\mu(\tau)$ of the test particle by the implicit
equation
\bb c\tau=\int_0^\tau\left[c^2\left(\,\frac{\de
t}{\de\tilde\tau}\,\right)^2-\,\frac{\de \vec x}{\de
\tilde \tau}\,\cdot\,\frac{\de \vec x}{\de
\tilde \tau}\,\right]^{1/2}\de \tilde \tau,\ee
or infinitesimally
\bb c^2 \de \tau^2=c^2 \de t^2-\de\vec x^2=: 
\eta_{\mu\nu}\de x^\mu\de x^\nu=:\de
x_\nu\,\de x^\nu.\ee
The Minkowski metric
\bb\eta_{\mu\nu}=\pp{1&0&0&0\cr 0&-1&0&0\cr 
0&0&-1&0\cr 0&0&0&-1}\ee
 and its inverse $\eta^{\mu\nu}$ are used to lower and
raise indices. We note that, without quantum
mechanics, only derivatives of the potential $A_\mu$
are measurable. They are called field strength or
curvature,
\bb F_{\mu\nu}= \pa_\mu A_\nu-\pa_\nu A_\mu,\qq
\pa_\mu:=\,\frac{\pa}{\pa x^\mu}\, .\ee
The field strength is an antisymmetric matrix made
up from the electric field $\vec E$  and the magnetic
field $\vec B$,
\bb F_{\mu\nu}=-\pp{
0&-E_x/c&-E_y/c&-E_z/c\cr 
E_x/c&0&B_z&-B_y\cr 
E_y/c&-B_z&0&B_x\cr 
E_z/c&B_y&-B_x&0}.\ee
The important property of the field strength is that it
is invariant under the gauge or phase
transformations, 
\bb \rho_V(\exp { i\theta})A_\mu=A_\mu+i
{\textstyle\frac{\hbar}{e}}\exp {
i\theta}\,\pa_\mu
\exp { -i\theta}=
A_\mu+{\textstyle\frac{\hbar}{e}}\pa_\mu\theta.
\ee 

The force law for test charges (\ref{test}) is nothing
but the Lorentz force describing `the coupling
between matter and the electromagnetic field'. In a
second stroke we want to generalize the static
differential equation (\ref{coudiff}) that tells us that
charge is the source of the electric
field.  We simply replace the potential $V$ by the four
potential $A$, the charge density $\rho$ by the four
current $j$ and the Laplace operator
$\Delta=\pa^2/\pa\vec x^2$ by its Lorentz invariant
extension, the wave or d'Alembert operator
\bb\Box={\textstyle\frac{1}{c^2}}\,\frac{\pa^2}{\pa
t^2}\,-\Delta=:\eta^{\mu\nu}\pa_\mu\pa_\nu=:
\pa^\nu\pa_\nu.\ee
Then we obtain Maxwell's equations 
\bb \Box A_\nu=\,\frac{1}{\epsilon_0 c^2}\,j_\nu\ee
in the Lorentz gauge $\pa_\mu A^\mu=0$. The gauge
invariant Maxwell equations read:
\bb\pa^\mu F_{\mu\nu}=\pa^\mu\pa_\mu
A_\nu-\pa_\nu\pa_\mu A^\mu=
\,\frac{1}{\epsilon_0 c^2}\,j_\nu.\label{eight}\ee
They make the electromagnetic field dynamical, it
propagates with the speed of light and therefore it is
often called radiation. In particular, Maxwell's
equation contains  Amp\`ere's  static law,
\bb {\rm rot}\vec B=\mu_0 \vec j,\ee
and we can identify
the static magnetic coupling constant
$\mu_0=1/(\epsilon_0 c^2)$.

Maxwell's equations derive from an action,
\bb S[A]=-\int_{\rr^4}
\left(\,{\frac{\epsilon_0c}{4}}\,
F_{\mu\nu}F^{\mu\nu}+{\textstyle\frac{1}{c}}j_\nu
A^\nu\right) \de^4 x.\ee 
If we measure electric charge in units of the electron
charge then we must replace $\epsilon_0$ by
$\epsilon_0/e^2$. The first term is manifestly gauge
invariant, the second, the minimal coupling to matter,
is gauge invariant thanks to charge conservation,
\bb \pa_\nu j^\nu=0.\ee 

Let us summarize our first geometric dreisatz: the
extension of Coulomb's static force law with its
coupling $\epsilon_0$ to Minkowskian geometry
characterized by the speed of light $c$ produces an
additional force, the magnetic force with feeble
coupling $\mu_0$. Maxwell's theory is
celebrated today as Abelian or should we say,
commutative Yang-Mills theory.
 Historically, the chronological
order was different. Both the static electric and
magnetic forces where known, Maxwell unified them
by rendering them dynamical. Plane waves came out
as particular dynamical solutions to his equations and
he found that the velocity of the waves, the speed of
light, was $c=(\epsilon_0\mu_0)^{-1/2}$. At his time
physicists still believed that the speed of light, like
any velocity, depended on the reference system. But
nobody really dared to object to Maxwell's relating the
speed of light to static constants, experimentally
Maxwell was right. Lorentz timidly introduced his
transformations to understand the puzzle. Only
Einstein dared to take the Lorentz transformations
serious. He operated a revolution on spacetime, e.g.
abolishing absolute time. His revolution is accessible to
experimental verification, without talking about
forces.

%% file: monsa2
\chapter{A technical interlude on differential forms}
\label{interlude}

The two dreis\"atze discussed so far call for
differential forms, which will also make the
generalization of Maxwell's theory to curved
spacetimes easy. Here is a crash course on the local
theory \cite{gs}.

\section{Vector fields}

\begin{figure}[hbt]
\hspace{2cm}
\def\epsfsize#1#2{0.8#1}
\epsfbox{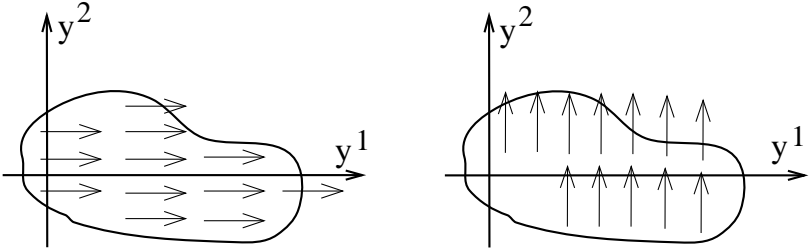}
\caption{The Cartesian vector fields ${\partial
/{\partial y^1}}$ and ${\partial /{\partial y^2}}$ }
\label{fields}
\end{figure}

Let $U$ be an open subset of $\rr^n$. A vector field $v$
on $U$ is a differentiable family $v(x)$ of vectors in
$\rr^n$ indexed by the points in $U$. (For us,
differentiable always means infinitely many times
differentiable.) For example, $U$ is a lake and
$v$ the wind. Note
that the `velocity' vectors $v(x)$ are not confined to
lie in a subset of $\rr^n$ as is the case for the points
$x$.
In Cartesian coordinates $y^\mu,
\mu=1,2,\cdots, n$, any vector field may be decomposed:
\bb v=\sum^n_{\mu=1}v^\mu(x){\partial\over{\partial
y^\mu}},\ee
 where ${\partial /{\partial
y^\mu}}$ are the vector fields with Cartesian
components $(0,\cdots, 0,1,0,\cdots,0)$. The one is the
$\mu$th entry. Figure \ref{fields} shows an example.
 Note that here ${\partial /{\partial
y^\mu}}$ is not a differential operator, but just a
symbol. Its mnemo-technical utility comes from the
definition in arbitrary coordinates
$x^\mu$:
\bb{\partial \over{\partial
x^\mu}}(x):=\sum_\nu{{\partial y^\nu}\over {\partial
x^\mu}}(x){\partial\over{\partial
y^\nu}}\label{def}\ee
 where ${{\partial
y^\nu}/{\partial x^\mu}}$is the Jacobian matrix
of the (general) coordinate transformation. We shall
consider explicitly the example of polar coordinates
later

\section{Differential forms}

By definition a (differential) $p$-form $\varphi$ is a
differentiable family of maps $\varphi_x$
\bb\varphi_x:\rr^n \times\cdots
\times\rr^n&\longrightarrow&\rr\cr
(v_1(x),\cdots,v_p(x))&\longmapsto&
\varphi_x(v_1(x),\cdots,
v_p(x)).\ee
 Each map $\varphi_x$  is required to be
multilinear (with respect to the real numbers) and
alternating, i.e.
\bb\varphi(\cdots,v_i,\cdots,v_j,\cdots)=-\varphi(\cdots,v_j,\cdots,
v_i
\cdots).\ee
 For convenience, we often
suppress the point $x$. We denote by $\Omega^pU$
the set of all $p$-forms on $U$. Note that if $p>n$ this
set only contains the zero element. For $p=0$ we define
$\Omega^0U$ to be the set of all (differentiable)
functions from $U$ into the real numbers.

\section{Wedge product}

The wedge product of a $p$-form with a $q$-form is
the $(p+q)$-form defined by:
\bb \wedge:\quad\Omega^pU\times\Omega^qU&
\longrightarrow&
\Omega^{p+q}U\cr
(\varphi,\psi)&\longmapsto&\varphi\wedge
\psi\cr
(\varphi\wedge\psi)(v,...,v_{p+q}):={1\over{p!q!}}
\sum_{\pi\epsilon S_{p+q}}&{\rm sig}\pi&
\varphi(v_{\pi(1)},...,v_{\pi(p)})
\psi(v_{\pi(p+1)},\cdots,v_{\pi(p+q)}),\ee
where the sum is over all permutations of $p+q$
objects and sig$\pi$ is the sign of the permutation
$\pi$.

The wedge product is bilinear, associative and graded
commutative, i.e.
\bb\varphi\wedge\psi=(-1)^{pq}\psi\wedge\varphi.
\label{gradcomm}\ee
In any coordinate system $x^\mu$ a $p$-form may now
be decomposed
\bb\varphi=\sum_{\mu_1,\cdots,\mu_p}
\varphi_{\mu_1,\cdots\mu_ p}(x)\de x^{\mu_1}
\wedge\cdots \wedge \de x^{\mu_ p},\ee
 where
for each $\mu=1, 2,\cdots, n, \de x^\mu$ is the 1-form
defined by
\bb\de x^\mu({\pa\over{\pa
x^\nu}})=\delta^\mu_\nu.\ee
 In tensor
language a vector field $v$ constitutes a contravariant
tensor $v^\mu$ of degree (rank) one while a $p$-form
constitutes a completely antisymmetric covariant
tensor $\varphi_{\mu_1
\cdots\mu_p}$ of degree $p$. The real number
obtained by evaluating a $p$-form on $p$ vector fields
corresponds to the complete contraction and the wedge
product corresponds to the antisymmetrized tensor
product of antisymmetric covariant tensors.

A collection of vector spaces $\Omega^pU, p=0, 1,\cdots,
n$, together with a bilinear, associative, graded
commutative product $\wedge$ is also called exterior
algebra or Grassmann algebra. Later, in order to
alleviate notations, we shall suppress  the wedge
symbol.

\section{Exterior derivative}

We define the exterior derivative of a form using a
coordinate system $x^\mu$:
\bb \de:\Omega^p
U&\longrightarrow&\Omega^{p+1}U\cr
\varphi&\longmapsto& \de \varphi\eee\bb
\de \varphi:=\sum_{\mu_1,\cdots, \mu_p,
\nu}({\partial\over{\partial
x^\nu}}\varphi_{\mu_1\cdots\mu_ p})
\de x^\nu\wedge \de x^{\mu_1}\wedge...\wedge
\de x^{\mu_ p}.\ee This definition does
not depend on the choice of the coordinate system
$x^\mu$.

The exterior derivative is a linear first order
differential operator. It obeys the Leibniz rule
\bb\de
(\varphi\wedge\psi)=(\de\varphi)\wedge\psi+
(-1)^p\varphi
\wedge \de\psi\ee and the so-called
co-boundary condition
\bb\de ^2=0.\ee
 In tensor language the
exterior derivative amounts to taking the gradient of
an antisymmetric covariant tensor and then
antisymmetrizing the covariant index of the gradient
with the others. The co-boundary condition is just the
statement that partial derivatives commute.

\section{Integration}

Let $\varphi$ be a $p$--form and $K$ a
$p$--dimensional sufficiently regular piece of $U$
parameterized by $x^1,x^2,...,x^p$, for example a cube.
Then we define the integral of $\varphi$ over $K$:
\bb\int_K\varphi:=\int_K\varphi_{12...p}\,\de
x^1\cdots\de x^p,\ee
 where the rhs is just the multiple Riemannian integral
of the coefficient function of
$\varphi$. The increasing order of the indices in the
coefficient function $\varphi_{12...p}$ means that we
suppose a fixed numbering of the coordinates of $K$,
i.e. an orientation. The definition of the integral of a
form does not depend on the choice of the coordinate
system. This is assured by the theorem that under a
change of coordinates the integrand in the
Riemannian integral changes with the absolute value
of the determinant of the Jacobian matrix.

Let us mention {\bf Stokes' theorem}: Let
$\varphi$ be a
$(p-1)$-form, $K$ a $p$--dimensional piece of $U$,
$\partial K$ its properly oriented boundary. Then
\bb\int_Kd\varphi=\int_{\partial
K}\varphi.\ee
 This theorem is useful
to derive field equations from an action.
Together with the Leibniz rule it allows to carry
out partial integrations. Finally, we remark that the
boundary of a boundary is empty,
\bb\partial\partial K=\emptyset,\ee
 which explains the
term co-boundary condition for $\de ^2=0$.

\section{Vector valued differential forms}

Let $W$ be a finite dimensional real vector space. Since
all operations introduced so far are linear, we can
generalize the values of differential forms from the
real numbers to vectors in $W$:
\bb\Phi_x:\rr^n\times...\times\rr^n\rightarrow W.\ee
We denote by $\Omega^p(U,W)$ the set of $p$--forms
on $U$ with values in $W$. In later applications $W$
will be a Lie algebra or a vector space carrying a
linear representation of some symmetry group. With
respect to a basis $T_a,a=1,2,...,\dim W$, any element
$w\in W$ can be written
\bb w=\sum^{\dim W}_{a=1}w^aT_a,\ee
 where
the $w^a$ are real numbers. Likewise any $p$--form
$\Phi$ with values in $W$ can be written as
\bb\Phi=\sum_a\varphi^aT_a,\ee
 where now
the $\varphi^a$ are real valued differential forms on
$U$. Of course, in order to define a wedge product in
this more general setting, $W$ must have a
multiplication law, i.e. $W$ must be an algebra. For
example, if $W$ is a Lie algebra, we define the the
commutator of a $p$-- form and a $q$--form, both with
values in the Lie algebra, by
\bb\lbrack\Phi,\Psi\rbrack(v_1,...,v_{p+q})={1\over{p!q!}}
\sum_{\pi\epsilon
S_{p+q}}{\rm
sig}\pi\lbrack\Phi(v_{\pi(1)},...,v_{\pi(p)}),
\Psi(v_{\pi(p+1)},...,v_{\pi(p+q)}\rbrack,\ee
or with respect to a basis $T_a$:
\bb\Phi=\sum_a\varphi_aT^a,&&\qq
\Psi=\sum_b\psi_bT^b,\cr
\lbrack\Phi,\Psi\rbrack&=&\sum_{a,b}\varphi_a\wedge\psi_b\lbrack
T^a,T^b\rbrack.\ee
 The commutator of forms
is graded commutative:
\bb\lbrack\Phi,\Psi\rbrack=-(-1)^{pq}
\lbrack\Psi,\Phi\rbrack,\ee
 where one
minus sign comes from the anticommutativity of the
commutator of two Lie algebra elements and the others
from equation (\ref{gradcomm}).

\section{Frames}

A frame on an open subset $U$ of $\rr^n$ is a set of
$n$ vector fields $b_1,b_2,...,b_n$ such that in each
point $x\in U$ the $n$ vectors $b_1(x),...,b_n(x)$ are
linearly independent. Other words used for frames are
tetrads (for $n=4$), vielbein or $n$--bein, rep\`ere
(mobile). If $x^\mu$ is a coordinate system, then
${{\partial}/{\partial x^\mu}},\mu=1,2,...,n,$ is a
frame. However, not every frame $b_i$ can be derived
from a coordinate system and we call a frame of the
particular kind ${{\partial}/{\partial x^\mu}}$
holonomic. Later we shall learn a recipe how to decide
whether a frame is holonomic.

Given two frames $b_i$ and $b'_i$ on $U$, we can
always at a given point
$x$ expand one in terms of the other:
\bb b'_i(x)=\sum_j(\gamma^{-1}(x))^j_{\phantom{j}i}
b_j(x),\label{exp}\ee
 where $\gamma^{-1}(x)$ is an
invertible $n\times n$ matrix:
\bb \gamma^{-1}(x)\in GL_n.\ee 
 Both frames
depend differentiably on $x$ and so does $\gamma^{-1}
(x)$, i.e. $\gamma ^{-1}$ is a differentiable function
from $U$ into
$GL_n$. The set of all such functions forms a group
where the multiplication is defined pointwise by the
matrix product. We call this group the $GL_n$ gauge
group
\bb ^UGL_n=\{\gamma:U\rightarrow
GL_n\}.\ee
 A dual frame (or simply frame,
when there is no risk of confusion) is a set of $n\
1$--forms $\beta^1,\beta^2,...,\beta^n$ such that for
every $x\in U$ $\beta^1(x),\beta^2(x),...,\beta^n(x)$
are linearly independent. A frame is called holonomic
if it is of the form $\de x^\mu$ where $x^\mu$ is a
coordinate system.

\noindent {\bf Theorem:} Let $U$ be simply connected.
Then the frame
$\beta^i$ is holonomic if and only if
\bb\de\beta^i=0\ee
 for $i=1,2,...,n$.

A dual frame $\beta^i$ is called dual to a frame $b_i$ if
\bb\beta^i(b_j)=\delta^i_j\ee
 for $i,j=1,2,...,n$ 
and a frame is holonomic if and only if its dual frame
is holonomic. If two frames $b_i$ and $b'_i$ are related
by the gauge transformation $\gamma^{-1}$,
equation (\ref{exp}), their corresponding dual frames are
related by the inverse transposed gauge
transformation:
\bb {\beta'}^i=\sum_j\gamma^i_{\phantom
{i}j}\beta^j,\label{dual}\ee 
transposed because of the
`wrong' order of the indices in equation (\ref{exp}).
Our convention is that the first index of a matrix
counts the rows, the second index the columns,
irrespective of whether the indices are upper or lower.

As an example let us consider three-dimensional polar
coordinates,
$U$ is $\rr^3$ without the $x-z$ half plane:
\bb U=\rr^3-\{(x,y,z),x\geq0,y=0\}.\ee
 Let
$b_i$ be the holonomic frame of Cartesian coordinates,
\bb b_1={\partial\over{\partial x}},\qq
b_2={\partial\over{\partial y}},
\qq b_3={\partial\over{\partial z}},\ee
 and
$b'_i$ the holonomic frame of polar coordinates,
\bb b'_1={\partial\over{\partial r}},\quad
b'_2={\partial\over{\partial\varphi}},\quad
b'_3={\partial\over{\partial\theta}},\ee
 with
\bb x&=&r\cos\varphi\sin\theta\label{polb}\\
y&=&r\sin\varphi\sin\theta\\
z&=&r\cos\theta.\label{pole}\ee
In order to calculate the
gauge transformation $\gamma$ relating the two
frames we use the definition (\ref{def}):
\bb {\partial\over{\partial r}}={{\partial
x}\over{\partial r}} {\partial\over{\partial
x}}+{{\partial y}\over{\partial r}}
{\partial\over{\partial y}}+{{\partial z}\over{\partial
r}} {\partial\over{\partial z}},\ee
 and two
similar identities; $\gamma^{-1}$ is just the Jacobian
matrix of equations (\ref{polb}-\ref{pole}):
\bb\gamma^{-1}=
\pmatrix{\cos\varphi\ \sin\theta& -r\ \sin\varphi
\ \sin\theta& r\ \cos\varphi\ \cos\theta\cr
\sin\varphi\  \sin\theta& r\ \cos\varphi\
\sin\theta& r\ \sin\varphi\ \cos\theta\cr
\cos\theta& 0& -r\ \sin\theta}.\ee
 The
corresponding holonomic dual frames are given by
\bb\beta^1=\de
x,\quad\beta^2=\de y,\quad\beta^3=\de z,\ee
and
\bb{\beta'}^1=\de r,\quad{\beta'}^2=\de \varphi,\quad
{\beta'}^3=\de \theta.\ee
 Using equation (\ref{dual}) we
then find
\bb \de x={{\partial x}\over{\partial r}}\de
r+{{\partial x}\over{\partial
\varphi}}\de \varphi+{{\partial
x}\over{\partial\theta}}\de \theta,\ee 
and similar equations for $\de y$ and $\de z$.

\section{Metrics on a vector space}

Let $V$ be an $n$--dimensional real vector space. A
(pseudo--)metric (or scalar product) on $V$ is a
bilinear form
\bb g:V\times V&\longrightarrow&\rr\cr
(v,w)&\longmapsto& g(v,w)\ee
which is symmetric:
\bb g(v,w)=g(w,v)\quad{\hbox{for all}}\quad v,w\in
V\ee
 and nondegenerate. The last
requirement means that only the zero vector has
vanishing scalar product with all vectors in $V$. If
$b_1,b_2,...,b_n$ is a basis of $V$, then due to the
bilinearity the metric $g$ is uniquely specified by the
$n\times n$ matrix of scalar products of the basis
vectors:
\bb g_{ij}:=g(b_i,b_j).\ee
 The symmetry and
nondegeneracy of $g$ imply that the matrix of
$g$ with respect to the basis is symmetric and
nondegenerate:
\bb g_{ij}&=&g_{ji},\cr
\det(g_{ij})&\not=&0.\ee
 The matrix
${g'}_{ij}$ of the metric $g$ with respect to a different
basis
${b_i}'$,
\bb
b'_i=\sum_i(\gamma^{-1})^j_{\phantom{j}i}b_j,
\ee
is given by
\bb
g'_{ij}:=g(b'_i,b'_j)=(\gamma^{-1T}g\gamma^{-1})_{ij}.
\label{metricch}\ee
Note here that we use $n\times n$ matrices to describe
a change of coordinates as well as a metric, two quite
different mathematical objects.

The following two theorems of linear algebra are of
fundamental importance for us.\hfil\break\noindent
{\bf Theorem (Gram \& Schmidt):} Any metric
has an orthonormal basis
$e_i$, i.e. a basis such that
\bb g(e_i,e_j)=\eta_{ij}:=
\pp{1&&&&&\cr 
&\ddots&&&&\cr 
&&1&&&\cr 
&&&-1&&\cr 
&&&&\ddots&\cr 
&&&&&-1}.
\ee
{\bf Theorem (Sylvester):} The number $r$ of
plus signs and the number $s$ of minus signs, $r+s=n$,
does not depend on the choice of the orthonormal basis
$e_i$.

From now on we shall reserve the letter $e$ for an
orthonormal basis. Of course, an orthonormal basis is
not unique, for instance
\bb e_1={1\choose 0},\quad e_2={0\choose
1},\ee
 and
\bb {e'}_1={{\textstyle\frac{1}{\sqrt2}}}{1\choose
1},\quad {e'}_2={{\textstyle\frac{1}{\sqrt2}}} {-1
\choose +1}\ee are both orthonormal for the Euclidean
metric of
$\rr^2$. In general, given an orthonormal basis $e_i$,
any other basis ${e'}_i$ with
\bb
{e'}_i=\sum_j(\Lambda^{-1})^j_{\phantom{j}i}e_j,
\quad\Lambda\in
GL_n,\ee
 is also orthonormal if and only if
\bb\eta=\Lambda^{-1T}\eta\Lambda^{-1}.\ee
The set of all $\Lambda$'s satisfying this condition
forms a subgroup of $GL_n$, the Lorentz group denoted
by $O(r,s)$. It is of dimension ${1\over 2}n(n-1)$.

There are two ways to parameterize all possible metrics
with given signature $(r,s)$ on $V$.\hfil\break {\bf
(i)} Choose a fixed basis $b_i$ of $V$. Then any metric
is parameterized by the symmetric $n\times n$ matrix
$g_{ij}$ of scalar products, that is
${1\over2}n(n+1)$ real numbers.\hfil\break {\bf (ii)}
Given any metric, choose an orthonormal basis $e_i$.
This basis characterizes the metric as well. With
respect to the fixed basis
$b_i$, the $e_i$ are parameterized by the $n\times n$
matrix $\gamma^{-1}$ consisting of $n^2$ numbers.
However, any other basis obtained from $e_i$ by a
Lorentz rotation describes the same metric. Therefore
we have to subtract from $n^2$ the number of
dimensions of the Lorentz group
${1\over2}n(n-1)$ yielding again
\bb
n^2-{\textstyle\frac{1}{2}}n(n-1)=
{\textstyle\frac{1}{2}}n(n+1).\ee
Being nondegenerate a metric $g$ on a vector space
$V$ induces a canonical metric $g^*$ on the dual
vector space
$V^*$: Let $\beta^i$ be the basis of $V^*$ dual to the
basis $b_i$:
\bb \beta^i(b_j)=\delta^i_j.\ee
 Define a metric on $V^*$ by
\bb g^*(\beta^i,\beta^j)=(g_{ij})^{-1}.\label{sroot}\ee
This metric is canonical, i.e. it does not depend on the
choice of the basis $b_i$.

It follows that the dual basis of an orthonormal basis
$e_i$ of $V$ is itself orthonormal with respect to $g^*$,
because $\eta$ is its own inverse. Attention, in the
following we denote an orthonormal basis of $V^*$ by
$e^i$, only the position of the index distinguishes basis
from dual basis.

\section{Metrics on an open subset of $\rr^n$}

We defined a vector field on an open subset $U$ of
$\rr^n$ as a differentiable family of vectors indexed
by the points $x$ of $U$. Likewise we now define a
metric $g$ on $U$ to be a differentiable family $g_x$ of
vector space metrics. With respect to a frame
$b_i(x)$ this family is described by the symmetric
$n\times n$ matrix
\bb g_{ij}(x):=g_x(b_i(x),b_j(x))\ee
 whose
elements are real valued functions on $U$. For
convenience we shall often suppress the $x$'s in the
following.

Since the orthonormalization procedure by Gram and
Schmidt only involves addition,  multiplication and
division, that is differentiable operations, it also
immediately guarantees the existence of orthonormal
frames $e_i(x)$,
\bb g_x(e_i(x),e_j(x))=\eta_{ij},\ee
 with $x$-independent rhs.

A frame may now have two nice properties: being
holonomic or being orthonormal. As often in life we
can have both only in trivial situations.\hfil\break
{\bf Theorem:} An open subset $U$ of $\rr^n$
admits a holonomic and orthonormal frame if and only
if it is flat.

We do not yet have a definition of flatness, but it is
sufficient to take the naive sense of the word, for
instance meaning that the angles of a triangle add up
to $180^{\rm o}$.

Let us return to our example of $\rr^3$ minus a half
plane and endow it with the Euclidean metric
\bb
g_{ij}=\left(\matrix{1&0&0\cr0&1&0\cr0&0&1\cr}
\right)\ee
with respect to the Cartesian holonomic frame, which
is therefore also orthonormal. On the other hand, the
polar holonomic frame is not orthonormal:
\bb g'_{ij}=(\gamma^{-1T}\,1\,\gamma^{-1})_{ij}=
\left(\matrix{1&0&0\cr 0&r^2\sin ^2\theta&0\cr
0&0&r^2\cr}
\right),\ee
 or in the dual frame
\bb g^{ij}=\left(\matrix{1&0&0\cr 0&1&0\cr
0&0&1\cr}\right)\ee
 and
\bb {g'}^{ij}=\left(\matrix{1&0&0\cr
0&r^{-2}\sin^{-2}\theta&0\cr
0&0&r^{-2}\cr}\right).\ee
 To have a non-flat
example consider a piece of the unit sphere, $r=1$. It is
an open subset of $\rr^2$ parameterized by $\varphi$
and $\theta$. Its metric is given by
\bb g^{ij}=\left(\matrix{{{\sin^{-2}\theta}}&0\cr
0&1\cr }\right)\ee
 with respect to the
holonomic frame $\de \varphi,\de \theta$. An
orthonormal frame is for instance
\bb e^1=\sin\theta \de \varphi\ ,\
e^2=\de \theta.\ee
 It is not holonomic:
\bb\de e^1=\de (\sin\theta d\de\varphi)=\cos\theta
\de \theta\wedge\de\varphi\not=0.\ee
 We will show in section \ref{farewell} that the sphere
is not flat and the above theorem then implies that
there is no holonomic and orhonormal frame on the
sphere. 
 
\section{Hodge star}

The Hodge star is a map turning a $p$--form into an
$(n-p)$--form. We define it in terms of a holonomic
frame:
\bb *:\Omega^p
U&\longrightarrow&\Omega^{n-p}U\cr
\varphi&\longmapsto&*\varphi\eee
\bb
*\varphi&:=&{1\over{(n-p)!}}
\sum_{\mu_{p+1}\cdots\mu_n}
\left[{1\over{p!}}\sum_{\mu_1..\mu_p}
\epsilon_{\mu_1\cdots\mu_n}\sqrt{\mid
\det g_{\cdot\cdot}\mid}\right.\cr &&\left. \times
\sum_{\nu_1\cdots\nu_p}\varphi_{\nu_1\cdots\nu_p}
g^{\mu_1\nu_1}\cdots
g^{\mu_p\nu_p}\right]
\de x^{\mu_{p+1}}\wedge\cdots\wedge
\de x^{\mu_n},\ee
 where
$\epsilon_{\mu_1}..._{\mu_n}$ is the completely
antisymmetric tensor with
\bb\epsilon_{1...n}=1.\ee
 Note that this
definition requires the choice of an orientation in
$\rr^n$, but does not depend on the particular
coordinate system used. Just as the wedge product the
Hodge star is a purely algebraic operation. It is linear
and its square is plus or minus the identity:
\bb**\varphi=(-1)^{p(n-1)+s}\varphi.\ee
Recall that $s$ is the number of minus signs in the
metric. Note that the Hodge star has a particularly
simple expression in an orthonormal frame.

\section{Coderivative and Laplace operator}

Just as the exterior derivative, the coderivative is a
linear first order differential operator which however
lowers the degree of a differential form by one unit:
\bb\delta:\Omega^p
U&\longrightarrow&\Omega^{p-1}U\cr
\varphi&\longmapsto&
\delta\varphi:=(-1)^{np+n+1+s}*
\de *\varphi.\ee
 It inherits nilpotency from the exterior derivative:
$\delta^2=0.$

If $U$ is `compact' and if the metric has Euclidean
signature, then
$\Omega U$ is a pre-Hilbert space with scalar product
\bb (\kappa,\varphi):= \int_U
\kappa\wedge*\varphi,
\ee for two differential forms $\kappa,\ \varphi$ of
equal degree. The scalar product vanishes if  the
degrees are not equal. In this situation, the
coderivative is the formal adjoint of the exterior
derivative.

In general, the Laplace operator is the linear second
order differential operator defined by:
\bb
\Delta:=-(d\delta+\delta
d)\,:\,\Omega^pU\rightarrow\Omega^pU.\ee
  If the metric is
Euclidean, the Laplace operator is Hermitean. If
the metric is indefinite, the Laplace operator is usually
called wave or d'Alembert operator and written as
$\Box$.

\section{Summary}

Before returning to physics, let us
summarize: We have recast a part of tensor
analysis in a coordinate free language using
differential forms. This serves two
purposes:
\begin{itemize}
\item
They carry less indices, making
some calculations more transparent.
\item 
Being coordinate independent they can easily be
generalized to more general spaces like manifolds.
\end{itemize}
May be, the
following dictionary can be useful:
\def\z{\noalign{\bigskip}}
$$\vbox{\tabskip=1cm
\halign{$#$\hfil&\hfil$#$\hfil\cr v&v^\mu\cr\z
\varphi\in\Omega^pU&\varphi_{\lbrack\mu_1\cdots\mu_
p\rbrack}\cr\z
\varphi(v_1, v_2,\cdots, v_p)&\sum
_{\mu_1\cdots\mu_p}\varphi
_{\lbrack\mu_1\cdots\mu_p\rbrack}v_1^{\mu_1}\cdots
v_p^{\mu_p}\cr\z
\varphi\wedge \psi&
\varphi_{\lbrack\mu_1\cdots\mu_p}
\quad \psi_{\mu_{p+1}\cdots\mu_q\rbrack}\cr\z
d\varphi&\partial_{\lbrack \mu_1} \varphi_{\mu_2
\cdots\mu_{p+1}
\rbrack}\cr \int_K\ \varphi&\int_K\ \varphi
_{1\cdots p}\ dx^1\cdots dx^p\cr\z g & g_{(ij)}\cr\z
g^*& g^{ij}=(g^{-1})_{ij}\cr\z
*\ \varphi&
\sum_{{\mu_1\cdots
\mu_p}\atop{\nu_1\cdots\nu_p}}\sqrt {\vert
\det g_{\cdot\cdot}\vert}\
\varphi_{\nu_1\cdots \nu_p} g^{\nu_1\mu_1}\cdots
g^{\nu_p\mu_p}
\epsilon_{\mu_1\cdots\mu_p,\mu_{p+1}\cdots\mu_n}\cr\z
-d* d*-* d* d& \Delta \cr\z}}$$

\section{Maxwell's equations}

Consider Minkowski space $U=\rr^4$ equipped
with the Minkowski metric of signature $+---$. We
subscribe again to Einstein's summation convention,
(summing over indices that appear twice). The
sources, electric charge and current densities, are
combined into a real valued 3-form:
\bb
j={1\over{3!}}\epsilon_{\mu\nu\lambda\rho}j^\mu
\de x^\nu
\wedge \de x^\lambda\wedge \de x^\rho\quad\in\
\Omega^3(\rr^4).\ee
 Integrating $j$ over a
3-dimensional space-like volume yields the total
charge inside that volume as a function of time.
Charge conservation reads
\bb\de j=0.\ee
 The field strength is a real valued 2-form
\bb
F={\textstyle\frac{1}{2}}F_{\mu\nu}\de x^\mu\wedge
\de x^\nu.\ee
Then
Maxwell's equations read:
\bb \de F&=&0,\label{maxh}\\  \delta
F&=&{\textstyle\frac{1}{\epsilon_0c^2}}\,*j.
\label{maxi}\ee
equation (\ref{maxi}) implies charge conservation.
Therefore only conserved currents,
$\de j=0$, may be coupled to the electromagnetic field.
Our spacetime being simply connected,
equation (\ref{maxh}) implies the existence of a potential, a
real valued 1-form $A$ such that
\bb F=\de A.\ee
 Expressed in terms of the
potential, equation (\ref{maxi}) can be obtained from
the action
\bb
S[A]=-\int_{\rr^4}\left(\,{\textstyle
\frac{\epsilon_0c}{2}}\,
F\wedge*F+\,{\textstyle\frac{1}{c}}\,
j\wedge A\right)\label{maxa}\ee
 upon variation of the
potential. This means we replace $A$ by $A+a$ in the
action, expand it and put the term linear in $a$ equal
to zero. Note that if spacetime was Euclidean, Maxwell's
action in the vacuum would be simply $S=
{\textstyle\frac{\epsilon_0}{2}}(F,F)$. This
will be Connes' starting point.

Writing Maxwell's theory with differential forms has
four advantages:
\begin{itemize}\item
Lorentz invariance is immediate; $SO(1,3)$, the
group of linear transformations preserving the metric
and the orientation of
$\rr^4$, also leaves the Hodge star and consequently
the Maxwell action (\ref{maxa}) invariant.
\item
In Maxwell's equations or in the action the flat
Minkowski metric may be replaced by any curved
metric. This tells us  how electromagnetism couples
to gravity.
\item
Gauge invariance now reads
\bb \rho_V(g)A=A+i{\textstyle\frac{\hbar}{e}}
g\de g^{-1}=A+
{\textstyle\frac{\hbar}{e}}\de\theta,\qq
g=\exp {i\theta}\ \in\  ^{\rr^4}U(1).\ee 
Its abelian group
$U(1)$  may easily be generalized to a non-Abelian,
compact Lie group. One then gets the celebrated
Yang-Mills theories.
\item
The invariance of  the action under diffeomorphisms
is manifest. They form a semidirect product with the
gauge group, here:
\bb {\rm Diff}(M)\semi\,^{\rr^4}U(1). \ee
 \end{itemize}

%% file: monsa3
\chapter{Yang-Mills-Higgs theories}

To get started we describe Yang-Mills-Higgs theories
as a black box or better as a slot machine. There are
 four slots for four bills. Once you have decided which
bills you choose and entered them, a certain number of
small slots will open for coins. Their number depends
on the choice of bills. You make your choice of coins,
feed them in, and the machine starts working. It
produces as output a complete particle
phenomenology: the particle spectrum with their
quantum numbers, cross sections, life times,
branching ratios. You compare the phenomenology to
experiment to find out whether your input wins or
loses.

\section{The bills}

The first bill is a finite dimensional, real, compact Lie
group $G$. The gauge bosons, spin 1, will live in its
adjoint representation whose Hilbert space is the
complexified of the Lie algebra $\gg$.

The remaining bills are three unitary
representations of $G$, $\rho_L,\ \rho_R,\ \rho_S$
defined on the complex Hilbert spaces, $\hh_L,\
\hh_R,\
\hh_S$. They classify the left- and right-handed
fermions, spin ${\textstyle\frac{1}{2}}$, and the
scalars, spin 0. The group $G$ is chosen compact to
ensure that the unitary representations are finite
dimensional, we want a finite number of different
Lego bricks.

\section{The coins}

The coins are numbers, coupling constants, more
precisely coefficients of invariant polynomials. We 
need an invariant scalar product on $\gg$. The set of
all these scalar products is a cone and the gauge
couplings are particular coordinates of this cone. If
the group is simple, say $G=SU(n)$, then the most
general, invariant scalar product is 
\bb (X,X')={\textstyle\frac{2}{g_n^2}}\t [X^*X'],\qq
X,X'\in su(n).\ee
If $G=U(1)$ we have
\bb (Y,Y')={\textstyle\frac{1}{g_1^2}}\bar YY',\qq
Y,Y'\in u(1).\ee
Mind the different normalizations, they are
conventional. The $g_n$ are positive numbers, {\it the
gauge couplings.} For every simple factor of $G$ there
is one gauge coupling. 

Then we need the Higgs potential $V(\varphi)$. It is
an invariant, fourth order, stable polynomial on
$\hh_S\owns\varphi$. Stable means bounded from
below. For $G=SU(2)$ and the Higgs scalar in the
fundamental or defining representation,
$\varphi\in\hh_S=\cc^2$, $\rho_S(g)=g$, we have
\bb V(\varphi)=\lambda\,(\varphi^*\varphi)^2-
{\textstyle\frac{1}{2}}\mu^2\,\varphi^*\varphi.\ee
The coefficients of the Higgs potential are the Higgs
couplings, $\lambda$ must be positive for stability. We
say that the potential breaks $G$ spontaneously if its
minimum is not a trivial orbit under $G$. In our
example, if $\mu$ is positive the minimum of 
$V(\varphi)$ is the 3-sphere $|\varphi|=v:=
{\textstyle\frac{1}{2}}\mu/\sqrt\lambda$. $v$ is
called vacuum expectation value and $SU(2)$ is said to
break down spontaneously to its little group $U(1)$.
The little group leaves invariant any given point of
the minimum, e.g. $\varphi=\,^t\!(v,0)$. On the other
hand if $\mu$ is purely imaginary, then the
minimum of the potential is the origin, no
spontaneous symmetry breaking.

Finally, we need the Yukawa couplings $g_Y$. They
are the coefficients of the most general trilinear
invariant on $\hh_L^*\,\ot\,\hh_R\,\ot\,(
\hh_S\op\hh_S^*)$. For every 1-dimensional
invariant subspace in the reduction of this tensor
representation, we have one complex Yukawa
coupling. 

We will see that, if the symmetry is broken
spontaneously, gauge and Higgs bosons acquire masses
related to the Higgs couplings, fermions
acquire masses related to the Yukawa couplings.

\section{The winner}

Physicists have spent some thirty years and billions of
Swiss Francs playing on the slot machine by
Yang-Mills \& Higgs. There is a winner, the standard
model of electro-weak and strong interactions. Its
bills are
\bb G &= & SU(3) \times SU(2) \times U(1) \cr \cr 
\hh_L &=& \bigoplus_1^3\lb
(3,2,{\textstyle\frac{1}{6}})\op
(1,2,-{\textstyle\frac{1}{2}})
\rb  \label{hl},\\  
 \hh_R& = &\bigoplus_1^3\lb 
(3,1,{\textstyle\frac{2}{3}})\oplus
(3,1,-{\textstyle\frac{1}{3}})\op (1,1,-1)
\rb, \label{hr}
\\ \cr    
 \hh_S &= &(1,2,-{\textstyle\frac{1}{2}})
\label{hs},\ee   
where $(n_3,n_2,y)$
denotes the tensor product of an $n_3$ dimensional
representation of $SU(3)$, an $n_2$ dimensional
representation of $SU(2)$ and the one dimensional
representation of $U(1)$ with hypercharge $y$:  
$\rho(\exp (i\theta)) = \exp (iy\theta) $. For
historical reasons the hypercharge is an integer
multiple of ${\textstyle\frac{1}{6}}$.
This is irrelevant: only the product of the
hypercharge by its gauge coupling is measurable.
 In the direct sum, we recognize the three
generations of fermions, the quarks are $SU(3)$
colour triplets, the leptons colour singlets. The basis
of the fermion representation is the periodic table,
\bb \pp{u\cr d}_L,\ \pp{c\cr s}_L,\ \pp{t\cr b}_L,\ 
\pp{\nu_e\cr e}_L,\ \pp{\nu_\mu\cr\mu}_L,\ 
\pp{\nu_\tau\cr\tau}_L\eee
\bb\matrix{u_R,\cr d_R,}\qq \matrix{c_R,\cr s_R,}\qq
\matrix{t_R,\cr b_R,}\qq  e_R,\qq \mu_R,\qq 
\tau_R\eee
The
parentheses indicate isospin doublets.

 We recognize the eight gluons in
$su(3)$. Attention, the $U(1)$ is not the one of electric
charge, it is called hypercharge, the electric charge
is a linear combination of hypercharge and weak
isospin, parameterized by the weak mixing angle
$\theta_w$ to be introduced below. This mixing is
necessary to give electric charges to the $W$ bosons.
The $W^+$ and $W^-$ are pure isospin states, while the
$Z^0$ and the photon are (orthogonal) mixtures of the
third isospin generator and hypercharge. 

 Because of the high
degree of reducibility in the bills, there are many
coins, among them 27 complex Yukawa couplings. Not
all of them have a physical meaning. They can be
converted into 18 physically significant, positive
numbers
\cite{data}, three gauge couplings, 
\bb g_3=1.218\,\pm\,0.026,\qq
g_2=0.6567\,\pm\,0.0007,\qq
g_1=0.3575\,\pm\,0.0001,\ee
two Higgs couplings, $\lambda$ and $\mu$, and 
13 positive parameters from the Yukawa couplings.
The Higgs couplings are related to the boson masses:
\bb m_W&=&{\textstyle\frac{1}{2}}g_2\,v
\,=\,80.33\,\pm\,.15\ {\rm GeV},\\
m_Z&=&{\textstyle\frac{1}{2}}\sqrt{g_1^2+g_2^2}\ v
=m_W/\cos\theta_w
\,=\,91.187\,\pm\,.007\ {\rm GeV},\\
m_H&=&2\sqrt 2\sqrt\lambda\,v\,>\,65\ {\rm GeV},\ee
with the vacuum expectation value
$v:={\textstyle\frac{1}{2}}\mu/\sqrt\lambda$ and the
weak mixing angle $\theta_w$ defined by
\bb\sin^2\theta_w:=g_2^{-2}/(g_2^{-2}+g_1^{-2})=
0.2315\,\pm\,0.0005.\ee
 For
the standard model, there is a one--to--one
correspondence between the physically relevant part
of the Yukawa couplings and the fermion masses and
mixings,
\bb m_e=0.51099906\pm 0.00000015\ {\rm MeV},&
m_u=5\pm 3\ {\rm MeV},&m_d=10\pm 5\ {\rm MeV},
\cr  m_\mu=0.105658389\pm 0.000000034\ {\rm GeV},&
m_c=1.3\pm 0.3\ {\rm GeV},&m_s=0.2\pm 0.1\ 
{\rm GeV},\cr 
m_\tau=1.7771 \pm 0.0005\ {\rm GeV},&
m_t=175\pm 6\ {\rm GeV},&m_b=4.3\pm 0.2\ 
{\rm GeV},\eee
Since the neutrinos are massless, the mixing only
occurs for quarks and is given by
a unitary matrix, the Cabibbo-Kobayashi-Maskawa
matrix 
\bb C_{KM}:=\pp{V_{ud}&V_{us}&V_{ub}\cr 
V_{cd}&V_{cs}&V_{cb}\cr  V_{td}&V_{ts}&V_{tb}}.\ee
For physical purposes it can be parameterized by
three angles $\theta_{12}$,  
$\theta_{23}$, $\theta_{13}$ and
one $CP$ violating phase $\delta$:
\bb C_{KM}=\pp{
c_{12}c_{13}&s_{12}c_{13}&s_{13}e^{-i\delta}\cr 
-s_{12}c_{23}-c_{12}s_{23}s_{13}e^{i\delta}&
c_{12}c_{23}-s_{12}s_{23}s_{13}e^{i\delta}&
s_{23}c_{13}\cr 
s_{12}s_{23}-c_{12}c_{23}s_{13}e^{i\delta}&
-c_{12}s_{23}-s_{12}c_{23}s_{13}e^{i\delta}&
c_{23}c_{13}},\eee
with $c_{kl}:=\cos \theta_{kl}$, 
$s_{kl}:=\sin \theta_{kl}$.
The
absolute values of the matrix elements are:
\bb \pp{
0.9753\pm 0.0006&0.221\pm 0.003&0.004\pm
0.002\cr 
0.221\pm 0.003&0.9745\pm 0.0007&0.040\pm 0.008\cr 
0.010\pm 0.006&0.039\pm 0.009&0.9991\pm 0.0004}.
\eee 
The physical meaning of the quark mixings is the
following: when a sufficiently energetic $W^+$ decays
into a $u$ quark, this
$u$ quark is produced together with a
$\bar d$ quark with probability $|V_{ud}|^2$, together 
with a
$\bar s$ quark with probability $|V_{us}|^2$, together
with a
$\bar b$ quark with probability $|V_{ub}|^2$. The
fermion masses and mixings together are an entity,
the fermionic mass matrix or the matrix of Yukawa
couplings multiplied by the vacuum expectation
value. 

Let us note
four important properties of the standard model.
\begin{itemize}\item
The gluons couple in the same way to left- and
right-handed fermions, the gluon coupling is
vectorial, strong interaction do not break parity.
\item
 The scalar is a colour singlet, the
$SU(3)$ part of $G$ does not suffer spontaneous break
down, the gluons remain massless.
\item
The $SU(2)$ couples only to left-handed fermions, its
coupling is chiral, weak interaction break parity
maximally.
\item
The scalar is an isospin doublet, the $SU(2)$ part
suffers spontaneous break down, the $W^\pm$ and the
$Z^0$ are massive.
\end{itemize}

\section{The rules} \label{rules}

It is time to open the slot machine and to see how it
works. Its mechanism falls into five pieces:

\medskip
\noindent 
{\bf The Yang-Mills action:}
\begin{itemize}\item
{ \qq Maxwell \ + \ non-Abelian gauge \ \
=
\ \ Yang-Mills.  }
\end{itemize}
 The actor in this piece is
$A$ called a connection, gauge potential, gauge bosons
or Yang-Mills field. It is a 1-form on spacetime $M$
with values in the Lie algebra $\gg$,
\bb A\in\Omega^1(M,\gg).\ee
We define its curvature or field strength,
\bb F:=\de A+{\textstyle\frac{1}{2}}[A,A]\ \in
\Omega^2(M,\gg),\ee
and the Yang-Mills action,
\bb
S_{YM}[A]=-{\textstyle\frac{1}{2}}\int_M(F,*F).\ee
The space of all connections carries an affine
representation $\rho_V$ of the gauge group
$^MG\owns g$:
\bb\rho_V(g)A=gAg^{-1}+g\de
g^{-1}.\label{inhom}\ee 
Restricted to $x$-independent
gauge transformation, the representation is linear,
the adjoint one. The field strength transforms
homogeneously under any gauge transformation,
\bb \rho_V(g)F=gFg^{-1},\ee 
and, as the scalar product $(\cdot,\cdot)$ is invariant,
the Yang-Mills action is gauge invariant,
\bb S_{YM}[\rho_V(g)A]=S_{YM}[A] \qq{\rm for\ all}
\ g\in ^MG.\ee
Note that a mass term for the gauge bosons,
\bb{\textstyle\frac{1}{2}}\int_M m^2_A(A,*A),\ee
is not gauge invariant because of the inhomogeneous
term in the transformation law of a connection
(\ref{inhom}). Gauge invariance forces the gauge
bosons to be massless. 

In the Abelian case $G=U(1)$, the Yang-Mills action is
nothing but Maxwell's action, quantum
electro-dynamics $(QED)$. Note however, that now the
vector potential is purely imaginary, while
conventionally, in Maxwell's theory it is chosen real.
For $G=SU(3)$ and $\hh_L=\hh_R=\cc^3$ we have
today's theory of strong interaction, quantum
chromo-dynamics $(QCD)$. 

\medskip
\noindent
{\bf The Dirac action:}
 Schr\"odinger's action is
non-relativistic. Dirac generalized it to be Lorentz
invariant, e.g. \cite{bd}. The price to be paid is
twofold. His generalization only works for
spin ${\textstyle\frac{1}{2}}$ particles and requires
that for every such particle there must be an
antiparticle with same mass and opposite charges.
Therefore Dirac's wave function $\psi(x)$ takes
values in $\cc^4$, spin up, spin down, particle,
antiparticle. Antiparticles have been discovered and
Dirac's theory was celebrated. Here it is in short for
(flat) Minkowski space of signature $+---$. Define the
four Dirac matrices,
\bb \gamma^0:=\pp{
1&0&0&0\cr 0&1&0&0\cr 0&0&-1&0\cr 0&0&0&-1},&&
\gamma^1:=\pp{
0&0&0&1\cr 0&0&1&0\cr 0&-1&0&0\cr -1&0&0&0},\\
\gamma^2:=\pp{
0&0&0&-i\cr 0&0&i&0\cr 0&i&0&0\cr -i&0&0&0},&&
\gamma^3:=\pp{
0&0&1&0\cr 0&0&0&-1\cr -1&0&0&0\cr 0&1&0&0}.\ee
They satisfy the anticommutation relations,
\bb\gamma^\mu\gamma^\nu+
\gamma^\nu\gamma^\mu=2\eta^{\mu\nu}1_4.\ee
In even spacetime dimensions, the chirality,
\bb \gamma_5:={\textstyle\frac{i}{4!}}\epsilon_{
\mu\nu\rho\sigma}\gamma^\mu\gamma^\nu
\gamma^\rho\gamma^\sigma=
i\gamma^0\gamma^1\gamma^2\gamma^3=\pp{
0&0&1&0\cr 0&0&0&1\cr 1&0&0&0\cr 0&1&0&0}\ee
is a natural operator and it paves the way to an
understanding of the chirality in weak interactions.
The chirality is a unitary matrix of unit square that
anticommutes with all four Dirac matrices.
$(1-\gamma_5)/2$ projects on the left-handed part,
$(1+\gamma_5)/2$ projects on the right-handed part.
The chirality applied to a left-handed spinor produces
its right-handed part. Similarly, there is the charge
conjugation, an anti-unitary operator of unit square,
that applied on a particle $\psi$ produces its
antiparticle 
\bb\psi^c:=i\gamma^2\psi^*.\ee
Here $\cdot^*$ denotes complex conjugation. 
The charge conjugation commutes with all four Dirac
matrices. In flat spacetime, the free Dirac operator is
simply defined by,
\bb \ddd:=i\hbar\gamma^\mu\pa_\mu.\ee
It is sometimes referred to as square root of the wave
operator because $\ddd^2=-\Box$.
 The coupling of the
Dirac spinor to the gauge potential $A=A_\mu\de
x^\mu$ is done via the covariant derivative, and called
minimal coupling. In order to break parity we write
left- and right-handed part independently:
\bb S_D[A,\psi_L,\psi_R]&=&
\int_M\bar\psi_L\left[\ddd+i\hbar
\gamma^\mu\tilde\rho_L(A_\mu)\right]
\,\frac{1-\gamma_5}{2}\,\psi_L\,\de^4x\cr 
&&+\int_M\bar\psi_R\left[\ddd+i\hbar
\gamma^\mu\tilde\rho_R(A_\mu)\right]
\,\frac{1+\gamma_5}{2}\,\psi_R\,\de^4x.\ee
The new actors in this piece are $\psi_L$ and $\psi_R$,
two multiplets of Dirac spinors or fermions, that is
vectors in the Hilbert spaces $\cc^4\ot\hh_L$ and
$\cc^4\ot\hh_R$. We use the notations,
$\bar\psi:=\psi^*\gamma^0$, where $\cdot^*$ denotes
the dual with respect to the scalar product in the
(internal) Hilbert space
$\hh_L$ or $\hh_R$. The $\gamma^0$ is needed for
energy reasons and for invariance of the
pseudo--scalar product of spinors under (covered)
Lorentz transformations. The
$\gamma^0$ is absent if spacetime is Euclidean. Then
we have a genuine scalar product and the square
integrable  spinors form a Hilbert space $\lll^2(\sss)$, the
infinite dimensional brother of the internal one. The
Dirac operator is then self-adjoint in this Hilbert
space. We denote by
$\tilde\rho_L$ the Lie algebra
representation in $\hh_L$.
 The covariant derivative,
$\dee_\mu:=\pa_\mu+\tilde\rho_L(A_\mu)$, deserves
its name,
\bb
\left[\pa_\mu+\tilde\rho_L(\rho_V(g)A_\mu)\right]
(\rho_L(g)\psi_L)=\rho_L(g)
\left[\pa_\mu+\tilde\rho_L(A_\mu)\right]\psi_L,\ee
for all gauge transformations $g\in^MG$. This
ensures that the Dirac action is gauge invariant.

If $\hh_L=\hh_R$ we may add a mass term
\bb -c\int_M\bar\psi_R\, m_\psi
\,\frac{1-\gamma_5}{2}\,\psi_L\,\de^4x\ -\ 
c\int_M\bar\psi_L\, m_\psi
\,\frac{1+\gamma_5}{2}\,\psi_R\,\de^4x
\ =\ -c\int_M\bar\psi\, m_\psi
\,\psi\,\de^4x\ee
 to the Dirac action. It gives identical masses to all
members of the multiplet. 
The fermion masses are gauge invariant if all
fermions in $\hh_L=\hh_R$ have the same mass.
Remember that gauge invariance forces gauge bosons
to be massless. Here it is parity {\it non-}invariance
that forces fermions to be massless.

 Let us conclude by reviewing 
briefly why the Dirac equation is the Lorentz
invariant generalization of the Schr\"odinger
equation. Take the free Schr\"odinger equation on
(flat)
$\rr^4$ it is a linear differential equation with
constant coefficients,
\bb \left(\,\frac{2m}{i\hbar}\,\frac{\pa}{\pa
t}\,-\Delta\right)\psi=0.\label{deB}\ee
We compute its polynomial following Fourier and de
Broglie,
\bb -\,\frac{2m}{\hbar}\,\omega+k^2=
-\,\frac{2m}{\hbar
^2}\,\left[E-\,\frac{p^2}{2m}\,\right].\ee
Energy conservation in Newtonian mechanics is
equivalent to the vanishing of the polynomial.
Likewise, the polynomial of the free, massive Dirac
equation $(\ddd -cm_\psi)\psi=0$ is 
\bb {\textstyle\frac{\hbar}{c}}\,\omega\gamma^0
+\hbar\,k_j\gamma^j-c\,m1.\ee
Putting it to zero implies energy conservation in
special relativity,
\bb ({\textstyle\frac{\hbar}{c}})^2\,\omega^2
-\hbar^2\,\vec k^2-c^2\,m^2=0.\ee
In short
\begin{itemize}\item
{ \qq Schr\"odinger \ + \ Minkowskian geometry \ \ =
\ \ Dirac.  }
\end{itemize}

So far we have seen the two noble pieces, Yang-Mills
and Dirac. Their noblesse has even convinced
mathematicians, Donaldson has used a
non-Abelian Yang-Mills theory to discover exotic
differential structures on $\rr^4$ and the Dirac
operator has been elected differential operator of the
decade by Atiyah \& Singer. I feel that these two
actions deserve the comparison with the circles of
planetary motion and we are ready for the epicycles,
the other three pieces are indeed cheap copies of the
circles with the gauge boson $A$ replaced by a scalar 
$\varphi$. We need these three epicycles to cure only
one problem, give masses to some gauge bosons and to
some fermions. These masses are forbidden by gauge
invariance and parity violation.  To simplify the
notation we will work from now on in units with
$c=\hbar=1$.

\medskip
\noindent
{\bf The Klein-Gordon action:}
The Yang-Mills action contains the kinetic term for
the gauge boson. This is simply the quadratic term,
$(\de A,\de A)$ that by Euler-Lagrange produces
linear field equations. We copy this for our new actor,
a multiplet of scalar fields or Higgs bosons, 
\bb \varphi\in\Omega^0(M,\hh_S), \ee
by writing the Klein-Gordon action,
\bb S_{KG}[A,\varphi]={\textstyle\frac{1}{2}}
\int_M (\dee \varphi)^**\dee \varphi,\ee
with the covariant derivative here defined with
respect to the scalar representation,
\bb\dee\varphi:=\de
\varphi+\tilde\rho_S(A)\varphi.\ee
Again we need this minimal coupling 
$\varphi^*A\varphi$ for gauge invariance. 

\medskip
 \noindent
{\bf The Higgs potential:}
The non-Abelian Yang-Mills action contains
interaction terms for the gauge bosons, a bounded,
invariant, fourth order polynomial, $2(\de
A,[A,A])+([A,A],[A,A])$. We mimic these interactions
for scalar bosons by adding the integrated Higgs
potential $\int_M*V(\varphi)$ to the action.

\medskip
\noindent
{\bf The Yukawa terms:}
We also mimic the (minimal) coupling of the gauge
boson to the fermions $\psi^*A\psi$ by writing all
possible trilinear invariants,
\bb S_Y[\psi_L,\psi_R,\varphi]:={\rm Re}
\int_M*\left(
\sum_{j=1}^ng_{Yj}\left(\psi_L^*,\psi_R,\varphi
\right)_j+\sum_{j=n+1}^mg_{Yj}\left(\psi_L^*,\psi_R,
\varphi^*\right)_j\right)
.\ee
In the standard model, there are 27 complex Yukawa
couplings, $m=27$.

\medskip

\begin{figure}[hbt]
\hspace{2 cm}
\def\epsfsize#1#2{0.6#1}
\epsfbox{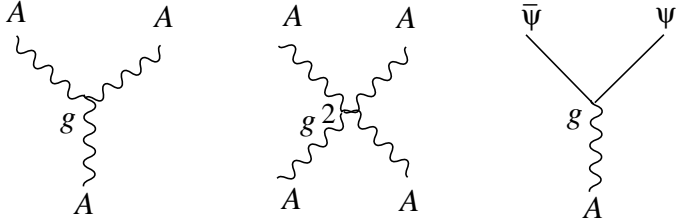}
\caption{The tri- and quadrilinear gauge couplings
and the minimal gauge coupling to fermions}
\label{couplings}
\end{figure}

The two circles, Yang-Mills and Dirac, contain three
types of couplings, a trilinear self coupling $AAA$, a
quadrilinear self coupling $AAAA$ and a the trilinear
minimal coupling $\psi^*A\psi$. The gauge self
couplings are absent if the group $G$ is Abelian, the
photon has no electric charge, Maxwell's
equations are linear. 

 The beauty of gauge
invariance is that if $G$ is simple, all these couplings
are fixed in terms of one positive number, the gauge
coupling $g$. To see this, take an orthonormal basis
$T_b,\ b=1,2,...\dim G$ of the complexified $\gg^\cc$ of the
Lie algebra with respect to the invariant scalar
product and an orthonormal basis $F_k,\
k=1,2,...\dim\hh_L$ of the fermionic Hilbert space, say
$\hh_L$, and develop the actors,
\bb A =: A_\mu^b T_b \de x^\mu,\qq
\psi=:\psi^kF_k.\ee 
Insert these expressions into the Yang-Mills and Dirac
actions, then you get the following interaction terms,
figure \ref{couplings},
\bb g\,
\pa_\rho A_\mu^aA_\nu^bA_\sigma^c\,f_{abc}\,
\epsilon^{\rho\mu\nu\sigma},\qq
g^2\,A_\mu^aA_\nu^bA_\rho^cA_\sigma^d\,{f_{ab}}^e
{f_{ecd}}\,
\epsilon^{\rho\mu\nu\sigma},\qq
g\,\psi^{k*}A^b_\mu\gamma^\mu\psi_\ell\,
{{t_b}_k}^\ell,\ee
with the structure constants ${f_{ab}}^e$,
\bb [T_a,T_b]=:{f_{ab}}^eT_e.\ee
The indices of the structure constants are raised and
lowered with the matrix of the invariant scalar
product in the basis $T_b$, that is the identity matrix. 
The ${{t_b}_k}^\ell$ is the matrix  of the operator
$\tilde\rho_L(T_b)$ with respect to the basis $F_k$.
The difference between the circles and the epicycles
is that the Higgs couplings, $\lambda$ and $\mu$ in
the standard model, and the Yukawa couplings
$g_{Yj}$ are arbitrary, are neither connected among
themselves nor connected to the gauge couplings
$g_i$. 

The standard model is the most painful humiliation of
physics today. The humiliation has four levels:
\begin{itemize}\item
The rules of the Yang-Mills-Higgs model building kit
contain three epicycles.
\item
The winning bills are unmotivated except for the
$U(1)$ coming from quantum mechanics.
\item
The winning coins are numerous, 18, and beg for an
understanding.
\item
The theory of gravity is completely different from the
Yang-Mills description of the electro-weak and strong
forces. The underlying group of gravity is the group
of diffeomorphisms of spacetime, Diff$(M)$, that
formalizes the coordinate transformations. This group
is not a Lie group. Any attempt to unify all four forces
has failed so far.
\end{itemize}
Nevertheless, and this makes the humiliation painful,
the standard model reproduces correctly millions of
experimental numbers that cost billions of Swiss
Francs. Every anomaly free Yang-Mills-Higgs model,
in particular the standard model, is renormalizable.
Renormalizable theories are rare and therefore
precious. Connes has shown that noncommutative
geometry eases the humiliation on all four levels.

\section{An example} \label{example}

We illustrate this chapter with the current model of
electro-weak interactions for one generation of
leptons. This is the Glashow-Salam-Weinberg model, a
submodel of the standard model. There are simpler
examples on the market, in particular models not
containing a $U(1)$ factor. Mathematically the $U(1)$
is so degenerate that it makes some computations
perfidious. 
\bb G&=&SU(2)\times U(1)\ \owns\  (a,b),\qq
\gg\ =\ su(2)\op u(1)\  \owns\ (X,Y),\\
((X,Y),(X',Y'))&=&{\textstyle\frac{2}{g_2^2}}\,\t
(X^*X') \ +\ {\textstyle\frac{1}{g_1^2}}\,\bar Y Y',\\
\hh_L&=&\cc^2\ \owns\ \psi_L,\qq
\rho_L(a,b)=ab^{y_L},\qq y_L=
-{\textstyle\frac{1}{2}},\\
\hh_R&=&\cc\ \owns\ \psi_R,\ \qq
\rho_R(a,b)=b^{y_R},
\qq y_R=-1,\\
\hh_S&=&\cc^2\ \owns\ \varphi,\qq
\rho_S(a,b)=ab^{y_S}, \qq
y_S=-{\textstyle\frac{1}{2}},\\
V(\varphi)&=&\lambda\,(\varphi^*\varphi)^2-
{\textstyle\frac{1}{2}}\mu^2\,\varphi^*\varphi,\\
\lll_Y&=& {\rm Re}[g_e(-\bar\psi_{1L}\bar\varphi_2+
\bar\psi_{2L}\bar\varphi_1)\psi_R].\ee
To see the physical content of the theory, we need 
orthonormal bases of the Hilbert spaces $\gg^\cc,\
\hh_L,\ \hh_R$ and $\hh_S$. 

A Cartan subalgebra of $\gg$ is spanned by the two
orthonormal vectors, `third isospin' and
`hypercharge',
\bb  I_3:=i\left(g_2\pp{1/2&0\cr 0&-1/2},0\right),\qq
Y:=i\left(0,g_1\right).\ee
The uncommitted choice for the electric charge 
generator $Q$ is:
\bb iQ:=i\left(g_2\sin\theta_w\pp{1/2&0\cr 0&-1/2}
,g_1\cos\theta_w\right),\ee
where $\theta_w$ is the weak mixing angle.
We complete $iQ$ to an orthonormal basis of
$\gg^\cc$ of eigenvectors of $[Q,\cdot]$
\bb \tilde Z&:=&
i\left(g_2\cos\theta_w\pp{1/2&0\cr 0&-1/2}
,-g_1\sin\theta_w\right),\cr \cr 
I^+&:=&i\left(\frac{g_2}{\sqrt 2}\pp{0&1\cr
0&0},0\right),\cr \cr 
 I^-&:=&i\left(\frac{g_2}{\sqrt
2}\pp{0&0\cr 1&0},0\right).\eee
The eigenvalues are 0 and $\pm
g_2\sin\theta_w=:\pm e$. The multiplet of gauge
bosons is now written as
 \bb A_\mu(x):= \gamma_\mu(x)\,iQ
+Z_\mu(x)\,\tilde
Z+\frac{1}{\sqrt{2}}\left(W_\mu(x)\,I^+
+W^*_\mu(x)\,I^-\right),\eee
where the photon $\gamma_\mu(x)$ and the
$Z_\mu(x)$ are real fields while the $W$
is complex. The kinetic term in the Yang-Mills
Lagrangian now has its standard form, a sum of three
pieces each of the form
\bb
-{\textstyle\frac{1}{2}}\pa_\mu W^*_\nu\pa^\mu
W^\nu+{\textstyle\frac{1}{2}}\pa_\mu W^{*\mu}
\pa_\nu W^\nu +
{\textstyle\frac{1}{2}}m_W^2W^*_\mu W^\mu.\ee
The mass term is absent from the Yang-Mills
Lagrangian because of gauge invariance. We will now
get it from the Klein-Gordon action by spontaneous
symmetry breaking.

Our group $SU(2)\times
U(1)$ is broken spontaneously down to $U(1)$. The
 former $U(1)$ defines the hypercharge. We will
identify the latter $U(1)$ with the electric charge. The
minimum of the Higgs potential is located at 
scalars $\varphi_0$ of norm
$|\varphi_0|=v$ where $v={\textstyle\frac{1}{2}}\mu/
\sqrt\lambda$ is the vacuum expectation value. Any
such minimal $\varphi_0$ is left invariant by a 
residual subgroup, the little group. Without loss of
generality let us choose $y_S=-{\textstyle\frac{1}{2}}$
and
$\varphi_0=\,^t\!(v,0)$ and let us compute the little
group. We are looking for elements $(a,b)\in G$ such
that 
\bb
\rho_S(a,b)\varphi_0=ab^{-1/2}\varphi_0=\varphi_0.
\ee
The solution is,
\bb a=\pp{\exp(i\theta/2)&0\cr 0&
\exp(-i\theta/2)},\qq
b=\exp(i\theta),\ee
the little group is $U(1)$ generated by $iQ$ if and only
if 
\bb g_1\cos\theta_w=g_2\sin\theta_w=e.\label{sin}\ee
Then
\bb \frac{1}{i}\tilde\rho_S(iQ)=\pp{0&0\cr 0&-e}.\ee
Next we compute the boson masses. We have to
develop the scalar field around a minimum of the
action, $\varphi=\varphi_0$ and not around
$\varphi=0$ which is not a minimum. To this end we
define 
\bb \varphi(x)=:\varphi_0\,+\, h(x).\ee 
Then the mass matrix of the gauge bosons is the
 term quadratic in $A$ contained in the Klein-Gordon
Lagrangian,
\bb
{\textstyle\frac{1}{2}}
\left|\tilde\rho_S(A_\mu)\varphi_0
\right|^2=
{\textstyle\frac{1}{2}}
m_Z^2Z_\mu Z^\mu
+{\textstyle\frac{1}{2}}
m_W^2 W^*_\mu W^{\mu} \ee
 with
\bb m_W=
{\textstyle\frac{1}{2}}g_2v\qq{\rm and}
\qq m_Z= {\textstyle\frac{1}{2}}\sqrt{g_1^2+g_2^2}\ v
=m_W/\cos\theta_w.\ee
The spontaneous symmetry breaking has given masses
to the $W$ and $Z$ bosons. Massless spin 1 particles
have two degrees of freedom, `the transverse modes',
the spin is orthogonal to the direction of
motion. A massive spin 1 particle has one more degree
of freedom, `the longitudinal mode', the spin is parallel
to the direction of motion. To become massive the
massless gauge boson takes this additional degree of
freedom from the Higgs field. In our example, we
parameterize the scalar as
\bb \varphi(x)=\varphi_0+\pp{H(x)+ih_Z(x)\cr 
h_W(x)},\ee
corresponding to an orthonormal basis of $\hh_S$.
The neutral $Z$ boson eats the neutral scalar field
$h_Z$ to become massive and the charged $W$ eats the
charged scalar $h_W$. There remains only one
physical scalar field $H$ which is neutral. Let us
compute its mass. To this end we must develop the
Higgs potential in terms of the fields $H,\ h_Z$ and
$h_W$,
\bb
V(\varphi(x))=V(\varphi_0)+
{\textstyle\frac{1}{2}}m_H^2H^2(x)+{\rm
terms\ of \ order\ 3\ and\ 4\  },\ee
 with
\bb m_H=2\sqrt2\sqrt\lambda\,v.\ee
The constant term $V(\varphi_0)$ is the energy of the
vacuum or cosmological constant. 

One defines the $\rho$-factor by 
\bb
\rho:=\frac{m_W^2}{\cos^2\theta_w \ m_Z^2}.
\ee
It is unit if the scalar sits in a doublet and it can take
any other real value with more complicated scalar
representations. Experimentally we have today
$\rho=1.0012\pm 0.0031$. 

Finally let us turn to the fermionic action. 
The spontaneous symmetry breaking also produces
the electron mass from the Yukawa term with
$\varphi=\varphi_0$. With respect to orthonormal
bases of
$\hh_L$ and $\hh_R$, we have
\bb \psi_L=\nu_L\pp{1\cr 0}+e_L\pp{0\cr 1},\qq
\psi_R=e_R,\ee
and the fermionic Lagrangian reads to second order:
\bb
\bar e\ddd e +\bar \nu\ddd\,\frac{1-\gamma_5}{2}\,
\nu+ m_e \bar e e,\ee
with \bb m_e={\rm Re}\, g_e\, v.\ee
The remaining terms are of order three, the minimal
couplings fermion-fermion-gauge boson and the
Yukawa couplings fermion-fermion-Higgs. They
describe interactions, terms giving rise to non-linear
field equations via Euler-Lagrange. For instance the
coupling of the photon to the neutrino $\bar\nu_L
\gamma\nu_L$ is,
\bb \pp{1& 0}{\textstyle\frac{1}{i}}\tilde\rho_L(iQ)
\pp{1\cr 0}=e\pp{1&0}\left[\pp{{\textstyle\frac{1}{2}}
&0\cr
0&-{\textstyle\frac{1}{2}}}-{\textstyle\frac{1}{2}}1_2
\right]\pp{1\cr 0}=0.\ee
The couplings of the photon to the left-handed
electron and of the photon to the
right-handed electron are both $-e$. The photon
coupling is vectorial, electromagnetism
preserves parity. On the other hand, the coupling of
the $W$ to the left-handed electron is $g_2$, to the
right-handed electron it vanishes, the $W$ coupling
is axial.  

%% file: monsa4
\chapter{Connes' first dreisatz}

Noncommutative geometry explains the Higgs field as
a magnetic field accompanying certain Yang-Mills
fields, among them the ones of the standard model.

\begin{itemize}\item
{ \qq Yang-Mills \ + \ noncommutative geometry \ \ =
\ \ Yang-Mills-Higgs.  }
\end{itemize}

The geometric noblesse of the two circles allows their
promotion to noncommutative geometries. The
promotion of the two circles to one of these, an almost
commutative geometry, produces the three epicycles
from the two promoted circles. 

To construct a Yang-Mills action $\int ( F,*F)$, we
need four ingredients, differential forms on
spacetime $M$, a Lie group $G$, `the internal space', a
scalar product on the space of differential forms
$\Omega M$ and an invariant scalar product on the
Lie algebra $\gg$ of the group $G$. To construct the
action which is a real number, we take the scalar
products of the field strength with itself. The first
scalar product involves the spacetime metric $g$
hidden in the Hodge star $*$,
$(\kappa,\varphi):=\int_M\kappa^**\varphi$,
$\kappa$ and $\varphi$ differential forms of same
degree. The second scalar
product is on the Lie algebra, e.g. for $G=SU(n)$, the
general invariant scalar product is
$(a,b)={\textstyle\frac{2}{g_n^2}}\t (a^*b)$, $a,b\in
su(n)$ and the coupling
constant $g_n$ is a positive number. Noncommutative
geometry in its almost commutative version unifies
spacetime and internal space and the two scalar
products are derived from one common scalar product.
At the same time coordinate transformations on
spacetime are unified with gauge transformations.
They are nothing but the automorphisms of the almost
commutative geometry. This last point will be the
starting point of the fourth geometric dreisatz
unifying Yang-Mills with gravity.

\section{Spectral triples}

Noncommutative geometry does to spacetime $M$,
what quantum mechanics did to phase space $\cal P$

\begin{itemize}\item
{ \qq Hamilton \ + \ noncommutative geometry \ \ =
\ \ Schr\"odinger.  }
\end{itemize}

An uncertainty relation is
introduced by allowing the commutative algebra of
functions
$\ccc^\infty(\cal P)$ to become noncommutative. Let
us call $\aa$ this new algebra that we suppose
defined over the real numbers, associative and
equipped with a unit and an involution.
$\aa$ is the algebra of quantum observables. Now on
spacetime $M$ we have a metric. But how  
define a distance on a space that has lost its
points? Following Connes
\cite{book}, we need a faithful representation
$\rho$ of
$\aa$ via bounded operators on a complex Hilbert
space
$\hh$, the space of fermions, and a selfadjoint `Dirac'
operator
$\dd$ on
$\hh$. Connes calls these three ingredients a spectral
triple,
$(\aa,\hh,\dd)$. They satisfy axioms. These axioms are
simply taken from the properties of the commutative
case,
$\aa=\ccc^\infty(M)$, where from now on we must
suppose that spacetime $M$ is Euclidean and compact.
The Hilbert space $\hh$ is the
space of ordinary, square integrable Dirac spinors. An
element $f$ of $\aa$ is a differentiable function on
spacetime, $f(x)$, and it acts on a spinor $\psi(x)$ by
multiplication $(\rho(f)\psi)(x):=f(x)\psi(x)$.
$\dd=\ddd$ is the ordinary Dirac operator. 
Only recently Connes has completed the list of axioms
\cite {grav} as to have a one-to-one correspondence
between commutative spectral triples and Riemannian
spin manifolds. To this end, he needed two other old
friends from particle physics, a chirality operator
$\chi$ and a real structure $J$. The chirality is a
unitary operator of square one that commutes with the
representation. Therefore $\chi$ decomposes the
representation space
 into a left-handed piece $(1-\chi)/2\,\hh$ and a
right-handed piece $(1+\chi)/2\,\hh$. In the
commutative case, of course $\chi=\gamma_5.$ The
real structure is an anti-unitary operator that in the
commutative case reduces to the charge conjugation
operator $C$. $J$ is of square plus or minus one,
depending on spacetime
 dimension and signature. Also depending on
spacetime dimension and signature, $J$ commutes or
anticommutes with $\chi$. 
The charge conjugation as well decomposes the
representation space into two pieces, particles and
anti-particles, all together
\bb \hh=\hh_L\op\hh_R\op\hh_L^c\op\hh_R^c.
\label{decom}\ee
Here are a few more properties from the commutative
case that become axioms
\begin{itemize}
\item
$\rho(a)$ commutes  with $J\rho(\tilde a)J^{-1}$, 
for all $a,\tilde a \ \in\ \aa,$
\item
$ \dd\chi=-\chi\dd,$ 
\item
$\dd J=+J\dd,$
\item
$[\dd,\rho(a)]$ is bounded for all $a$ in $\aa$,
\item
$[\dd,\rho(a)]$ commutes with  $J\rho(\tilde a)J^{-1}$,
for all
$a,\tilde a$ in $\aa$. 
\end{itemize}
The last axiom is called first order, because in the
commutative case, it just says that the Dirac operator is
a first order differential operator. The dimensionality
of $M$ can be recovered from the spectrum of the
Dirac operator. Indeed for compact manifolds, the
spectrum is discrete and the eigenvalues $\lambda_n$
grow like $n^{1/\dim M}$. This motivates the name
spectral triple. Let us mention two more axioms. The
orientability axiom relates the chirality to the volume
form, a differential form of maximal degree. The
Poincar\'e duality on manifolds is promoted to an
axiom in quite an abstract form. We anticipate that, in
the case of the standard model, this Poincar\'e duality
will prohibit right-handed neutrinos \cite{tresch}.

\noindent {\bf Warning:} My presentation of
noncommutative geometry is that of a modest
physicist. For a precise account the reader is referred
to Joe V\'arilly's beautiful lectures at this School
\cite{joe}.

Since we are now in Euclidean signature let us spell
out again the case of a four dimensional spacetime.
A spinor has four square integrable components,
\bb   \psi = \pmatrix{
\psi_1(x) \cr \psi_2(x) \cr \psi_3(x) \cr \psi_4(x)
}\ \in\ \lll^2(\sss).\ee   
The (flat) Dirac operator is 
\bb   \ddd\psi := i\gamma^\mu
{\partial\over{\partial x^\mu}}\psi.\ee  
We choose the gamma
matrices self adjoint, 
\bb   \gamma ^0 :=  \pmatrix {1
& 0 & 0 & 0 \cr
 0 & 1 & 0 & 0 \cr
 0 & 0 & -1 & 0 \cr
 0 & 0 & 0 & -1} \qquad \gamma ^1 :=  \pmatrix {0 & 0 &
0 & i \cr
 0 & 0 & i & 0 \cr
 0 & -i & 0 & 0 \cr
 -i & 0 & 0 & 0}\label{g0eucl}\ee  
 \bb   \gamma ^2 :=  \pmatrix {0 & 0 & 0 &
1 \cr
 0 & 0 & -1 & 0 \cr
 0 & -1 & 0 & 0 \cr
 1 & 0 & 0 & 0} \qquad \gamma ^3 :=  \pmatrix {0 & 0 & i
& 0 \cr
 0 & 0 & 0 & -i \cr
 -i & 0 & 0 & 0 \cr
 0 & i & 0 & 0}.\label{g2eucl}\ee   
They satisfy the anticommutation
relation 
\bb   \gamma ^\mu \gamma ^\nu +\gamma ^\nu
\gamma ^\mu =  2\eta ^{\mu \nu }1\ee  
 with the flat
Euclidean metric 
\bb   \eta = \pmatrix {1 & 0 & 0 & 0 \cr
 0 & 1 & 0 & 0 \cr
 0 & 0 & 1 & 0 \cr
 0 & 0 & 0 &1 }.\ee  
 The chirality operator is by definition
\bb    \gamma_5 :=
\gamma^0\gamma^1\gamma^2\gamma^3 = 
 \pmatrix {0 & 0 & -1 & 0 \cr
 0 & 0 & 0 & -1 \cr
 -1 & 0 & 0 & 0 \cr
 0 & -1 & 0 & 0}. \label{g5eucl}\ee   
It is unitary and of unit square as
postulated. Since it anticommutes with all other
gamma matrices
\bb   \gamma^\mu\gamma_5+\gamma_5\gamma^\mu =
0,\ee   
the Dirac operator is odd 
\bb   \ddd\gamma_5+\gamma_5\ddd = 0.\ee 
The charge conjugation is
\bb \psi^c=\gamma^0\gamma^2\psi^*.\ee
Let us note that in four dimensional Euclidean
spacetime, the chirality commutes with charge
conjugation,
\bb (\psi_L)^c=(\psi^c)_L=:\psi_L^c.\ee
In the following we will take advantage of this
notational simplification. Attention, in Minkowskian
signature, the notation $\psi_L^c$ is ambiguous,
because there the two operators anti-commute. 
Finally, we abbreviate the representation of a
function $f$ on a spinor
$\psi$ by
$\rho(f)=:\ul f$,
$(\ul f\psi)(x)= f(x)\psi(x).$

\section{Differential forms}

Our next aim is to construct differential forms starting
from a spectral triple. In the commutative case, we
want this construction to reproduce de Rham's
differential forms, $\Omega M$. 

 We start with an auxiliary differential algebra 
$\Omega \aa,$
the universal differential envelope of $\aa$:
$ \Omega^0\aa := \aa$.   
$\Omega^1\aa$ is
generated by symbols $\delta a$, $ a \in \aa$ with
relations 
$\delta 1 = 0,$   
$\delta(aa') = (\delta a)a'+a\delta a'.$
 $\Omega^1\aa$ consists
of finite sums of terms of the form $a_0\delta a_1,$ and
likewise for higher degree $p$,
\bb   \Omega^p\aa = \left\{ \sum_j
a^j_0\delta a^j_1...\delta a^j_p ,\quad a^j_q\in
\aa\right\}.\ee   
 The differential $\delta$ is defined by
\bb   \delta(a_0\delta a_1...\delta a_p) := 
   \delta a_0\delta a_1...\delta a_p.\ee   
The involution $^*$ is
extended from the algebra $\aa$ to 
 $\Omega^1\aa$ by putting
$(\delta a)^* := \delta(a^*) =:\delta a^*$ and to the
entire differential envelope by
$(\kappa\varphi)^*=\varphi^*\kappa^*$. 
The next step is to extend the
representation $\rho$  from the  algebra $\aa$   to
its envelope $\Omega\aa$. This
extension deserves a new name: 
\bb  \pi :
\Omega\aa  \longrightarrow \bigoplus_p {\rm
End}(\hh) 
   \eee 
 \bb\pi(a_0\delta a_1...\delta a_p) :=
(-i)^p\rho(a_0)[\dd,\rho(a_1)] ...[\dd,\rho(a_p)]
\label{pi}.\ee
$\pi$ is a representation of $\Omega\aa$ as graded
involution algebra. Note the $(-i)^p$ on the rhs which
 is not uniform in the literature. 
We are tempted to define also a
differential, again denoted by $\delta$,
 on $\pi(\Omega\aa)$ by
$\delta\pi(\hat\varphi):=\pi(\delta\hat\varphi).$
However, this definition does not make sense because
there are forms
$\hat\varphi\in\Omega\aa$ with
$\pi(\hat\varphi)=0$ and  $\pi(\delta\hat\varphi)
\not= 0$. By dividing out these unpleasant forms,
we arrive at the desired differential algebra
$\Omega_\dd\aa$,
\bb   \Omega_\dd\aa :=
{{\pi\left(\Omega\aa\right)}\over {\cal J}},\qq {\rm
with}
\qq {\cal J} := \pi\left(\delta\ker\pi\right) =:
\bigoplus_p {\cal J}^p,\ee
(${\cal J}$ for junk).  
On the quotient,
the differential is now well defined. Degree by
degree we have:  \bb   \Omega_\dd^0\aa =\rho(\aa)\ee   
because ${\cal J}^0=0$ , 
\bb   \Omega_\dd^1\aa = \pi(\Omega^1\aa)\ee   
because $\rho$ is faithful,
and in degree $p\geq2$ 
\bb   \Omega_\dd^p\aa = 
{{\pi(\Omega^p\aa)}\over
{\pi(\delta(\ker\pi)^{p-1})}}.\ee  
In the commutative case, $\delta=\de$,
$\Omega_\ddd\ccc^\infty(M)
$ is isomorphic to de Rham's differential algebra
$\Omega M$ with
\bb\pi(f_0\de f_1\de f_2...\de f_p)\cong
f_0\gamma^
{\mu_1}\left(\frac{\partial}{\partial
x^{\mu_1}}f_1\right)\gamma^
{\mu_2}\left(\frac{\partial}{\partial
x^{\mu_2}}f_2\right)...\gamma^
{\mu_p}
\left(\frac{\partial}{\partial
x^{\mu_p}}f_p\right).
\label{iso}\ee
Dividing out the junk renders the lhs
graded commutative.

Let us illustrate this isomorphism for 1- and 2-forms
on a four dimensional spacetime $M$.
We need the
commutator  
\bb [\ddd,{\ul f}]\psi &=&
i\gamma^\mu{\partial\over{\partial x^\mu}} (f\psi)-
if\gamma^\mu{\partial\over{\partial x^\mu}}\psi
\cr  &=& i\lb\gamma^\mu{\partial\over{\partial
x^\mu}}f\rb\psi.\ee   
 Therefore 
\bb   [\ddd,{\ul f}] =
i\gamma^\mu{\partial\over{\partial x^\mu}}f  =:
i\gamma(\de f)\ee   
with 
\bb   \de f =
\lb{\partial\over{\partial x^\mu}}f\rb\de x^\mu.\ee   
At
this point, we see that the restriction to flat
spacetime  can be dropped. Let us anticipate the Dirac
operator on curved manifolds
\bb   i\gamma^\mu(x)\lb{\partial\over{\partial
x^\mu}}+ \omega_\mu\rb.\ee  
It differs from the flat one in two
respects, the gamma matrices are $x$  dependent, no
problem in the above commutator, and an additional 
algebraic term, a spin connection $\omega = 
\omega_\mu\de x^\mu$ valued in $so(4)$ appears but
drops out  from the commutator. Since the Dirac
operator only shows up in commutators, Connes'
algorithm works on any Riemannian spin manifold. 

The representation of functions
by multiplication on spinors is  faithful, of course, and
\bb   \Omega^1_\ddd\aa \cong
\pi(\Omega^1\aa).\ee    A
general element of the rhs is a finite sum of terms 
\bb    
\pi(f_0\de f_1),\quad f_0,f_1\in\aa.\ee   
It is identified
with the differential 1-form on $M$ 
\bb     f_0\de f_1
\quad \in\Omega^1M.\ee   
For 2-forms the situation is
less trivial, we must compute the junk 
$\jj^2=\pi(\de(\ker\pi)^1)$. Consider  
\bb   h^{-1}\de
h+h\de h^{-1}\ee   
an element in $\Omega^1\aa$ where
$h\in\aa$ is a non-vanishing  function,
$h^{-1}(x)=1/h(x)$. As
$\Omega\aa$ is not  graded commutative this
element does not vanish! 
\bb   h^{-1}\de h+h\de h^{-1}
\ne h^{-1}\de h+(\de h^{-1})h =  \de (h^{-1}h) = \de 1 =
0.\ee   
Its image under $\pi$ however does vanish
\bb   \pi(h^{-1}\de h+h\de h^{-1}) &=& 
\gamma(h^{-1}\de h+h\de h^{-1})\cr &=& 
\gamma(h^{-1}\de h+(\de h^{-1})h) = 0.\ee   
Therefore the considered element is in
$(\ker\pi)^1$ and the  corresponding element in
$\pi(\de(\ker\pi)^1)$ is
 \bb \pi(\de
h^{-1}\de h+\de h\de h^{-1}) &=&  \gamma(\de
h^{-1})\gamma(\de h) +\gamma(\de h)\gamma(\de
h^{-1})\cr
 &=& \gamma^\mu\left({\partial\over{\partial
x^\mu}}h^{-1}\right)
     \gamma^\nu{\partial\over{\partial x^\nu}}h
  +\gamma^\nu\left({\partial\over{\partial x^\nu}}h
\right)
    \gamma^\mu{\partial\over{\partial
x^\mu}}h^{-1}\cr 
 &=&
\lb\gamma^\mu\gamma^\nu+\gamma^\nu\gamma^
\mu\rb
       \left( {\partial\over{\partial x^\mu}}h^{-1}\right)
        {\partial\over{\partial x^\nu}}h\cr
 &=& -\left({2\over{h^2}}g^{\mu\nu}
           \left({\partial\over{\partial x^\mu}}h\right)
           {\partial\over{\partial x^\nu}}h\right) 1.\ee  
 By
linear combination we get the junk,
\bb   \pi(\de(\ker\pi)^1) =
\left\{f1,\ \ f\in\aa \right\}.\ee   
On the other hand 
\bb   \pi(\Omega^2\aa) =  \left\{
f_{\mu\nu}\gamma^\mu\gamma^\nu, \ \
f_{\mu\nu}\in\aa\right\}\ee   
and 
\bb   \pi(\de f_1\de
f_2+\de f_2\de f_1) = 
\left(2g^{\mu\nu}{\partial\over{\partial
x^\mu}}f_1
           {\partial\over{\partial x^\nu}}f_2\right) 1\ee  
After dividing out the junk, $\pi(\de f_1)$ and 
$\pi(\de f_2)$ anticommute whereas they did not
anticommute in $\pi(\Omega^2\aa)$. We
may now identify a general  element 
\bb   \pi(f_0\de
f_1\de f_2)\ \ \in\Omega^2_\ddd\aa\ee  
 with the differential
2-form on $M$ 
\bb   f_0\de f_1\de f_2\ \ \in\Omega^2M.\ee  
Note that we have treated the quotient space like
a subspace  which is legitimate only in presence of an
appropriate scalar  product. This scalar product
will be defined in terms of the  involution and a trace
in the next section.

The involution that $\Omega M$
inherits from $\Omega_\ddd\aa$ via the  sketched
isomorphism is with our conventions 
\bb   (f_0\de f_1\de
f_2...\de f_p)^* =  (-1)^{(1/2)p(p-1)}\bar f_0\de\bar
f_1\de\bar f_2...\de\bar f_p.\ee    

 The orientability axiom alluded to above
is motivated from this isomorphism, $\de x^1
\de x^2\de
x^3\de x^4\cong(\det
g_{\cdot\cdot})^{1/2}
\gamma^1\gamma^2\gamma^3\gamma^4=
(\det g_{\cdot\cdot})^{1/2}\gamma_5.$

\section{The scalar products in noncommutative
geometry} \label{scalarp}

To play the Yang-Mills game, we need a scalar product
for differential forms. In the noncommutative
context, the scalar product has another utility. It
allows us to interpret the differential forms in
$\Omega_\dd\aa$  not as classes but as concrete
operators on the Hilbert space $\hh$: degree by
degree, we embed
$\Omega^p_\dd$ in $\pi(\Omega^p\aa)$ as orthogonal
complement of $\jj^p$. If $\hh$ was
finite dimensional, we would naturally take as scalar
product of two operators $\kappa$ and $\varphi$, 
$<\kappa,\varphi>={\rm Re\,}\t(\kappa^*\varphi)$.
For infinite dimensional Hilbert spaces $\hh$, like
the space of spinors
$\lll^2(\sss)$, we have to regularize and
 we use the Dirac operator to do so. Thanks to the
asymptotic behavior of its spectrum, ${\rm
Re\,}\t[\kappa^*
\varphi\,|\dd|^{-\dim M}]$ only diverges
logarithmically. The Dixmier trace $\tt$ gets rid of
this divergence
\cite{dix}: For any bounded, positive
operator $Q$ on $\hh$  we define the {\it Dixmier
trace} $\tt$ by 
 \bb   \tt(Q|\dd|^{-\dim}) :=
\lim_{N\rightarrow\infty}
                                {1\over{\log
N}}\sum_{n=1}^N\lambda_n,\label{dix}\ee  
 where the $\lambda_n$ are the eigenvalues of
 $Q|\dd|^{-\dim}$
 arranged in a decreasing sequence discarding the zero
modes of the Dirac operator. 
Now we proceed
as in the finite  dimensional case ($\dim M=0$) and
define a scalar product on  $\pi(\Omega\aa)$ by
\bb   <\kappa,\varphi> := {\rm Re\,}\tt
(\kappa^*\varphi|\dd|^{-\dim}), \quad 
\kappa, \varphi \in \pi(\Omega^p\aa).\ee   
Note that $\kappa$ and $\varphi$ are bounded
because 
$[\dd,\rho(a)]$ are by axiom. In the commutative case,
 for a four dimensional spacetime $M$, this scalar
product can be computed to be 
\bb   \label{one}
<\kappa,\varphi>= 
{\textstyle\frac{1}{32\pi^2}}\,{\rm
Re}\int_M\t_4
\lb\kappa^*\varphi\rb\de^4x.\ee
It is independent of $M$.   
$\t_4$
denotes the trace over the gamma matrices.
 With this scalar product  $\Omega_\dd\aa$ is a
subspace of
$\pi(\Omega\aa)$, by  definition orthogonal
to the junk. As subspace  
$\Omega_\dd\aa$ inherits a scalar product, that we
denote by 
\bb   (\kappa,\varphi) = <\kappa,\varphi>,
\quad \kappa, \varphi \in \Omega_\dd^p\aa.\ee   
In the commutative case in four dimensions,
thanks to well known results for $\t_4\lb
\gamma^{\mu_1}v_{\mu_1}...
\gamma^{\mu_q}v_{\mu_q}\rb$
this scalar product vanishes for forms with
different degree. By the isomorphism (\ref{iso})
between 
$\Omega_\ddd\aa$ and $\Omega M$ the corresponding
scalar product on differential forms, still denoted by $
(\cdot,\cdot )$, is 
 \bb   (\kappa,\varphi) = 
{\textstyle\frac{1}{8\pi^2}}\,{\rm Re}
\int_M\kappa^**\varphi , \quad \kappa, \varphi \in 
\Omega^pM.\ee    

Let us illustrate this by a simple example on the flat
four torus with all circumferences measuring $2\pi$.
 Denote by $\psi_B(x)$, $B=1, 2, 3, 4$,
the four components of the spinor. The Dirac
operator is
\bb \ddd = \pmatrix{
i{\partial/{\partial x^0}}&0
&-{\partial/{\partial x^3}}&
-{\partial/{\partial x^1}}+
i{\partial/{\partial x^2}}\cr
0&i{\partial/{\partial x^0}}&
-{\partial/{\partial x^1}}-
i{\partial/{\partial x^2}}&
{\partial/{\partial x^3}}\cr
{\partial/{\partial x^3}}&
{\partial/{\partial x^1}}-
i{\partial/{\partial x^2}}&
-i{\partial/{\partial x^0}}&0\cr
{\partial/{\partial x^1}}+
i{\partial/{\partial x^2}}&
-{\partial/{\partial x^3}}&0&
-i{\partial/{\partial x^0}} }
.\ee
After a Fourier transform
\bb \psi_B(x)\ =:\ \sum_{j_0,...,j_3\in\zz}
\hat\psi
_B(j_0,...,j_3)\exp(-ij_\mu x^\mu),\quad B=1,2,3,4\ee
the eigenvalue equation $\ddd\psi=\lambda\psi$ reads
\bb \pmatrix{
j_0&0&ij_3&ij_1+j_2\cr
0&j_0&ij_1-j_2&-ij_3\cr
-ij_3&-ij_1-j_2&-j_0&0\cr
-ij_1+j_2&ij_3&0&-j_0}
\pmatrix{\hat\psi_1\cr \hat\psi_2\cr \hat\psi_3\cr
\hat\psi_4}\ =\
\lambda
\pmatrix{\hat\psi_1\cr \hat\psi_2\cr \hat\psi_3\cr
\hat\psi_4}. \ee
Its characteristic equation is
$ \lb \lambda^2-(j_0^2+j_1^2+j_2^2+j_3^2)^2\rb^2=0$
and for fixed $j_\mu$, each eigenvalue
$ \lambda=\pm\sqrt{j_0^2+j_1^2+j_2^2+j_3^2}$
has multiplicity two. Therefore asymptotically for large
$\Lambda$ there are 
$ 4B_4\Lambda^4$ eigenvalues (counted with their
multiplicity) whose absolute values are smaller than 
$\Lambda$. 
$ B_4=\pi^2/2$
denotes the volume of the unit ball
in $\rr^4$. Let us arrange the absolute values of the
eigenvalues in an increasing
sequence. Taking due account of their multiplicities
we have for large $n$
\bb |\lambda_n|\approx
\left({n\over{2\pi^2}}\right)^{1/4}\ee
and we can check the Dixmier trace in equation 
(\ref{one}) for instance with $\kappa=\varphi
=1\ \in \pi(\Omega^0\aa)$
\bb <1,1>&=&\t_\omega(|\ddd|^{-4})=
\lim_{N\rightarrow\infty}
                                {1\over{\log
N}}\sum_{n=1}^N|\lambda_n|^{-4}\cr
&=&\lim_{N\rightarrow\infty}
                                {1\over{\log
N}}\sum_{n=1}^N{{2\pi^2}\over n}=
\lim_{N\rightarrow\infty}{1\over{\log
N}}\int_{1}^N{{2\pi^2}\over n}\,\de n\cr
&=&2\pi^2={1\over{32\pi^2}}\int_M\t_4[1]\de^4x.
\ee

In the commutative case the following two scalar
products 
\bb   <\kappa,\varphi> &:=& {\rm Re\,}\tt
(\kappa^*\varphi|\dd|^{-\dim}),
\qq\kappa, \varphi \in \pi(\Omega^p\aa),\\
<\kappa,\varphi>& :=&{\textstyle\frac{1}{2}} {\rm
Re\,}\tt ([\kappa+J\kappa J^{-1}]^*
[\varphi+J\varphi J^{-1}]\,
|\dd|^{-\dim})\label{sp},\ee
are identical. This is not true in general. We
anticipate that the generalization of the principle of
general relativity to noncommutative geometry,
Connes' second dreisatz, will exclude the first scalar
product.

\section{The commutative Yang-Mills action}

\label{commym}
The message of this section is that the commutative
spectral triple of spacetime $M$ is a natural tool to
reconstruct Maxwell's theory: this reconstruction
unifies spacetime with internal space, $G=U(1)$. The
first sign for this unification comes from the group of
unitaries of $\aa$. Remember that $\aa$ is the algebra
of complex valued functions on $M$ with involution
just complex conjugation. The group of unitaries
$U(\aa):=\{u\in\aa,\ uu^*=u^*u=1\}$ for this algebra
is the group of functions from spacetime into $U(1)$
and this is Maxwell's gauge group, $U(\aa)=\,^MU(1)$.
Maxwell's four potential $A\in\Omega_\ddd^1\aa$ is a
 1-form that we take {\it anti}-Hermitean now in
order to harmonize the abelian and non-Abelian case.
A gauge transformation or unitary
$u=\exp i\theta$ acts affinely on the gauge
potential by
\bb
\rho_V(u)A:=\rho(u)A\rho(u^{-1})+\rho(u)\de\rho(u^{-1})
=A-i\de\theta.\ee
The field strength
\bb F:=\de A+A^2=\de A\qq\in\Omega_\ddd^2\aa\ee
transforms homogeneously under unitaries
and is even gauge invariant in the commutative case, 
\bb \rho_V(u)F=\rho(u)F\rho(u^{-1})=F.\ee
The obviously gauge invariant
Maxwell action can be written,
\bb S_{\rm Maxwell}[A]&=&z(F,F)=
z{\rm Re\,}\tt( F^*F|\ddd|^{-4})=
{\textstyle\frac{z}{8\pi^2}}\int_M F^**F\cr &=&
{\textstyle\frac{z}{16\pi^2}}\int_M F_{\mu\nu}^*
F^{\mu\nu}(\det g_{\cdot\cdot})^{1/2}\de^4x=:
{\textstyle\frac{\epsilon_0}{4e^2}}\int_M 
F_{\mu\nu}^*
F^{\mu\nu}(\det g_{\cdot\cdot})^{1/2}\de^4x,\ee
where $z=\pi/\alpha_{\rm em}$ is the
fine-structure constant or gauge coupling
$\alpha_{\rm em}:=e^2/(4\pi
\epsilon_0\hbar c)$. 
The commutative pure Yang-Mills theory is linear and
to justify the word coupling, we have to add
matter, say an electron $\psi$. The Dirac operator acts
on it defining its kinetic energy, unitaries act on it by
\bb \rho_{\rm spinor}(u)\psi=\rho(u)\psi,\qq u\in
U(\aa),\qq
\psi\in\hh, \ee
and we define the minimal coupling by the covariant
Dirac operator
$\ddee:=\ddd-\pi(A)$.
We have already noted that the gravitational field
$\omega$ drops out when we construct the differential
forms. The same is true for the electromagnetic field,
$\Omega_\ddd\aa=\Omega_\ddee\aa$.
 The Dirac action then reads
\bb S_D[\psi,A]=\int_M\psi^*\ddee\psi\,
|\det g_{\cdot\cdot}|^{1/2}\de^4x,\ee
 A mass
term $m_\psi\psi^*\psi$ may be added.

Let us stress again that in Connes' formulation, the
gauge coupling, that is the invariant scalar
product in internal space, is induced from the scalar
product of differential forms over spacetime.

\section{Almost commutative geometries}
\label{acg}

One way to see the above commutative example is to say
that the associative algebra of the spectral triple is
$\aa_t=\ff\ot\aa_f$, a tensor product of the
 commutative, infinite dimensionsal algebra of
 {\it real} valued functions $\ccc^\infty(M)$ on
spacetime and the commutative, finite dimensional,
{\it real} algebra
$\aa_f=\cc$. The gauge group then is Abelian,
$G=U(1)\subset\aa_f$. It is natural to try
noncommutative algebras for $\aa_f$ to get
   non-Abelian gauge groups $\ $\cite{cl}. In this
spirit we  consider tensor products of entire spectral
triples, and the message of this section is that if the
fermionic representation breaks parity, the Higgs
scalar and the symmetry breaking potential come free
of charge. We call almost commutative geometry this
cheap tensor product of the commutative, infinite
dimensional spectral triple of a spacetime with a
noncommutative finite dimensional spectral triple of
a matrix algebra \cite{other}. Remember that the
spinning particle in quantum mechanics is also such a
cheap tensor product, of an ordinary wave function
with a vector in a representation space of $SU(2)$.

Let us denote by $(\ff,\lll^2(\sss),\ddd,\gamma_5,C)$
the commutative spectral triple of a four dimensional
spacetime and by
$(\aa_f,\hh_f,\dd_f,\chi_f,J_f)$, $\cdot_f$ for finite,
the one of a (zero dimensional) internal space. Note
that our  $C$ is anti-unitary.
According to the rules of noncommutative geometry
the tensor product of these two spectral triples
$(\aa_t,\hh_t,\dd_t,\chi_t,J_t)$, $\cdot_t$ for tensor,
is:
\bb
\aa_t\,=\,\ff\ot\aa_f,\qq
&\hh_t\,=\,\lll^2(\sss)\ot\hh_f,\qq&
\dd_t\,=\,\ddd\ot 1\,+\,\gamma_5\ot\dd_f,\cr 
&\chi_t\,=\,\gamma_5\ot\chi_f,\qq &J_t\,=\,C\ot
J_f.\ee 
Before turning the crank, we must talk about
the internal Dirac operator $\dd_f$. From the axioms,
we infer that with respect to the decomposition
(\ref{decom}) of the fermionic Hilbert space $\hh_f$
the internal Dirac operator has the form:
\bb \dd_f=\pp{0&\mm&0&0\cr 
\mm^*&0&0&0\cr 
0&0&0&\bar\mm\cr 
0&0&\bar\mm^*&0}\qq{\rm or}\qq
\dd_f=\pp{0&\mm&0&0\cr 
\mm^*&0&0&0\cr 
0&0&0&0\cr 
0&0&0&0},\ee
where $\mm$ is the fermionic mass matrix. This is
another manifestation of the unification of spacetime
and internal space, the naked Dirac operator $\ddd$
and its mass matrix obey the same axioms. 

As in the commutative case, we start by identifying
the gauge group, the functions from spacetime into
the finite dimensional Lie group $G=U(\aa_f)$.  It is
represented affinely on the bosonic fields. They are
anti-Hermitean 1-forms.  But now,
\bb \Omega^1_{\dd_t}\aa_t&=&
\Omega^1_\ddd\ff\ot\Omega^0_{\dd_f}\aa_f\,\op\,
\Omega^0_\ddd\ff\ot\Omega^1_{\dd_f}\aa_f\cr 
&\cong&
\Omega^1(M,\aa_f)\,\op\,
\ff\ot\Omega^1_{\dd_f}\aa_f\ \owns A_t=:(A,H).\ee
From the anti-Hermiticity of $A_t$, it
follows that
$A$ is in fact a Lie algebra valued 1-form on spacetime,
$A\in\Omega^1(M,\gg)$, i.e. a Yang-Mills potential.
$\gg:=u(\aa_f):=\{X\in\aa_f,\ X+X^*=0\}$ is the Lie
algebra of the group of unitaries
$G=U(\aa_f)$.  On the other hand,
 the Higgs scalar $H$ is a 0-form on spacetime, valued
in a representation of the Lie group
$G$. The inhomogeneous transformation law,
\bb \rho_{tV}(u)A_t&:=&
\rho_t(u)A_t\rho_t(u^{-1})+\rho_t(u)\delta_t\rho_t(
u^{-1}) = (\rho_V(u)A,\rho_S(u)H),\\
\rho_V(u)A&=&\rho_f(u)A\rho_f(u)^{-1}+\rho_f(u)\de \rho_f(u)^{-1},\\ 
\rho_S(u)H&=&\rho_f(u)H\rho_f(u^{-1})+
\rho_f(u)\delta_f\rho_f(u^{-1}),
\ee
determines according
to which {\it group} representation $\rho_S$ the
Higgs scalar transforms and this depends on the
details of the internal spectral triple. We denote by
$\rho_t$ the representation of
$\aa_t$ on $\hh_t$, by $\rho_f$ the representation of
$\aa_f$ on $\hh_f$, by $\delta_t$ the differential of
$\Omega_{\dd_t}\aa_t$ and so forth.
Next we define the field strengh,
\bb F_t =
\delta_tA_t+A_t^2\qq\in\Omega^2_{\dd_t}\aa.\ee 
To decompose the field strength, it is comfortable to
change scalar variables,
\bb \Phi(x)\,:=\,H(x)-i\dd_f\,=\,-\Phi^*\ \in
\Omega^0(M,\Omega^1_{\dd_f}\aa_f).\ee 
This change of variables is well defined within
$\Omega^0(M,\Omega^1_{\dd_f}\aa_f)$ thanks to the
orientability axiom \cite{tk}. $\Phi$ has the good
taste to transform homogeneously under a gauge
transformation $u$ and we can define its covariant
exterior derivative, 
\bb\dee\Phi:= \de\Phi+\left[\rho_f(A),\Phi\right]
\ \in\Omega^1(M,\Omega^1_{\dd_f}\aa_f)
\label{covder}.\ee
The field strength decomposes as
\bb F_t=(F,C-\alpha C,-\dee\Phi\gamma_5),\ee
with
\bb F&=&\de A+A^2\ \in\Omega^2(M,\gg),\\
C&=&\delta_f H+H^2
\in\Omega^0(M,\Omega^2_{\dd_f}\aa_f).\ee
The
internal field strength $C$, $C$ for curvature, should
not be confused with the $C$ of charge conjugation. 
$\alpha C\in\Omega^0(M,\Omega^2_{\dd_f}\aa_f+
{\cal J}^2_f)$ is the tricky piece of the computation, it
comes from the interference in degree two of
spacetime junk and internal junk. The former is
isomorphic to $\Omega^0 M$, a happy
circumstance that allows us to
compute $\alpha C$ pointwise \cite{sz}. For fixed $x$,
$C\in\Omega_{\dd_f}\aa_f\subset {\rm End}\hh_f$
and
$\alpha C\in
\pi(\Omega\aa_f)\subset {\rm End}\hh_f$ are finite
dimensional operators, i.e. matrices. Let us denote by 
$<\kappa,\varphi>\ =
{\textstyle\frac{1}{2}}\re\t
[(\kappa+J_f\kappa J_f^{-1})^*(\varphi+J_f
\varphi J_f^{-1})]$ the finite dimensional scalar
product. Then $\alpha C$ is uniquely determined by
the linear equations
\bb <r,C-\alpha C>\ =0&\qq {\rm for\ all}\ &
r\in\rho_f(\aa_f),\label{r}\\ 
<j,C-\alpha C>\ =0&\qq {\rm for\ all}\ &
j\in\jj^2_f,\label{j}\ee
where the trace is over the finite dimensional Hilbert
space $\hh_f$. 
Under a gauge transformation $u(x)$, the field
strength transforms homogeneously and we can
define, as before, the Yang-Mills action,
\bb S_{\rm YM}[A_t]=z(F_t,F_t) =z {\rm Re\,}\tt(
 F_t^*F_t|\dd_t|^{-4}).\ee 
The differential algebra
contains the Lie algebra as 0-forms and the scalar
product $(\cdot,\cdot)$ 
 restricted to the Lie algebra is an
invariant scalar product. Therefore this action is
gauge invariant. Let us decompose it,
$S_{\rm YM}[A_t]=S_{\rm YM}[A,H]$: 
\bb S_{\rm YM}[A,H]=
{\textstyle\frac{z}{8\pi^2}}\int_M( F,*F)+\,
{\textstyle\frac{z}{8\pi^2}}\int_M( \dee\Phi,*
\dee\Phi)+\,
{\textstyle\frac{z}{8\pi^2}}\int_M*V(H),
\label{ymh3}\ee
with
\bb V(H)=\ <C-\alpha C,C-\alpha C>\ 
=(C,C)\,-<\alpha C,\alpha C>.\ee
The first term, a non-Abelian Yang-Mills action, is no
surprise. The second, a Klein-Gordon action,
propagates the Higgs scalar. The Higgs potential
$V(H)$ breaks the gauge group spontaneously, if the
fermions break parity. As we shall see, the
computation of the Higgs sector, representation and
potential, will be intricate even though it follows from
a simple geometric definition,
$S_{\rm YM}[A_t]=z(F_t,F_t)$. 

To end this section, we mention the Dirac
Lagrangian, $\lll_{\rm Dirac}=\psi^*\dd_{t,{\rm
cov}}\psi$. The total, covariant Dirac operator is
\bb \dd_{t,{\rm cov}}=\dd_t-\pi_t(A_t)+
J_t(\dd_t-\pi_t(A_t))J_t^{-1}.\label{covdir}\ee
It is covariant with respect to the {\it group}
representation,
\bb \rho_{\rm
spinor}(u)\,\psi\ =\
\rho_t(u)\,J_t\rho_t(u)J^{-1}_t\
\psi,\qq u\in U(\aa_t)=\,^MU(\aa_f).\ee
 Note the appearance of charge conjugation that will
be crucial. The decomposition of this Lagrangian is:
\bb\lll_{\rm
Dirac}=\psi^*(2\ddd+i\rho_f(\aaa)+J_ti\rho_f(\aaa)
J_t^{-1})
\psi+
\psi^*(i\Phi\gamma_5+J_ti\Phi\gamma_5J_t^{-1})\psi.
\label{dirac3}\ee 
In words: almost commutative geometry promotes the
Higgs scalar to a connection and thereby unifies the
gauge couplings hidden in $\rho_f(\aaa)$ with
the Yukawa couplings hidden in $\Phi$.

The general Poincar\'e duality of noncommutative
geometry is beyond the scope of this introduction. In
the almost commutative case the Poincar\'e duality
can be stated  easily. Since it holds in commutative
geometry we only have to worry about the finite
dimensional internal space. Let $p_j$ be a set of
minimal projectors of $\aa_f$ and define the
intersection form $\cap$ to be the matrix
\bb \cap_{ij}:=\t
\left[\chi_f\,\rho_f(p_i)\,J_f\rho_f(p_j)J_f^{-1}
\right].\ee
Poincar\'e duality holds if and only if the intersection
form is non-degenerate, $\det \cap\not= 0$. Note that
in the finite dimensional case, the Poincar\'e duality
does not involve the Dirac operator.

\section{A minimax example}

\label{mima}
It is time for an example. To the best of my
knowledge, the simplest, nontrivial example -- a
maximum of pleasure with a minimum of effort --
is quite complicated. Strange enough, it resembles
the standard model of electro-weak forces, the
example section \ref{example}. 

We just learned that all computations can be done in
the finite dimensional, internal space. Therefore we
drop the subscript $\cdot_f$. Consider the internal
spectral triple,
\bb \aa&=&\hhh\op\cc\ \owns \ (a,b),\\
\hh&=&\hh_L\op\hh_R\op\hh_L^c\op\hh_R^c\ =\ 
(\cc^2\op\cc\op\cc^2\op\cc)\,\ot\cc^N,\\
\rho(a,b)&=&\pp{\rho_L(a)&0&0&0\cr 
0&\rho_R(b)&0&0\cr 
0&0&\bar\rho_L^c(b)&0\cr 
0&0&0&\bar\rho_R^c(b)}\cr &=&\pp{
a\ot 1_N&0&0&0\cr 
0&\bar b 1_N&0&0\cr 
0&0& b1_2\ot 1_N&0\cr 
0&0&0&b1_N},\\
\dd&=&\pp{0&\mm&0&0\cr 
\mm^*&0&0&0\cr 
0&0&0&0\cr 
0&0&0&0},\qq \mm=\pp{0\cr 1}\ot M_e,\qq
M_e=\pp{m_e&0\cr 0&m_\mu},\\
\chi&=&\pp{-1_{2N}&0&0&0\cr 0&1_N&0&0\cr 
0&0&-1_{2N}&0\cr 0&0&0&1_N},\\
J&=&\pp{0&1_{3N}\cr 1_{3N}&0}\circ {\rm complex
\ conjugation}.\ee
We denote by $\hhh$ the $real$, four dimensional
algebra of quaternions. We write its elements as
complex
$2\times 2$ matrices,
\bb a=\pp{x&-\bar y\cr y&\bar x},\qq x,y\in\cc.\ee
The involution in $\hhh$ is Hermitian conjugation
and  the group of unitaries of $\hhh$ is $SU(2)$. The
algebra $\cc\owns b$ is also taken as real, two
dimensional algebra. The physical basis of the {\it
complex} fermionic Hilbert space consists of an
electron and its left-handed neutrino in the first
generation and a muon and its left-handed neutrino
in the second generation. Of course there are also the
anti-particles,
\bb 
\pp{\nu_e\cr e}_L,\ \pp{\nu_\mu\cr\mu}_L,\ 
\qq  e_R,\qq \mu_R,\qq\qq
\pp{\nu_e\cr e}^c_L,\ \pp{\nu_\mu\cr\mu}^c_L,\ 
\qq  e^c_R,\qq \mu^c_R. \ee
$N$ counts the number of generations, $N=2$. 

We are ready to turn the crank and start with the
commutator
\bb [\dd,\rho(a,b)] &=& \pmatrix
{0&\mm\rho_R(b)-\rho_L(a) \mm&0&0\cr
\mm^*\rho_L(a)-\rho_R(b)
\mm^*&0&0&0\cr 0&0&0&0\cr 0&0&0&0}.\ee   
We take advantage of the following simplification in
our model,
\bb\mm\rho_R(b)=\rho_L\pp{b&0\cr 0&\bar b}\mm=
:\rho_L(B)\mm \label{simp}\ee
to compute a general 1-form. It is a sum of terms 
\bb   
\pi((a_0,b_0)\delta(a_1,b_1)) =
-i \pmatrix
{0&\rho_L(a_0(B_1-a_1))
\mm&0&0\cr
\mm^*\rho_L(B_0(a_1-B_1)) &0&0&0
\cr 0&0&0&0\cr 0&0&0&0}\ee  
 and as
vector space 
\bb   \Omega_\dd^1\aa = \left\{i\pmatrix
{0&\rho_L(h)\mm&0&0\cr \mm^*\rho_L(\tilde
h^*)&0&0&0
\cr 0&0&0&0\cr 0&0&0&0},\ h,\tilde h\in
\hhh\right\}.\ee
The Higgs being an anti-Hermitian 1-form
\bb H=  i\pmatrix
{0&\rho_L(h)\mm&0&0\cr 
\mm^*\rho_L(h^*)&0&0&0
\cr 0&0&0&0\cr 0&0&0&0},\qq
h=\pp{h_1&-\bar h_2\cr h_2&\bar h_1}\in \hhh\ee
is parameterized by one complex doublet
\bb \pp{h_1\cr h_2},\qq h_1,h_2\in\cc.\ee
  Likewise a general element in
$\pi(\Omega^2\aa)$ is 
\bb
&&\pi((a_0,b_0)\delta(a_1,b_1)\delta(a_2,b_2))=\ee
$$\pmatrix
{\rho_L(a_0)\rho_L(a_1-B_1)
\mm\mm^*\rho_L(a_2-B_2)&0&0&0\cr
0&\mm^*\rho_L(B_0)
\rho_L(a_1-B_1)
\rho_L(a_2-B_2)\mm&0&0
\cr 0&0&0&0\cr 0&0&0&0}.$$
We rewrite the (1,1) matrix element,
\bb(1_2\ot\Sigma) \rho_L(a_0)
\rho_L(a_1-B_1)
\rho_L(a_2-B_2) 
+(1_2\ot\Delta)
\rho_L(a_0)
\rho_L(a_1-B_1)(\sigma_3\ot1_N)
\rho_L(a_2-B_2),\eee 
 where we have
used the decomposition 
\bb   \mm\mm^* = \pmatrix
{0&0\cr 0&M_eM_e^*} = 1_2\ot\Sigma
+\sigma_3\ot\Delta\ee   with   
\bb   \sigma_3 :=
\pmatrix {1&0\cr 0&-1},\qq
\Sigma={\textstyle\frac{1}{2}}M_eM_e^*,\qq
\Delta=-{\textstyle\frac{1}{2}}M_eM_e^*.\ee   
A general element in
$(\ker\pi)^1$ is a finite sum of the form 
\bb    \sum_j
(a^j_0,b^j_0)\delta (a^j_1,b^j_1) \ee  
 with the two
conditions 
\bb   \left[ \sum_j\rho_L( a^j_0)
\rho_L(a^j_1-B^j_1)\right]\mm
= 0,\qq \mm^*\left[ \sum_j \rho_L(B^j_0)
\rho_L(a^j_1-B^j_1)\right] =
0.\ee   
Therefore the corresponding general element
in
$\pi(\delta (\ker\pi)^1)$ has only one nonvanishing
matrix element in position (1,1):  
\bb    (1_2\ot\Sigma)\sum_j
\rho_L(a^j_0-B^j_0)\rho_L(a^j_1-B^j_1)+
(1_2\ot\Delta)\sum_j\rho_L(a^j_0-B^j_0)
(\sigma_3\ot 1_N)\rho_L(a^j_1-B^j_1)\ee  
 still subject to the two conditions and we have  the
following inclusion 
\bb   \pi(\delta(\ker\pi)^1) \supset 
\left\{i\pmatrix {(1_2\ot\Delta)\sum_j
 \rho_L(a^j_0(i\sigma_3)
a^j_1)&0&0&0\cr 0&0&0&0\cr 0&0&0&0\cr 0&0&0&0
},\quad \sum_j a^j_0a^j_1=0 \right\}.\ee
Note that $\rho_L$ is faithful and that 
\bb\left\{\sum_j
 a^j_0(i\sigma_3)
a^j_1,\quad \sum_j a^j_0a^j_1=0 \right\}\eee
is an ideal in $\hhh$. This ideal is not zero, take for
example
\bb a_0^1= \pp{1&0\cr 0&1},\qq
a_1^1=\pp{0&-1\cr 1&0},\qq
a_0^2=\pp{0&i\cr i&0},\qq
a_1^2=\pp{i&0\cr 0&-i},\eee
with
\bb \sum_j a^j_0(i\sigma_3) a^j_1=-2\pp{0&i\cr
i&0}.\ee
The quaternions being a simple algebra,
the ideal is the entire algebra and the junk is
\bb \jj^2=   \pi(\delta(\ker\pi)^1) 
 = \left\{ i\pmatrix {j\ot\Delta &0&0&0\cr 0&0&0&0
\cr 0&0&0&0
\cr 0&0&0&0}, \quad j \in
\hhh\right\}.\ee  
 Next we have to project out the junk using the scalar
product,
\bb \Omega_\dd^2\aa=\left\{\pp{
\tilde c\ot \Sigma &0&0&0\cr 
0&\mm^*\rho_L(c)\mm&0&0
\cr 0&0&0&0\cr 0&0&0&0},\qq 
\tilde c,c\in\hhh\right\}.\ee
Since $\pi$ is a
homomorphism of involution algebras, the product  in
$\Omega_\dd\aa$ is given by matrix multiplication
followed by the orthogonal projection $P$ and
the involution  is given by transposition complex
conjugation. In order to
calculate the  differential $\delta$, we go
back to the universal differential envelope. The result
is 
\bb\delta :
\Omega_\dd^1\aa&\longrightarrow&
\Omega_\dd^2\aa\cr
\nobreak\cr 
i\pmatrix
{0&\rho_L(h)\mm&0&0\cr \mm^*\rho_L(\tilde
h^*)&0&0&0\cr 0&0&0&0\cr 0&0&0&0}
 &\longmapsto& \pp{
\tilde c\ot \Sigma&0&0&0\cr 
0&\mm^*\rho_L(c)\mm&0&0
\cr 0&0&0&0\cr 0&0&0&0}\ee    
with
$\tilde c=c=h+\tilde h^*.$

We are now in position to compute the curvature:
\bb C:=\delta H+H^2=
\left(1-|\varphi|^2\right)\pp{1_2\ot\Sigma&0&0&0
\cr 0&\mm^*\mm&0&0
\cr 0&0&0&0\cr 0&0&0&0}\ee 
with the homogeneous scalar
variable
\bb \Phi:= H-i\dd=:i\pmatrix
{0&\rho_L(\phi)\mm&0&0\cr
\mm^*\rho_L(\phi^*)&0&0&0
\cr 0&0&0&0\cr 0&0&0&0},\qq
\phi=\pp{\varphi_1&-\bar \varphi_2\cr
\varphi_2&\bar \varphi_1}\in \hhh,\label{ffi}\ee
 \bb \varphi=\pp{\varphi_1\cr \varphi_2},\qq
 |\varphi|^2=|\varphi_1|^2+|\varphi_2|^2.\ee
In the example of section \ref{example}
we also had two useful parameterizations of the scalar
field, $\varphi$ and $h$. They coincide precisely with
the two parameterizations here, only they appear in
opposite chronology. 
 The computation of $\alpha C$ is long but
presents no difficulty. In this example there is no junk
component:
\bb\alpha C= (1-|\varphi|^2)\,\rho(\alpha 1_2,\beta).
\ee
The real numbers $\alpha$ and $\beta$ are
determined by the two linear equations
\bb\begin{array}{rcrcl}
N \alpha&+&N\beta
&=&
{\textstyle\frac{1}{2}}(m_e^2+m_\mu^2),\cr \cr 
N\alpha&+&3N \beta&=&
{\textstyle\frac{3}{2}}(m_e^2+m_\mu^2).
\end{array}\label{sys}\ee
Their solution is
\bb \alpha=0,\qq\beta={\textstyle\frac{1}{2N}}
(m_e^2+m_\mu^2),\ee
and the Higgs potential is,
\bb V(H)
=(C,C)\,-<\alpha C,\alpha
C>={\textstyle\frac{z}{8\pi^2}}
{\textstyle\frac{3}{2}}(1-|\varphi|^2)^2[m_e^4+
m_\mu^4-{\textstyle\frac{1}{N}}(m_e^2+m_\mu^2)^2].
\ee
Now we can explain why our minimax model must
contain at least $N=2$ generations of leptons with
distinct masses. Otherwise the Higgs potential
vanishes. 

Next we compute the Klein-Gordon action,
\bb {\textstyle\frac{z}{8\pi^2}}\int_M( \dee\Phi,*
\dee\Phi)=
{\textstyle\frac{z}{8\pi^2}}2(m_e^2+m_\mu^2)\int_M
\dee\varphi^**\dee\varphi.\ee
The covariant derivative with respect to the gauge
potential
$A=(^{(2)}\!A,\ {\textstyle\frac{1}{2}}\,^{(1)}\!A)
\in\Omega^1(M,su(2)
\op u(1))$ follows from the group representation
carried by the scalar doublet,
 equations (\ref{covder}) and
(\ref{ffi}),
\bb \dee\varphi=\de\varphi\,+\,^{(2)}\!A\varphi\,+\,
{\textstyle\frac{1}{2}}\,^{(1)}\!A\varphi.\ee
The factor ${\textstyle\frac{1}{2}}$ in front of 
$^{(1)}\!A$ is conventional: we want the hypercharge
of the Higgs scalar to be one half. To put the scalar
Lagrangian into conventional form,
\bb
{\textstyle\frac{1}{2}}\dee_\mu\varphi_{\rm ph}^*
\dee^\mu\varphi_{\rm ph}\,+\,
\lambda|\varphi_{\rm ph}|^4\,-\,
{\textstyle\frac{1}{2}}\mu^2|\varphi_{\rm ph}|^2,\ee
we renormalize the scalar field,
\bb |\varphi_{\rm ph}|^2:=
\left[{\textstyle\frac{z}{8\pi^2}}\,4\,(m_e^2+m_\mu^2)
\right]\,|\varphi|^2.\ee
The physical scalar $\varphi_{\rm ph}$ now has the
correct dimensions of a mass and we will drop the
subscript $\cdot_{\rm ph}$. For the scalar couplings
we get,
\bb \lambda\,=\,{\frac{3\pi^2}{4z}}\,
\left[\,\frac{m_e^4+m_\mu^4}{(m_e^2+m_\mu^2)^2}\,
-\,{\frac{1}{N}}\,\right],\qq
\mu^2\,=\,{\textstyle\frac{3}{2}}
\left[\,\frac{m_e^4+m_\mu^4}{(m_e^2+m_\mu^2)}\,
-\,{\frac{1}{N}}\,(m_e^2+m_\mu^2)\right].\ee
The energy of the vacuum or cosmological constant
$V(\varphi_0)$ vanishes automatically.  The vacuum
expectation is,
\bb |\varphi_0|^2= v^2={\textstyle\frac{z}{2\pi^2}}
(m_e^2+m_\mu^2).
\ee
and the group of unitaries $SU(2)\times
U(1)$ is broken spontaneously down to $U(1)$. 
To avoid any misunderstanding, the miracle is not the
symmetry breaking. This symmetry breaking is
introduced by hand with the masses for chiral
fermions. The miracle is that this explicit symmetry
breaking produces a Higgs field and that this
Higgs field promotes the symmetry breaking from
explicit to spontaneous. The spontaneous symmetry
breaking in turn produces the gauge boson masses.
In other
words, in almost commutative geometry the invariance
group of the fermionic mass matrix is necessarily
equal to the invariance group of the mass matrix of
the gauge bosons, the little group. This is not
true in a general Yang-Mills-Higgs model, but it is
true in the standard model. 

We compute the Yang-Mills action,
\bb {\textstyle\frac{z}{8\pi^2}}\int_M( F,*F)=
{\textstyle\frac{z}{8\pi^2}}\int_M
\left[N\t ^{(2)}\!F^**\,^{(2)}\!F\,+\,
{\textstyle\frac{3}{2}}N\,^{(1)}\!F^**\,^{(1)}\!F
\right].\ee
Comparing with the action in conventional form,
\bb
{\textstyle\frac{1}{2}}
\int_M
\left[{\textstyle\frac{2}{g_2^2}}\t
^{(2)}\!F^**\,^{(2)}\!F\,+\,
{\textstyle\frac{1}{g_1^2}}\,^{(1)}\!F^**\,^{(1)}\!F,
\right]\ee
we get the gauge couplings,
\bb
g_2^2=\,\frac{8\pi^2}{Nz}\,,\qq
g_1^2=\,\frac{8\pi^2}{3Nz}\,.\ee
The weak mixing angle $\theta_w$ is therefore fixed,
\bb\sin^2\theta_w:=\,\frac{g_1^2}{g_1^2+g_2^2}\,=
{\textstyle\frac{1}{4}}.\ee
Also fixed is the $\rho$-factor, 
\bb
\rho:=\frac{m_W^2}{\cos^2(\theta_w) \ m_Z^2}=1.
\ee
It is unit because the scalar sits in a doublet.
 
Noncommutative geometry unifies the gauge, Higgs
and Yukawa  couplings, in the same way that gauge
invariance unifies the tri- and quadri-linear self
couplings of the gauge bosons and the minimal
couplings of the gauge bosons to fermions and scalars:
\bb
\lambda&=&{\textstyle\frac{3}{32}}(N-1)g_2^2\,+\,
O(m_e^2/m_\mu^2)\,g_2^2,\\
g_e^2&=&{\frac{m_e^2}{v^2}}\ =\
{\textstyle\frac{2\pi^2}{z}}\,\frac{m_e^2}
{m_e^2+m_\mu^2}\ =\ O(m_e^2/m_\mu^2)\,g_2^2,\\
g_\mu^2&=&{\frac{m_\mu^2}{v^2}}\ =\
{\textstyle\frac{2\pi^2}{z}}\,\frac{m_\mu^2}
{m_e^2+m_\mu^2}\ =\ {\textstyle\frac{N}{4}}g_2^2\,+
\,O(m_e^2/m_\mu^2)\,g_2^2.
\ee

A general lesson that we learn from our minimax
example is the link between parity break down and
spontaneous gauge symmetry break down. They go
together in almost commutative models. Indeed take
any vectorial model, that means $\rho_L=\rho_R$ and
a mass matrix $\mm$ commuting with this
representation $\rho_L$. Examples of vectorial
theories are the parity preserving electromagnetic
and strong forces. For these models, the internal
differential forms vanish identically except in degree
zero. Consequently there is no Higgs scalar, no
spontaneous symmetry break down and the gauge
bosons, e.g. the photon and the gluons remain
massless. The Yang-Mills-Higgs model building kit on
the other hand allows for spontaneous symmetry
break down of any model, parity violating or
vectorial. 

The last item we have to discuss is Poincar\'e duality.
There are two minimal projectors, $p_1=(1_2,0)$ and
$p_2=(0,1)$. The intersection form is computed easily,
\bb \cap_{ij}:=\t
\left[\chi\,\rho(p_i)\,J\rho(p_j)J^{-1}\right]=
-2N\pp{0&1\cr 1&-1}.\ee
It is non-degenerate and  Poincar\'e duality holds.

Before we leave our minimax model we must talk about
its short coming, 
the quarks with their electric charges are difficult to fit in.
This problem will be cured by the inclusion of
strong interactions.

\section{The standard model from Connes' first
dreisatz}\label{smconnes}

The strong interactions being vectorial their addition
to the minimax example is not difficult and we
go quickly over the calculations
\cite{book}\cite{cl}\cite{tresch}\cite{reviews}.
The finite dimensional algebra is chosen to reproduce 
$SU(2)\times U(1)\times SU(3)$,
\bb \aa=\hhh\op\cc\op M_3(\cc)\,\owns\,(a,b,c).\ee
The fermionic Hilbert spaces are copied from the
standard model,
\bb \hh_L&=&
\left(\cc^2\ot\cc^N\ot\cc^3\right)\ \op\ 
\left(\cc^2\ot\cc^N\ot\cc\right), \\
\hh_R&=&\left((\cc\op\cc)\ot\cc^N\ot\cc^3\right)\ 
\op\ \left(\cc\ot\cc^N\ot\cc\right).\ee
 In each summand, the first factor
denotes weak isospin doublets or singlets, the second
denotes
$N$ generations, $N=3$, and the third denotes color
triplets or singlets.
Let us choose the following basis
of 
$\hh=\cc^{90}$: 
\bb
& \pp{u\cr d}_L,\ \pp{c\cr s}_L,\ \pp{t\cr b}_L,\ 
\pp{\nu_e\cr e}_L,\ \pp{\nu_\mu\cr\mu}_L,\ 
\pp{\nu_\tau\cr\tau}_L;&\cr \cr 
&\matrix{u_R,\cr d_R,}\qq \matrix{c_R,\cr s_R,}\qq
\matrix{t_R,\cr b_R,}\qq  e_R,\qq \mu_R,\qq 
\tau_R;&\cr  \cr 
& \pp{u\cr d}^c_L,\ \pp{c\cr s}_L^c,\ 
\pp{t\cr b}_L^c,\ 
\pp{\nu_e\cr e}_L^c,\ \pp{\nu_\mu\cr\mu}_L^c,\ 
\pp{\nu_\tau\cr\tau}_L^c;&\cr\cr  
&\matrix{u_R^c,\cr d_R^c,}\qq 
\matrix{c_R^c,\cr s_R^c,}\qq
\matrix{t_R^c,\cr b_R^c,}\qq  e_R^c,\qq \mu_R^c,\qq 
\tau_R^c.&\eee
The representation $\rho$ acts on $\hh$ by
\bb \rho(a,b,c):=\pp{\rho_{w}(a,b)&0\cr 0&\bar\rho_{s}(b,c)} := 
\pp{\rho_{wL}(a)&0&0&0\cr 
0&\rho_{wR}(b)&0&0\cr 
0&0&{\bar\rho_{sL}(b,c)}&0\cr 
0&0&0&{\bar\rho_{sR}(b,c)}}\ee
with
\bb\rho_{wL}(a)&:=&\pp{
a\ot 1_N\ot 1_3&0\cr
0&a\ot 1_N&},\qq
\rho_{wR}(b)\ :=\ \pp{
B\ot 1_N\ot 1_3&0\cr
0&\bar
b1_N},\\ \cr &&
B:=\pp{b&0\cr 0&\bar b},
\\ \cr    
  \rho_{sL}(b,c)&:=&\pp{
1_2\ot 1_N\ot c&0\cr
0&\bar b1_2\ot 1_N},\qq
\rho_{sR}(b,c)\ :=\ \pp{
1_2\ot 1_N\ot c&0\cr
0&\bar b1_N}.   
\ee
The chosen representation $\rho$ takes into
account weak interactions $\rho_w(a,b),\ a\in\hhh,\ 
b\in\cc$, and strong interactions $\rho_s(b,c),\ c\in
M_3(\cc)$, $c$ for color. This choice discriminates
between leptons (color singlets) and quarks (color
triplets). The role of $b\in\cc$ appearing in both
weak interactions $\rho_w(a,b)$ and strong
interactions $\rho_s(b,c)$ is crucial to make
$\rho(a,b,c)$ a representation of $\aa$ and is crucial
for weak hypercharge computations. There is an
apparent asymmetry between particles and
anti-particles, the former are subject to weak, the
latter to strong interactions. However, since particles
and anti-particles are permuted in the covariant Dirac
operator (\ref{covdir}) by
\bb J=\pp{0&1_{15N}\cr 1_{15N}&0}\circ 
\ {\rm complex\ conjugation},\ee
the theory is invariant under charge conjugation.
 For
completeness, we record the chirality as matrix
\bb \chi=\pp{-1_{8N}&0&0&0\cr 0&1_{7N}&0&0\cr
 0&0&-1_{8N}&0\cr 0&0&0&1_{7N} }.\ee
The
third item in the spectral triple is the Dirac operator
\bb \dd=\pp{0&\mm&0&0\cr 
\mm^*&0&0&0\cr 
0&0&0&0\cr 
0&0&0&0}.\ee
The fermionic mass matrix of the standard model is
\bb\mm=\pp{
\pp{1&0\cr 0&0}\ot M_u\ot 1_3\,+\,
\pp{0&0\cr 0&1}\ot M_d\ot 1_3
&0\cr
0&\pp{0\cr 1}\ot M_e},\ee
with
\bb M_u:=\pp{
m_u&0&0\cr
0&m_c&0\cr
0&0&m_t},&& M_d:= C_{KM}\pp{
m_d&0&0\cr
0&m_s&0\cr
0&0&m_b},\\ M_e:=\pp{
m_e&0&0\cr
0&m_\mu&0\cr
0&0&m_\tau}.&&\ee
 All indicated fermion masses are supposed positive and
different. The
Cabibbo-Kobayashi-Maskawa matrix  $C_{KM}$
is supposed non-degenerate in the sense that there is
no simultaneous mass and weak interaction
eigenstate.

Let us turn the crank and record a few
intermediate steps:
\bb   \Omega_\dd^1\aa = \left\{i\pmatrix
{0&\rho_{wL}(h)\mm&0&0\cr \mm^*\rho_{wL}(\tilde
h^*)&0&0&0\cr 
0&0&0&0\cr 0&0&0&0}
,\ h,\tilde h\in \hhh\right\}.\ee
The Higgs being an anti-Hermitian 1-form
\bb H=  i\pmatrix
{0&\rho_{wL}(h)\mm&0&0\cr 
\mm^*\rho_L(h^*)&0&0&0\cr 
0&0&0&0\cr 0&0&0&0},\qq
h=\pp{h_1&-\bar h_2\cr h_2&\bar h_1}\in \hhh\eee
is parameterized by one complex doublet
\bb \pp{h_1\cr h_2},\qq h_1,h_2\in\cc.\ee
The internal junk in degree two is again
\bb \jj^2=\left\{i\pp{j\ot\Delta&0&0&0\cr 
 0&0&0&0\cr 
 0&0&0&0\cr 0&0&0&0},\qq
j\in\hhh\right\}\ee
with
\bb\Delta:={\textstyle\frac{1}{2}}\pp{
\left(M_uM_u^*-M_dM_d^*
\right)\ot 1_3&0\cr 
0&-M_eM_e^*}.\ee
also containing the quark masses.
The homogeneous scalar variable is:
\bb \Phi:= H-i\dd=:i\pmatrix
{0&\rho_{wL}(\phi)\mm&0&0\cr
\mm^*\rho_{wL}(\phi^*)&0&0&0
\cr 0&0&0&0\cr 0&0&0&0},\qq
\phi=\pp{\varphi_1&-\bar \varphi_2\cr
\varphi_2&\bar \varphi_1}\in \hhh,\ee
and with
$\varphi\ :=\ ^t\!(\varphi_1,\varphi_2)$,
the internal field strength is:
\bb C&:=&\delta H+H^2=
\left(1-|\varphi|^2\right)\pp{1_2\ot\Sigma&0&0&0
\cr 0&\mm^*\mm&0&0
\cr 0&0&0&0\cr 0&0&0&0},  \\ \cr 
&&\Sigma:={\textstyle\frac{1}{2}}\pp{
\left(M_uM_u^*+M_dM_d^*
\right)\ot 1_3&0\cr 
0&M_eM_e^*}.\ee

Again
$\alpha C$ has no junk component, 
\bb\alpha C=(1-|\varphi|^2)\,\rho(\alpha
1_2,\beta,\gamma 1_3).\ee
To compute the real numbers $\alpha,\ \beta,\
\gamma$, we neglect all fermion masses with respect
to the top mass. This approximation is 
good to $m_b^2/m_t^2= 0.0006$ and we have the
three linear equations:
\bb\begin{array}{rcrcrcr}
4N\, \alpha&+&N\, \beta&+&3N\, \gamma
&=&
{\textstyle\frac{3}{2}}m_t^2\,\cr 
2N\, \alpha&+&12N\, \beta&+&
6N\, \gamma&=&3m_t^2\,\cr 
3N\, \alpha&+&3N\,\beta&+&
6N\,\gamma&=& 3m_t^2,
\end{array}\ee
with solution
\bb \alpha=0,\qq \beta=0,\qq \gamma=\,\frac{1}{2N}\,
m_t^2.\ee
The Higgs and Yukawa couplings follow:
\bb \mu^2&=&\left(\,\frac{3}{2}\,-\,\frac{1}{N}\,
\right)\,m_t^2,\\
\lambda&=&\,\frac{\pi^2}{6z}\,
\left(\,\frac{3}{2}\,-\,\frac{1}{N}\,\right),\\
g_t^2&=&\,\frac{m_t^2}{v^2}\ =\
\frac{2\pi^2}{3z}\,.\ee
 Before 
computing the gauge couplings, we face a problem.
The group of unitaries $SU(2)\times U(1) \times U(3)$
is too big by one $U(1)$ factor. Indeed there is no
associative algebra with $SU(3)$ as unitary group.
Howerver there is an encouraging miracle, the
representation of a linear combination of the two
$u(1)$s coincides with the representation of the
hypercharge $Y$ in the standard model. This miracle
needs three colours and vectorial couplings of the
$U(3)$. These vectorial couplings, in turn, are an
immediate consequence of the first order condition in
spectral triples together with the maximal parity
violation of weak interactions
\cite {tresch}\cite{becca}\cite{tkmz}. All four  {\it ad
hoc} features of the standard model, 
\begin{itemize}\item
gluons couple vectorially,
\item
gluons are massless,
\item
the $W$ couples axially,
\item
the $W$ is massive,
\end{itemize}
are rigidly tied together by the axioms of the spectral
triple. 
To obtain the standard model we can 
project out the other, unwanted linear combination of
 the two $u(1)$s in $\gg$ by imposing the so-called
unimodularity condition, 
\bb\t \left[P\left(\rho(a,b,c)+J\rho(a,b,c)J^{-1}
\right)\right]=0,\ee
 where $P$ is the projection on $\hh_L\op\hh_R$, the
space of particles. 
 We note
that this condition is equivalent to the condition of
vanishing gauge anomalies \cite{anom}. Nevertheless
the unimodularity condition is at this stage an artefact.
Connes second dreisatz will improve this situation
\cite{grav}. 

The computation of the gauge couplings is now
straightforward,
\bb g_3^{2}=\,\frac{2\pi^2}{Nz}\,,\qq
g_2^{2}=\,\frac{2\pi^2}{Nz}\,,\qq 
g_1^{2}=\,\frac{6\pi^2}{5Nz}\,.\qq\ee
In particular we have 
\bb\sin^2\theta_w=
{\textstyle\frac{3}{8}},\qq  g_3=g_2,\label{grand}\ee 
as in grand
unified theories and from the geometric unification
of gauge and Higgs bosons,
\bb \lambda={\textstyle\frac{3N-2}{24}}g_2^2,\qq
 g_t^2={\textstyle\frac{N}{3}}g_2^2.\label{petit}\ee
The confrontation of these four constraints with
experiment calls for the renormalization group flow
to be discussed in the next chapter. 

Before leaving the standard model, we must verify its
Poincar\'e duality. We have now three minimal
projectors, 
\bb p_1=(1_2,0,0),\qq p_2=(0,1,0), \qq p_3=
\left(0,0,\pp{1&0&0\cr 0&0&0\cr 0&0&0}\right).\ee
Note that $1_3$ is not minimal in $M_3(\cc)$ because
it is a sum of three projectors of rank one,
\bb \pp{1&0&0\cr 0&1&0\cr 0&0&1}=
\pp{1&0&0\cr 0&0&0\cr 0&0&0}+
\pp{0&0&0\cr 0&1&0\cr 0&0&0}+
\pp{0&0&0\cr 0&0&0\cr 0&0&1}. \ee
All three are unitarily equivalent.
The intersection form,
\bb \cap=-2N\pp{0&1&1\cr 1&-1&-1\cr 1&-1&0},\ee 
is non-degenerate. However if we add right-handed
neutrinos to the standard model, massive or not, then
the intersection form,
\bb \cap=-2N\pp{0&1&1\cr 1&-2&-1\cr 1&-1&0},\ee 
is degenerate and Poincar\'e duality fails.

\section{Necessary conditions}

 We have become accustomed to see 
supersymmetric versions of any theory or model
already on the market, supersymmetric
quantum mecanics, supersymmetric Yang-Mills
theories, supersymmetric $\sigma$-models, super
gravities, super strings,... You should not believe that
you can put noncommutative in front of any theory,
not even in front of any Yang-Mills theory. It
remains a miracle that the standard model is in the
tiny, priviledged class of Yang-Mills theories allowing
a noncommutative generalization and that putting
almost commutative in front of the standard model
produces its correct Higgs sector. The purpose of the
present section is to assess this miracle. Needless to
say that we call it a miracle because we do not have
the slightest explanation for it today.

Recall the input bills of a Yang-Mills theory, a finite
dimensional, real, compact Lie group $G$ and two
unitary representations $\rho_L$ and $\rho_R$. The
classification of all such groups teaches us that its Lie
algebras $\gg$ are direct sums of simple Lie algebras
from the three classical series $so(n)$, $su(n)$,
$sp(n)$, and of the five exceptional Lie algebras
$G_2$, $F_4$, $E_6$, $E_7$, $E_8$. Each of the simple
Lie algebras has an infinite number of irreducible
representations, for example $u(1)=so(2)$ has one
irreducible representation for any charge $y\in\rr$
and $su(2)=sp(1)$ has an irreducible representation of
any dimension $d\in\nn$ corresponding to spin
$j=(d-1)/2$. 

The input bills of an almost commutative Yang-Mills
theory are a finite dimensional, real, associative
involution algebra with unit, $\aa$ and two faithful
representations $\rho_L$ and $\rho_R$. The
classification of these algebras is easier than the one
for groups. Any such algebra is a direct sum of
finite algebras from the three series,
$M_n(\rr)$, $M_n(\cc)$, $M_n(\hhh)$, the $n\times
n$ matrices with real, complex and quaternionic
entries. The corresponding groups of unitaries have
Lie algebras $so(n)$, $su(n)$, $sp(n)$. Therefore the
exceptional Lie groups are unsuitable for Connes'
dreisatz, not a great loss. Things are more exciting
concerning the representations. Any associative
algebra representation induces a Lie algebra
representation but only very few Lie algebra
representations can be extended to a representation of
the ambient associative algebra. The tensor product
of two $\gg$ representations is a $\gg$
representation. The tensor product of two $\aa$
 representations is not an $\aa$ representation. The
only irreducible representations of $M_1(\cc)$ have
charge 1 and $-1$, the only irreducible
representation of $M_1(\hhh)$ is on $\cc^2$. In
general $M_n(\rr)$ has only one irreducible
representation, the fundamental one, on $\rr^n$,
$M_n(\cc)$ has two, the fundamental one, on $\cc^n$,
and its conjugate, and $M_n(\hhh)$ has one, the
fundamental one, on $\cc^{2n}$. Note that the
fermions of the standard model only contain colour
triplets and singlets and isospin doublets and singlets.
The singlets are admitted thanks to the real structure
$J$. The hypercharges may deviate from
$\pm 1$ thanks to the unimodularity condition. The
above general conditions on the group and its
fermionic representations exclude already all popular
grand unified models from almost commutative
geometry. 
 The axioms of the
spectral triple contain further restrictions on the
fermionic representations, the first order axiom
and Poincar\'e duality. The complete classification
of almost commutative geometries is given in
\cite{tkmz}. The standard model is in this
classification, the first order axiom implies that 
strong interactions are vectorial, Poincar\'e duality
excludes right-handed neutrinos. 

\begin{figure}[hbt]
\hspace{3.3cm}
\def\epsfsize#1#2{0.75#1}
\epsfbox{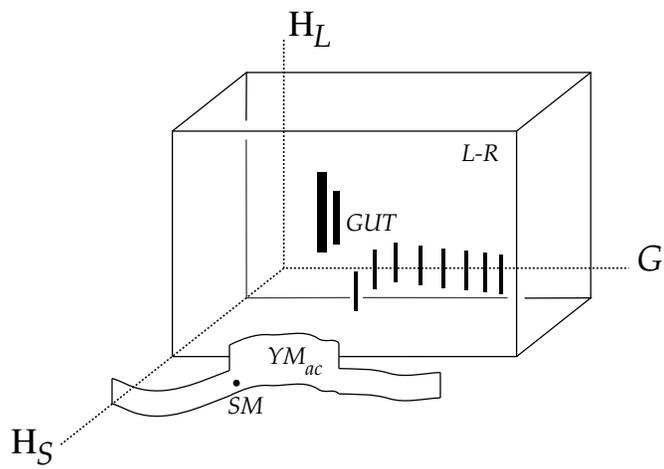}
\caption{ An artist's partial view of the space of bills of all
Yang-Mills-Higgs models and some of its subsets.
$GUT$ stands for `Grand Unified Theories',
 $L-R$ stands for left-right symmetric models,
 $SM$ stands for standard model
and $YM_{ac}$ for almost commutative Yang-Mills
models.}
\label{artist}
\end{figure}

 Concerning the
coins, the Yang-Mills input is any invariant scalar
product on the Lie algebra. In almost commutative
geometry, this scalar product is the restriction of a
scalar product on the entire space of differential
forms. However, anticipating on Connes' second
dreisatz we have not taken the most general such
scalar product on forms, but we have picked one of the
two simplest, (\ref {sp}). It involves only one positive
constant, $z$, and consequently the three gauge
couplings of the standard model are related by two
constraints, equations (\ref{grand}). Finally, in
almost commutative geometry all parameters in the
fermionic mass matrix are input coins. The scalar
representation is only a group representation and is
computed to be a subrepresentation of the tensor
product
$\hh_L^*\ot\hh_R$ and its conjugate \cite{versus}.
This subrepresentation depends on the details of the
fermionic mass matrix. The inclusion is however
sufficient to exclude all left-right symmetric models
from almost commutative geometry \cite{lr}. In
left-right symmetric models parity violation is
spontaneous, induced from the mass matrix of the
gauge bosons. Finally the Higgs couplings are also
computed as a function of the fermionic mass matrix,
equations (\ref{petit}) for the standard model.
The induced mass matrix of the gauge bosons has
the same invariance as the fermionic mass matrix. 
As the minimax example shows, the computation of the
Higgs representation and couplings is involved.
The most modest Yang-Mills-Higgs model
beyond the standard model has the group
$SU(3)\times SU(2) \times U(1) \times U(1)$. Any
model in almost commutative geometry yielding this
group, like for instance the standard model without
the unimodularity condition, is incompatible with
experiment \cite{beyond}.
At present we have no complete classification of all
Yang-Mills-Higgs models accessible to almost
commutative geometry. Figure \ref{artist} tries to give
an impression of the situation. 

%% file: monsa5
\chapter{Running coupling constants}

Quantum field theory teaches us that coupling
constants are functions of the energy used to measure
them. Today this energy dependence is accessible to
accelerator experiments. Physically it can be
understood in analogy with the screening effect from
condensed matter physics. The computational origin of
this energy dependence lies in divergent Feynman
diagrams.

Consider an electric charge $Q$ placed in a dielectric
medium, like water. The water molecules carry an
electric dipole moment. These dipoles orient
themselves around the charge such that the effective
charge seen from far away is smaller than $Q$: the
cloud of dipoles  surrounding the charge partially
screens $Q$. By convention we keep the charge
constant and say that the effective coupling constant 
$(e^2/\epsilon)^{1/2}$ has decreased. In the vacuum 
$\epsilon_0=8.85\cdot  10^{-12}{\rm
s^2\,C^2/(m^3\,kg)}$ is also called vacuum
permittivity, otherwise $\epsilon$ is the permittivity
of the dielectric medium, $\epsilon= 699\cdot 
10^{-12}{\rm s^2\,C^2/(m^3\,kg)}$ for water.

\begin{figure}[hbt]
\hspace{2cm}
\def\epsfsize#1#2{0.8#1}
\epsfbox{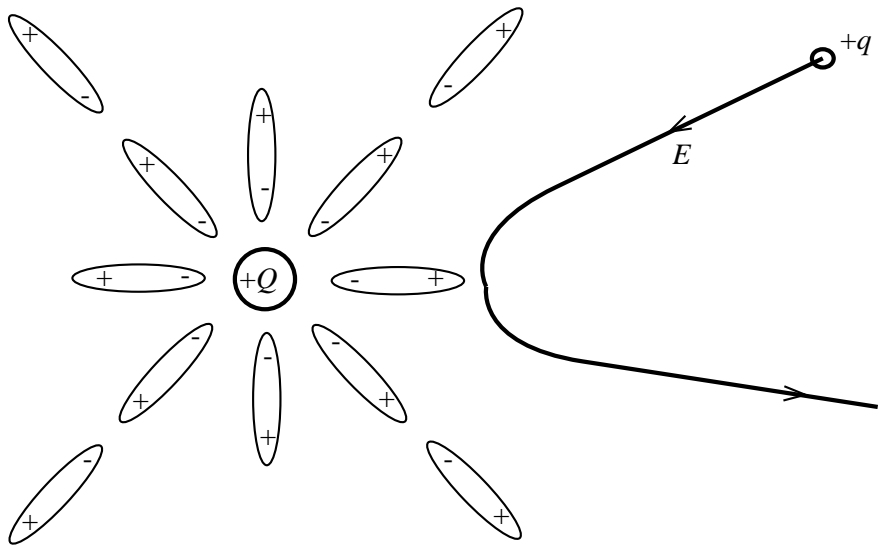}
\caption{Vacuum polarization}
\label{dipoles}
\end{figure}

Let us now place the central charge $Q$ in the
vacuum and let us measure the effective
coupling dynamically by scattering a test charge $q$
off the central charge $Q$ with an energy $E$, figure
\ref{dipoles}. Dirac tells us that electron positron pairs
are created, dipoles that will screen the central
charge like the dipole moments of the water molecules
before. With increasing energy the test charge
penetrates deeper into the dipole cloud and we
measure an increasing effective charge or
equivalently an increasing effective coupling
constant squared
$e^2/\epsilon_0(E)$. The vacuum permittivity is a
decreasing function of energy, for instance
$\epsilon_0=8.27\cdot 10^{-12}{\rm
s^2\,C^2/(m^3\,kg)}$ at $E=m_Z$. The effect is now
called vacuum polarization or running coupling
constant. 

\begin{figure}[hbt]
\hspace{3.66cm}
\def\epsfsize#1#2{0.8#1}
\epsfbox{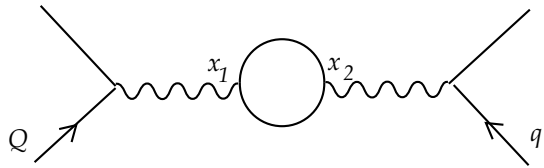}
\caption{A 1-loop graph}
\label{1-loop}
\end{figure}

\begin{figure}[hbt]
\hspace{3.66cm}
\def\epsfsize#1#2{0.8#1}
\epsfbox{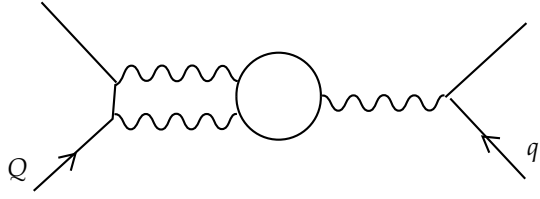}
\caption{A 2-loop graph}
\label{2-loop}
\end{figure}

The quantitative treatment of the running coupling is
cumbersome. So far we only have perturbative
calculations. The cross section is computed as a power
series in the fine structure constant
$e^2/(4\pi\epsilon_0\,\hbar c)$ at a fixed energy.
Even for small couplings this power series diverges
and physicists take a pragmatic point of view. As the
computation of the higher order terms is exceedingly
complicated, the power series is truncated at first (or
second) order. One talks about 1-loop contributions,
this means that the photon exchanged between the
central charge and the test charge produces one
particle antiparticle pair only, figure \ref{1-loop}.  In
2-loop one admits the possibility that one of the
particles of the pair may in turn produce new
particles e.g. via bremsstrahlung, figure
\ref{2-loop}. Even at 1-loop, one has to live with
divergent integrals, essentially short distance or
ultra-violet divergences. For example consider figure
\ref{1-loop}, call $x_1$  the point of
pair creation and $x_2$ the point of pair annihilation.
The integral over $x_1$ and $x_2$ diverges for short
distances between the two points. This divergence
has to be regularized to get a finite cross
section, a delicate man\oe uvre trusted only in
renormalizable models. Even in renormalizable models,
there
are different regularization schemes leading to
different cross sections. Fortunately the scheme
dependence is weak and again physicists take a
pragmatic point of view. This point of view is backed
by an impressive agreement between the computed 
and measured numbers and adds to the humiliation of
the standard model. Let us note that 95 \% of physics is
described without the use of loops. Renormalization is
needed to fit the experimental numbers with higher
precision.


One major motivation for noncommutative geometry
in particle physics is that a spacetime
uncertainty naturally cures short distance
divergences. In fact most of these divergences
are logarithmic and resemble the divergences
encountered under the Dixmier trace (\ref{dix}).

To cut a long story short, the running of the
couplings is governed order by order by differential
equations, the renormalization group equations,
\bb \,\frac{\de g}{\de t}\,=\beta_g,\qq t:=\log\,
E/\Lambda.\ee
$\Lambda$ is the energy cut off from the
regularization. The rhs of the differential equation is 
called the $\beta$ function of the coupling $g$. For
the standard model with $N=3$ generations, in 1-loop
`approximation', neglecting threshold effects and
neglecting all fermion masses with respect to the top
mass, the $\beta$ functions are \cite{jones}
\bb 
\beta_1&=&\ {\textstyle\frac{1}{16\pi^2}}\,
{\textstyle\frac{41}{6}}\ g_1^3\label{bet1}\\
\beta_2&=&-{\textstyle\frac{1}{16\pi^2}}\,
{\textstyle\frac{19}{6}}\ g_2^3\\
\beta_3&=&-{\textstyle\frac{1}{16\pi^2}}\,
7\ g_3^3\\
\beta_t&=&\ {\textstyle\frac{1}{16\pi^2}}\,
(9g_t^3-8g_3^2g_t-
{\textstyle\frac{9}{4}}g_2^2g_t-
{\textstyle\frac{17}{12}}g_1^2g_t),\\
\beta_\lambda&=&\  {\textstyle\frac{1}{16\pi^2}}\,
(96\lambda^2+24\lambda g_t^{2}-6g_t^4
-9\lambda g_2^2-3\lambda g_1^2+
{\textstyle\frac{9}{32}}g_2^4+
{\textstyle\frac{3}{32}}g_1^4+
{\textstyle\frac{3}{16}}g_2^2g_1^2),\\
\beta_{\mu^2}&=&\ {\textstyle\frac{1}{16\pi^2}}\,
\mu^2(48 \lambda+12g_t^2-
{\textstyle\frac{9}{2}}g_2^2
-{\textstyle\frac{3}{2}}g_1^2).\label{betmu}\ee

\begin{figure}[hbt]
\hspace{2.5cm}
\def\epsfsize#1#2{0.6#1}
\epsfbox{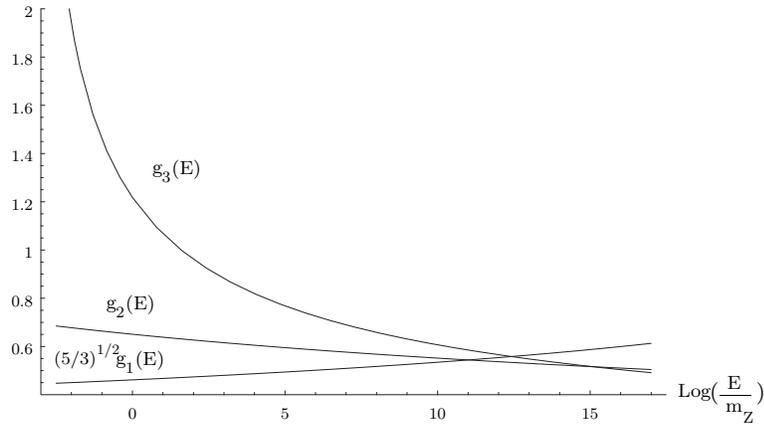}
\caption{The evolution of the three gauge couplings}
\label{gauge}
\end{figure}

With the cited approximations, the three gauge
couplings
$g_i$ decouple from the Yukawa and Higgs couplings
and can be solved immediately,
\bb
g_i^{-2}(t)=g_i^{-2}(0)-{\textstyle\frac{1}{8\pi^2}}
c_it,\qq \beta_i=:{\textstyle\frac{1}{16\pi^2}}\,
c_i\,g_i^{3}.\ee
Figure \ref{gauge} shows the
logarithmic running of the three gauge couplings
with experimental initial values,
$g_3=1.207 ,\ g_2=0.6507,\ g_1=0.3575$ at $E=m_Z$.
In agreement with our hand waving argument, the
abelian coupling
$g_1$ increases with energy. The non-Abelian ones,
the weak and strong couplings decrease with energy.
This is called asymptotic freedom and has rendered
non-Abelian Yang-Mills theories popular. At energies
below  1 GeV the curve of the strong coupling constant
loses all meaning because it  leaves the perturbative
regime. This is taken as evidence for confinement. 
On the other side, the curves have been extrapolated to
science fiction energies of $10^{19}$ GeV with the
insolent hypothesis of the big desert. I.e. we pretend
that from presently accessible energies of $10^2$ GeV
all the way up to $10^{19}$ GeV, energies that will
never be accessible to man, no new forces, no new
particles exist. This hypothesis was invented in the
seventies together with grand unified theories. To
ease somewhat the humiliation of the standard model,
some physicists were looking for a simple Lie group
like $SU(5)$ that contains $SU(3)\times SU(2)\times
U(1)$. As a simple Lie group only has one coupling
constant, this idea constrains the three gauge
couplings:
\bb g_3=g_2,\qq
g_1=\sqrt{{\textstyle\frac{3}{5}}}\,g_2.\ee The
picture was that at the unification energy
$\Lambda$ of around $10^{15}$ GeV $SU(5)$ breaks
spontaneously down to $SU(3)\times SU(2)\times
U(1)$. The gauge bosons that acquire a mass of the
order of $\Lambda$ are called lepto-quarks because
they mediate transitions between leptons and quarks
 rendering the proton unstable with a life time of the
order of $\hbar\Lambda^{4}/  m_p^{5}$,
 some $10^{29}$ years. $m_pc^2=0.938$ GeV is the
proton mass. At energies
$E$ below
$\Lambda$ the lepto-quarks decouple leaving  the standard model
with its three couplings $g_i$ running as in figure
\ref{gauge}. In the seventies the
experimental initial conditions and uncertainties
were different, such that the three curves could
cross in one single point. Furthermore the
experimental lower limit on the life time of the proton
was $10^{28}$ years, compatible with the theoretical
value. It was also clear that the lower limit could be
improved by several orders of magnitude within a few
years falsifying grand unification or discovering new
physics. The former happened, today the proton life
time is longer than
$10^{32}$ years. 

But grand unification also implied constraints on the
Yukawa and Higgs couplings and therefore on the
top and Higgs masses, 
\bb m_t=2\,g_t/g_2\ m_W,\qq m_H=4\sqrt 2\,\sqrt
\lambda/g_2\ m_W.\ee
The $\beta$ functions (\ref{bet1}-\ref{betmu}) have
been computed with dimensional regularization and
the modified minimal subtraction scheme where only
logarithmic divergences are kept. With Wilson's
lattice regularization, $\mu$ has in addition to its
logarithmic divergence a quadratic one that modifies
its $\beta$ function. To avoid this ambiguity we note
that, thanks to its dimensionality, $\mu$ decouples
from the other couplings which are dimensionless. If
we identify the pole masses $m_p=m(m_p)$ 
 with their running masses at the $Z$ mass $m(m_Z)$,
and only compute mass ratios we never need the
ambiguous renormalization behaviour of $\mu$.

\begin{figure}[hbt]
\hspace{2.5 cm}
\def\epsfsize#1#2{0.6#1}
\epsfbox{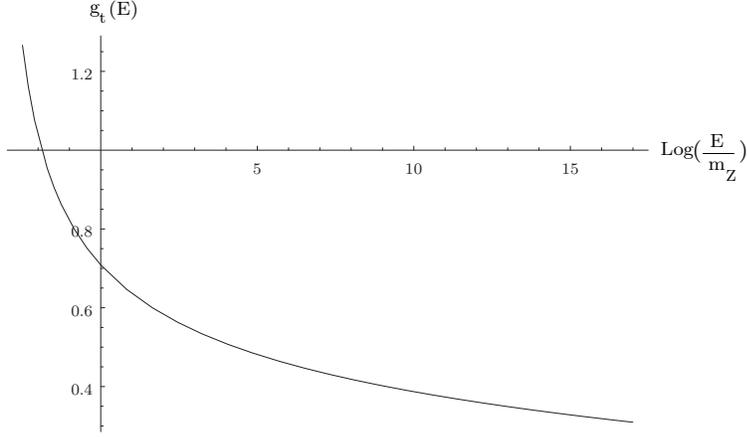}
\caption{The top coupling}
\label{top}
\end{figure}
\begin{figure}[hbt]
\hspace{2.5cm}
\def\epsfsize#1#2{0.6#1}
\epsfbox{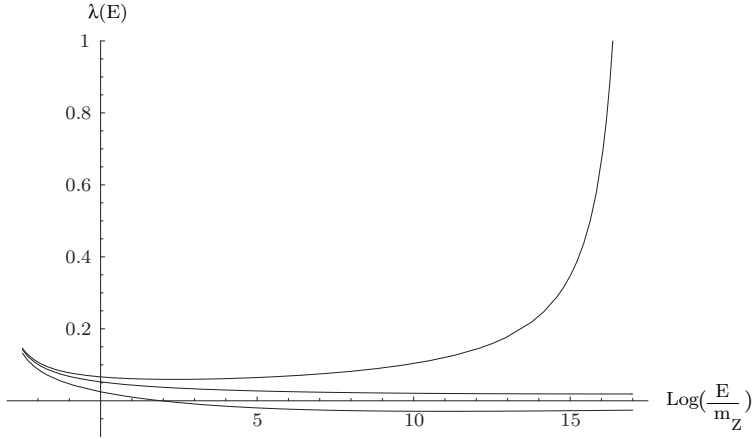}
\caption{The Higgs selfcoupling for $m_H(m_Z) = 120$ (lower graph), 
$160$ and $180$ Gev (upper graph) for $m_t(m_Z) = 175$ GeV}
\label{Higgs}
\end{figure}
\begin{figure}[hbt]
\hspace{2.5 cm}
\def\epsfsize#1#2{0.6#1}
\epsfbox{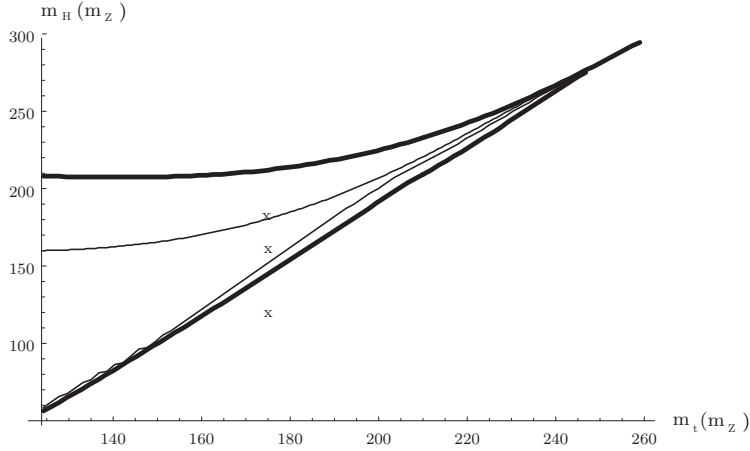}
\caption{Two allowed domains of initial values for 
$\Lambda= 10^{10}$ GeV (fat lines) and 
$\Lambda=10^{19}$ GeV (thin lines)}
\label{allow}
\end{figure}

In 1-loop the Yukawa coupling decouples from the
Higgs couplings. Figure \ref{top} shows its
energy dependence with an initial value of
$m_t(m_Z)=175$ GeV. All initial conditions not
mentioned are set to their central experimental
values. Finally figure \ref{Higgs} shows the
Higgs coupling
$\lambda$ with three initial values $m_H(m_Z)=120,\ 
160$ and 180 GeV (upper curve) and with
$m_t(m_Z)=175$ GeV. We see that two catastrophes may 
happen while traversing the big desert from $m_Z$ to
$\Lambda$ \cite{cab}: the Higgs coupling may become
negative, rendering the Higgs potential unstable or
the Higgs coupling may become too large and ruin the
perturbative computation. Figure \ref{allow} shows
the allowed domain of initial values $m_t(m_Z),\
m_H(m_Z)$ that avoid both catastrophes up to
$\Lambda=10^{10}$ GeV, thin lines, and up to
$\Lambda=10^{19}$ GeV, fat lines.  The upper curves
limit perturbation, the lower curves limit stability.
The three points indicate the initial conditions of
figure \ref{Higgs}. 

Let us now discuss the constraints from
almost commutative Yang-Mills for the standard model,
\bb g_3=g_2, &&
g_1=\sqrt{{\textstyle\frac{3}{5}}}\,g_2\cr \cr  
g_t=g_2,&&
\lambda={\textstyle\frac{7}{24}}\,g^2_2.\ee
Again we suppose that they hold at some energy
scale $\Lambda$ which immediately implies that we
must swallow the big desert. In grand unification
$\Lambda$ characterizes new gauge interaction, here
it characterizes a new spacetime geometry. $\Lambda$
measures the spacetime uncertainty like
$\hbar$ measures the phase space uncertainty. Today
the experimental values of the three gauge couplings
do not allow to fit the constraints at one energy,
figure \ref{gauge},
$\Lambda=10^{13}-10^{17}$ GeV. The corresponding
mismatch in the gauge couplings is on the 10 \% level.
We expect that the new uncertainty will explain this
mismatch. Indeed at energies close to the cut off
$\Lambda$ the $\beta$ functions computed from the
ultra-violet divergences cannot be trusted together
with noncommutative geometry. But so far we do not
have a quantum field theory on noncommutative
spacetimes.  Nevertheless we cannot refrain from
computing the numbers produced by the other two
constraints:
\bb m_t=187\pm 14\ {\rm GeV}, \qq
m_H=197\pm 9\ {\rm GeV}. \ee
The low value of $\Lambda$ produces the low top
mass and the low Higgs mass.  All masses are
automatically compatible with a stable
and perturbative Higgs coupling. 

We believe that almost commutative geometry is just a
low energy mirage of a truly noncommutative
geometry on the high energy side of the big dessert.
In grand unification, the direct product of groups was
replaced by one group at the scale $\Lambda$. In the
new picture, the tensor product of algebras should be
replaced by one algebra at the scale $\Lambda$. We
find it encouraging that this scale is close to, but lower
than the Planck mass, 
\bb m_P=\sqrt{\,\frac{\hbar c}{G}\,}\,\cong 
10^{19}\ {\rm GeV}.\ee
Indeed there is an old hand waving argument
combining Heisenberg's uncertainty relation of phase
space with the Schwarzschild horizon to find an
uncertainty relation in spacetime with a scale
$\Lambda$ smaller than the Planck mass:
To measure a position with a precision $\Delta x$ we
need, following Heisenberg, at least a momentum 
$\hbar/\Delta x$ or, by special relativity, an energy
$\hbar c/\Delta x$.  According to general relativity,
such an energy creates an horizon of size $G\hbar
c^{-3}/\Delta x$. If this horizon exceeds $\Delta x$ all
information on the position is lost. The best we can do
is resolve positions with $\Delta x$ such that $\Delta x=
G\hbar c^{-3}/\Delta x$, that is $\Delta x
=\hbar/(m_Pc)$,  the Planck length. The problem
with this argument is that we do not have a consistent
quantum theory in curved spaces. Despite many
efforts no renormalizable quantum field theory of
gravity is known. Even the pragmatic physicists
cannot agree on the energy dependence of the
gravitational coupling $G$. The numerical static
value, used in the hand waving argument does not
seem reasonable.

It is time to talk about gravity.

%% file: monsa6
\chapter{The Riemannian dreisatz}

\begin{itemize}\item
{ \qq Newton \ + \ Riemannian geometry \ \ = \ \
Einstein-Hilbert.}
\end{itemize}
Einstein was a passionate sailor. We speculate that
this was no accident. The subtle harmony between
geometries and forces becomes palpable to the sailor,
he sees the curvature of the sail and feels the force
that it produces. Before Einstein, it was generally
admitted that forces are vector fields in an Euclidean
space, $\rr^3$, the scalar product being necessary to
define work and energy. Einstein generalized
Euclidean to Minkowskian and Riemannian geometry
and found special and general relativity with
invariance groups, the Lorentz group $SO(1,3)$ and
the diffeomorphism group Diff$(M)$. These groups
define the principles of special and general relativity.

\section{First stroke}

Newton's universal, static law of gravity,
\bb F=G\,\frac{mM}{r^2}\, ,\ee 
the proportionality constant being Newton's constant
$G=6.670\cdot 10^{-11}\,{\rm m^3/(s^2 kg)}$,
resembles Coulomb's law. However there is a subtle
difference, the electric charge is Lorentz invariant,
the mass is not. Minkowskian geometry is the
geometry of a flat spacetime, with the flat Minkowski
metric $\eta$. Riemannian geometry is the geometry
of curved spacetimes, with an arbitrary metric $g$.
Riemannian geometry also suggests the principle of
general relativity, invariance under general
coordinate transformations whereas in Minkowskian
geometry or special relativity we only had
invariance under the Lorentz group, under those
special transformations that map inertial coordinates
(holonomic, orthonormal frames) into inertial
coordinates. 

As for Maxwell, the extension of Newton's law is
done in two strokes and starts with the trajectory of a
test mass $m$,
\bb m\,\frac{\de ^2\vec x}{\de t^2}\,=0,\ee
in inertial coordinates, a straight line. In arbitrary
coordinates, still in flat spacetime this becomes the
geodesic equation,
\bb m\,\frac{\de^2 x^\lambda}{\de p^2}&=&
-m{\Gamma^\lambda}_{\mu\nu}
\,\frac{\de x^\mu}{\de p}\,
\,\frac{\de x^\nu}{\de p}\, ,\label{geod}\\
&&{\Gamma^\lambda}_{\mu\nu}:={\textstyle\frac{1}{2}}
\eta^{\lambda\kappa}\left[
\pa_\mu\eta_{\kappa\nu}+\pa_\nu\eta_{\kappa\mu}
-\pa_\kappa\eta_{\mu\nu}\right].\ee
 The Christoffel symbols $\Gamma$ are first
derivatives of the matrix $\eta_{\mu\nu}$ of the flat
metric in the coordinates $x^\mu$, defined by
equation (\ref{metricch}). The geodesic equation  and
the definition of the Christoffel symbols are to be
compared to their Maxwell brothers,
\bb m\,\frac{\de^2 x^\mu}{\de \tau^2}\,&=&q
{F^\mu}_\nu \,\frac{\de x^\nu}{\de
\tau}\,,\\ 
&& F_{\mu\nu}:= \pa_\mu A_\nu-\pa_\nu A_\mu.\ee
The geodesic equation simply describes the straight
line in non-Cartesian coordinates. Nevertheless it
already contains a lot of physics. If the $x^\mu$ are
the coordinates of the rotating disk the geodesic
equation is nothing but centrifugal and Coriolis
forces. We can also repeat the above argument
replacing the free particle of Newton's mechanics
with the free particle of Schr\"odingers quantum
mechanics,  i.e. a plane wave. Then choosing for
$x^\mu$ oscillating coordinates, we understand some
observed interference patterns of neutrons
\cite{bonse}. 

The equivalence principle says that in absence of 
friction with air, a down falls as fast as a marble.
In other words, inertial and gravitational masses are
equal. This suggests to use a non-flat metric $g$ to
describe the trajectory of the marble in a
non-vanishing gravitational field.  The mass on
the lhs of the geodesic equation is inertial, on the rhs
the mass is gravitational and the two masses still
cancel by virtue of the equivalence principle. The
Christoffel symbols,
\bb {\Gamma^\lambda}_{\mu\nu}:={\textstyle\frac{1}{2}}
g^{\lambda\kappa}\left[
\pa_\mu g_{\kappa\nu}+\pa_\nu g_{\kappa\mu}
-\pa_\kappa g_{\mu\nu}\right],\label{chris}\ee
are the gravitational field, the underlying metric $g$
is the gravitational potential. The electromagnetic
potential $A$ can only be measured partially as
integral over a closed curve and this only via quantum
effects, the Aharonov-Bohm effect. The metric can be
measured classically, but again only as integral over a
curve, the proper time,
\bb c^2 \de \tau^2= g_{\mu\nu}\de x^\mu\de
x^\nu.\ee
Let us emphazise that the geodesic equation and the
proper time are invariant under general coordinate
transformations.

\section{Second stroke}

In the second stroke, Einstein used the full power of
the principle of general relativity to
 derive the dynamics of the gravitational field.
The source of  electromagnetism is charge,
\bb \left(\dd_{\rm Maxwell}A\right)_\mu=
-{\textstyle\frac{1}{\epsilon_0c^2}}j_\mu.\ee
We know the coupling constant from Coulomb's law
and we know that the differential operator $\dd_{\rm
Maxwell}$ must reduce to the Laplace operator in the
static case. The source of gravity is mass or -- with
special relativity -- energy,
\bb \left(\dd_{\rm Einstein}g\right)_{\mu\nu}=
{\textstyle\frac{8\pi G}{c^4}}\tau_{\mu\nu}.\ee
The energy-momentum tensor has the good taste to be
symmetric, $\tau_{00}$ is the energy density,
$\tau_{0i}$ are the energy currents, $\tau_{i0}$ are
the momentum densities and $\tau_{ij}$ their
currents. Newton's law fixes the coupling constant
$G$ and from the $1/r^2$ fall-off we know that
$\dd_{\rm Einstein}$ is second order. Covariance
under general coordinate transformations and
energy-momentum conservation then determine the
differential operator uniquely:
\bb
\left(\dd_{\rm
Einstein}g\right)_{\mu\nu}=G_{\mu\nu} -\Lambda_C
g_{\mu\nu},\ee where
$G_{\mu\nu}=R_{\mu\nu}-
{\textstyle\frac{1}{2}}Rg_{\mu\nu}$ is the Einstein
tensor.
\bb {R^\lambda}_{\mu\nu\kappa}:=
\pa_\nu {\Gamma^\lambda}_{\mu\kappa}
-\pa_\kappa {\Gamma^\lambda}_{\mu\nu}
+{\Gamma^\eta}_{\mu\kappa}{\Gamma^\lambda}_
{\nu\eta}
-{\Gamma^\eta}_{\mu\nu}{\Gamma^\lambda}_
{\kappa\eta}
\ee
is the Riemann tensor, $R_{\mu\kappa}:=
{R^\lambda}_{\mu\lambda\kappa}$ is the Ricci tensor
and $R:=R_{\mu\nu}g^{\mu\nu}$ is the curvature
scalar. $\Lambda_C$ is the cosmological constant
that we discard for phenomenological reasons.
Maxwell's differential operator, equation
(\ref{eight}), is linear and has eight terms. Einstein's
operator is non-linear and has roughly 80 000 terms.
Otherwise the two theories are very similar. As light
from Maxwell's equation, Einstein's equation has plane
wave solutions, `gravitational waves'. They too travel
at the speed of light. `Gravito--magnetic' forces with
feeble couplings are contained in Einstein's equations
and have been measured, as the advance of perihelia,
the curvature of light in a gravitational field, radar
delay, or spin precession.

Einstein's equation derives from an action, the
Einstein-Hilbert action
\bb S[g]={\textstyle\frac{-c^3}{16\pi G}}
\int_{\rr^4} R\,|\!\det g_{\cdot\cdot}|^{1/2}\de ^4x 
\ +\ {\rm matter},\ee
where $R$ is the curvature scalar. The energy
momentum tensor $\tau_{\mu\nu}$ is the variation of
the matter action with respect to $g^{\mu\nu}$.
 
\section{The principle of general relativity}

Connes' second dreisatz will unify Yang-Mills
theories with general relativity. To understand
Yang-Mills theories in terms of noncommutative
geometry, it was very useful to formulate them with
differential forms. The same is true for general
relativity. The remaining sections of this chapter
continue the technical interlude of chapter
\ref{interlude}. We will use the local concepts of
chapter \ref{interlude} to construct general
relativity in presence of spinors. Spacetime is an open
subset
$U$ of $\rr^4$ with signature $+---$ for concreteness.
The generalization to any dimension and signature is
immediate. The outcome of this construction will be a
gauge theory based on the Lorentz group $SO(1,3)$ or
its spin cover and the coupling of the gravitational
field to matter will be minimal, i.e. a covariant
derivative.

General relativity promotes the spacetime
metric to a dynamical field describing
gravity. Therefore we look for differential
equations determining the metric. By definition the
metric is a differentiable family of bilinear symmetric
forms, and we do not know what differential equations
for such objects are. We have seen that any metric
can be described using a frame of 1-forms. For 1-forms
we know differential operators. Einstein has used
holonomic frames. The principle of general relativity
requires that the metric and only the metric
generates gravitational interaction. Therefore we
want field equations that do not depend on the
particular coordinate system used to define the
holonomic frame. In the following, we use 
orthonormal frames of 1-forms to parameterize all
metrics. The
principle of general relativity now requires that the
particular orthonormal frame chosen to describe a
given metric is irrelevant. Our task therefore is to find
differential equations for the orthonormal frames
$e^i$ which are covariant under gauge
transformations $\Lambda$:
\bb {e'}^i =\Lambda^i_j e^i, \quad \Lambda\in\
^USO(1,3).\ee
 We restrict ourselves to orientation
preserving Lorentz transformations because we 
 use the Hodge star. It is sometimes
convenient to consider the orthonormal frame $e^i$
as a 1-form $e$ with values in the fundamental
representation of $SO(1,3)$. To be more precise, we
must add the restriction that the $e^i$ be linearly
independent which is compatible with the gauge
transformation
$e'=\Lambda e.$ To get gauge covariant
field equations for $e$ we use the Yang-Mills trick:
We introduce a connection, write down an invariant
action and obtain the desired field equations by
variation. In Yang-Mills theories the connection
actually represents new physical fields like the
photon or the weak bosons $W^\pm,Z$. Here we just
signed the principle prohibiting the
introduction of new fields. A natural solution of this
dilemma will show up automatically, and for the
moment we allow for a new field, the connection
$\omega$, a 1-form with values in the Lie algebra of
$SO(1,3)$
\bb\omega\in\Omega^1(U, so(1,3)),\ee
also
called spin connection. As a connection it is supposed
to transform under gauge transformations according
to
\bb\omega'=\Lambda\omega\Lambda^{-1} +\Lambda
\de\Lambda^{-1}.\ee
 As before we define the curvature
\bb R:=\de\omega+{\textstyle\frac{1}{2}}
\left[ \omega,
\omega\right]\ \in\
\Omega^2(U, so(1,3)).\ee
This definition is
known as Cartan's second structure equation. Again
we have immediately the homogeneous
transformation property of the curvature:
\bb R'=\Lambda R\Lambda^{-1}.\ee
We define
torsion by Cartan's first structure equation
\bb T:=\de
e+\omega e=\dee e\in\Omega^1(U,
\rr^4).\label{torsion}\ee
As a covariant derivative, also
the torsion transforms homogeneously under gauge
transformations:
\bb T'=\Lambda T.\ee
From $\de^2=0$ and the Jacobi identity, we obtain the
Bianchi identities:
\bb \dee R&=&\de R+\left[\omega,
R\right]=0,\\
\dee T&=&\de T+\omega T=R e.\ee

\section{The Einstein-Cartan equations}

For a Yang-Mills theory without matter the cheapest
gauge invariant action is quadratic in the curvature:
\bb S_{YM}[A]=-\,\frac{1}{g^2}\,\int {F^{*a}}_b *
F^b_{\phantom{b}a}.\label{ymac}\ee
 For the moment the pure gravitational
field is coded into two fields $e$ and $\omega$.
Consequently, we have an invariant action already
linear in the curvature,
\bb S_{EH}[e,\omega]=\,{\frac{1}{32\pi G}}\,\int
R^a_{\phantom{a}b}* (e^be_a),\label{ehac}\ee
where indices
are raised and lowered with $\eta^{ab}$ and
$\eta_{ab}$, and $e$ always denotes orthonormal
frames of 1-forms. Equation (\ref{ehac}) is the
Einstein-Hilbert action. 
Using the definition of the Hodge star in four
dimensions the Einstein-Hilbert action can also be
written,
\bb S_{EH}[e,\omega]=\,{\frac{-1}{32\pi G}}\,\int
R^{ab} e^c e^d\epsilon_{abcd}.\ee
 We introduce matter by adding a
functional $\int \lll_M$ depending on the matter
fields and on $e$ and $\omega$,
\bb S[e,\omega]=\,{\frac{-1}{32\pi G}}\,\int R^{ab} e^c
e^d\epsilon _{abcd}+ \int
\lll_M[e,\omega,...].\label{ac}\ee 
For example, the
matter could be a Yang-Mills action (\ref{ymac}) with
$A$ now considered as matter field. This
particular matter action depends only on $e$ (through
the Hodge star) and not on
$\omega$.

Let us derive the field equations following from
(\ref{ac}).\hfill\break
{\bf Variation of $e$:}\qq We call
$\tau$ the variation of the matter Lagrangian with
respect to $e$:
\bb\lll_M[ e+f]-\lll_M[e]=:-f^c \tau_c+O(f^2),\ee
where $\tau$ is a 3-form with values in $\rr^4$,
\bb\tau\in\Omega^3(U, \rr^4),\ee
the
`energy momentum tensor'. Integrating $\tau$ over
a 3-dimensional volume yields the energy momentum
contained in that volume. E.g. for pure
electromagnetic radiation,
\bb\lll_M=-{\textstyle\frac{\epsilon_0}{2}}F^**F,\ee
we obtain after a lengthy calculation
\bb\tau_{00}={\textstyle\frac{\epsilon_0}{2}}(\vec
{E^2}+\vec{B^2})\ee
with
\bb *\tau_c=:\tau_{ca}e^a.\ee
Variation of
the total action (\ref{ac}) with respect to $e$ gives
immediately the Einstein equations:
\bb R^{ab} e^d\ \epsilon_{abcd}=-16\pi
G\tau_c.\ee
For given energy momentum
$\tau$, they are non-linear first order differential
equations for the connection. They are also linear
equations for the curvature, `energy is the source of
curvature'. Despite the algebraic nature of the
equations curvature propagates in four dimensions:
Vanishing $\tau$ does not imply vanishing curvature
as is illustrated, for example, by Schwarzschild's
solution. This comes from the fact that the curvature
has
$6\times6=36$ independent coefficients
$R^{ab}_{\phantom{ab}\mu\nu}$ (antisymmetric in
$\mu$ and $\nu$ because $R$ is a 2-form,
antisymmetric in $a$ and $b$ because $R$ takes values
in the Lorentz algebra) while Einstein's equation,
being an equation for 3-forms with values in $\rr^4$
contains only $4\times4=16$ linear equations. In two-
and three-dimensional space times the counting is
different and curvature does not
propagate.\hfill\break
{\bf Variation of $\omega$:}\qq We
define the spin density
\bb \sss\in \Omega^3(U, so(1,3))\ee
by
\bb\lll_M[\omega+\chi]-\lll_M
[\omega]=:- {\textstyle\frac{1}{2}}
\chi^{ab} \sss_{ab}+O(\chi^2).\ee
Of
course, the spin density is zero for the Yang-Mills
action (\ref{ymac}). It is non-vanishing, for instance,
for the Dirac action describing spin
${\textstyle\frac{1}{2}}$ fields, which motivates the
name spin density. Varying $\omega$ in the total
action (\ref{ac}) yields, after an integration by parts,
the equation
\bb T^c e^d\ \epsilon_{abcd}=-8\pi\, G\, \sss_{ab}.
\label{fetor}\ee
`Spin is the source of
torsion'. If we now count the number of linear
equations and unknowns, we find them to match in
any dimension. Torsion does not propagate: Vanishing
spin density implies vanishing torsion.

\section{A farewell to $\omega$} \label{farewell}

We now come to the promised elimination of the spin
connection as an independent field. There are two
possible routes.
\hfill\break
{\bf Einstein's point of view:}\qq
Einstein puts torsion to zero right from the beginning.
By virtue of equation (\ref{torsion}),
\bb 0=T=\de e+\omega e \label{zerotor}\ee
is a covariant
constraint and therefore it does not spoil the
covariance of Einstein's equation. Let us consider this
constraint as a system of linear equations with
the components of the spin connection
$\omega^{ab}_{\phantom{ab}\mu}$ as unknowns.
Since $\omega$ is
$so$(1,3)-valued, it is antisymmetric in the indices $a$
and $b$ and there are $6\times 4$ unknowns. On the
other hand, (\ref{zerotor}) is an equation for
$\rr^4$-valued 2-forms and has $4\times 6$
components
$T^a_{\phantom{a}\mu\nu}$. Consequently, there
exists (for any signature and dimension) a unique
solution expressing the spin connection as a function
of the frame and its first derivatives. This solution is
called Riemannian connection. Its explicit form is
most conveniently written down expanding $\omega$
with respect to the orthonormal frame $e$:
\bb
\omega^a_{\phantom{a}b}=
\omega^a_{\phantom{a}bc}e^c.\ee 
Then the Riemannian connection is given by
\bb \omega^a_{\phantom{a}bc}=
{\textstyle\frac{1}{2}}(C^a_{\phantom{a}bc}-
C^{\phantom{b}a}_{b\phantom{a}c}-
C^{\phantom{c}a}_{c\phantom{a}b}),\label{riecon}\ee
where the functions $C$ are defined by
\bb\de e^a=:
{\textstyle\frac{1}{2}}C^a_{\phantom{a}bc}\
e^be^c.\ee
Substituting the Riemannian
connection $\omega(e,\partial e)$ into Einstein's
equations they become non-linear second order
differential equations for the orthonormal frame.
Alternatively they can be obtained by substituting
first the Riemannian connection into the
Einstein-Hilbert action and then varying with respect
to the frame, `second order formalism'.

Let us make the link between the Riemannian
connection with respect to the orthonormal
frame $e^a$, the $so(1,3)$ valued 1-form $\omega$
and the {\it same} Riemannian connection 
with respect to a holonomic frame $\de x^\mu$,
the $gl_4$ valued 1-form $\Gamma$. The
link between the two frames is a $GL_4^+$ gauge
transformation:
\bb e^a={\gamma^a}_\mu\de x^\mu, \qq\gamma\in
^UGL_4^+,\ee
and consequently the link between the the two
expressions of the Riemannian connection with
respect to the two frames is:
\bb \omega=\gamma\Gamma\gamma^{-1}+
\gamma\de\gamma^{-1}.\ee
In holonomic components this last equation reads:
\bb\,\frac{\pa}{\pa
x^\nu}\,{\gamma^a}_\mu-{\gamma^a}_\alpha
{\Gamma^\alpha}_{\mu\nu}+
{\omega^a}_\nu{\gamma^b}_\mu=0.\ee The $^UGL_4^+$
element ${\gamma^a}_\mu$ is often denoted
${e^a}_\mu$ and called vierbein. (Attention, the lhs of
the last equation is often called covariant derivative of
the vierbein and the equation is confused with the
metricity property of the Riemannian connection by
calling the vierbein a square root of the metric,
$g_{\mu\nu}(x)=
\eta_{ab}\,{e^a}_\mu(x){e^b}_\nu(x)$.)
The coefficients of the $gl_4$ valued 1-form
${\Gamma^\alpha}_{\mu\nu}\de x^\nu$ of the
Riemannian connection with respect to the
orthonormal frame are the Christoffel symbols,
equation (\ref{chris}).

\smallskip
\noindent{\bf Cartan's point of view:}\qq
Cartan keeps $\omega$ as an independent field which
eliminates itself at the end through its own
(algebraic) field equation (\ref{fetor}): $\omega =
\omega(e, \partial e, \sss)$. Therefore in this so-called
Einstein-Cartan theory Riemannian geometry is
only valid outside matter with spin. Only there it is
verified experimentally. Furthermore the observed
spin density in the universe is small and torsion
couples to it via the universal coupling constant $G$
implying that although different in principle
Einstein's and Einstein-Cartan's theories are presently
indistinguishable experimentally.

It can be shown \cite{Yates} that the Einstein-Hilbert
action is the unique action that leads to vanishing
torsion in the vacuum as field equation, unique of
course up to terms containing no spin connection, 
the cosmological term
\bb{\textstyle\frac{\Lambda_C}{4!}}\int e^a
e^b e^c e^d
\ \epsilon_{abcd}.\ee

As promised we now show that a piece of the
2-dimensional unit sphere (chapter \ref{interlude})
cannot have a holonomic and orthonormal
frame.\hfill\break {\bf Theorem:} An open subset $U$
of $\rr^n$ with a metric $g$ admits a holonomic and
orthonormal frame if and only if its Riemannian
connection has everywhere vanishing curvature.

We use equation (\ref{riecon}) to calculate the
Riemannian connection from
\bb e^1=\sin \theta\ \de\varphi\ ,
\quad e^2 =d\theta\ee and
\bb \de e^1=-\,{\frac{\cos\theta}{\sin\theta}}\,e^1
e^2,\qq \de e^2=0.\ee
Therefore
\bb
{C^1}_{\phantom{1}12}=-{C^1}_{\phantom{1}21}=
-\,{\frac{\cos\theta}
{\sin\theta}},\ee
and all other $C'$s vanish. Consequently, the
Riemannian connection is
\bb\omega^1_{\phantom{1}2}=\,
{\frac{\cos\theta}{\sin\theta}}\,e^1=\cos\theta\,
\de\varphi \ee
and its curvature
\bb R^1_{\phantom{1}2}=e^1 e^2\ee
is different from zero.

To conclude, following Cartan we have presented
general relativity using orthonormal frames. This
may be somewhat unfamiliar because Einstein
formulated his theory with the help of holonomic
frames. Of course, both approaches have advantages
and inconveniences. Two major shortcomings of
holonomic frames are: Their invariance
group is $GL_4$ which does not admit spinor
representations \cite{Cartan} therefore excluding
fields with half integer spin.
Holonomic frames break the gauge invariance of
general relativity, ignoring today's belief that all
fundamental interactions are described by gauge
theories.

\section{The Dirac operator}

Quantum mechanical experiments with neutrons teach
us that interference patterns repeat themselves only
 after a rotation through $720^o$ of one of the two
neutrons \cite{rauch}. Mathematically this means that
the relevant group for spin ${\textstyle\frac{1}{2}}$ is
not the rotation group
$SO(3)$ put its universal cover $SU(2)$. In relativistic
theories the rotation group is embedded in the Lorentz
group $SO(1,3)$ and we need its universal cover, the
Clifford group $Spin(1,3)$. The Dirac spinor is a vector
in the fundamental representation of the Clifford
group. In curved spacetime the Lorentz group is
gauged and so we must gauge the Clifford group in
order to define the Dirac operator there. You will not be
surprised that in the gauged case, we need a covariant
derivative. The connection takes values in the Lie
algebra of the group, here the Clifford group. By
definition the Lie algebra of a Lie group is the same as
the Lie algebra of its universal cover. This is the short
cut that we use to avoid developing the theory of
Clifford algebras and groups. All we need is the
representation of an infinitesimal Lorentz
transformation ${X^a}_b\in so(1,3)$ on a spinor $\psi$:
\bb \tilde\rho(X)\,\psi={\textstyle\frac{1}{4}}X_{ab}
\gamma^a\gamma^b \psi=
{\textstyle\frac{1}{8}}X_{ab}[\gamma^a,\gamma^b]
\psi.\ee
This transformation law tells us that the Dirac spinor
has spin ${\textstyle\frac{1}{2}}$ and this is the
transformation law that we should have given already
in section \ref{rules} to prove the Lorentz invariance
of the Dirac equation.
We recall that we use the flat metric $\eta_{aa'}$ to 
lower latin indices and that
$X_{ab}={X^{a'}}_b\eta_{aa'}$ is antisymmetric. The
$\gamma$ matrices with latin indices are the
$x$-independent Dirac matrices introduced in
section \ref{rules}. To write down the Dirac operator
we need partial derivatives. They are calculated in a
holonomic frame. On the other hand we need an
orthonormal frame to represent Lorentz
transformations. The link between the two frames is a
$GL^+_4$ gauge transformation, the (inverse) vierbein
${e^c}_\mu(x)$:
\bb \de x^\mu ={e^\mu}_c e^c.\ee
We use it to define $x$-dependent $\gamma$ matrices,
\bb \gamma^\mu (x):={e^\mu}_c(x)\,\gamma^c.\ee
We are ready to define the Dirac operator,
\bb\ddd\psi\ :=\ i\gamma^\mu (x)\left(\,\frac{\pa}
{\pa x^\mu}\,+\tilde\rho(\omega_\mu)\right)\,\psi\
=\ 
\gamma^\mu (x)\left(\,\frac{\pa}
{\pa x^\mu}\,+{\textstyle\frac{1}{4}}
\omega_{ab\mu}\gamma^a\gamma^b\right)\,\psi.\ee
In flat Minkowski space with inertial coordinates
$x^\mu$, the holonomic frame is orthonormal,
${e^\mu}_c={\delta^\mu}_c$, the spin connection
$\omega$ vanishes and we retrieve the flat Dirac
operator.

\section{The Dirac action}

To derive the Dirac equation from an action principle
we need a pseudo scalar product on the space of spinors,
invariant under the Clifford group. At this point the
signature of spacetime matters. With Minkowskian
signature and unitary Dirac matrices, this product is,
\bb (\psi,\chi)=\bar\psi\chi=\psi^*\gamma^0\chi,\ee
where here the star $\cdot^*$ denotes the transposed,
complex conjugate. With Euclidean signature, we have
a genuine scalar product,
\bb (\psi,\chi)=\psi^*\chi.\ee
In both signatures, the Dirac action reads:
\bb S_{\rm Dirac}[e,\omega,\psi]=\int*
(\psi,\ddd\psi)=
{\textstyle\frac{1}{3!}}\int(\psi,\gamma^a\dee \psi)\,
e^be^ce^d\epsilon_{abcd},\ee with the exterior
covariant derivative,
\bb \dee\psi=\de\psi+\tilde\rho(\omega)\psi=
\de\psi+{\textstyle\frac{1}{4}}
\omega_{ab}\gamma^a\gamma^b.\ee
Two remarks are in order. 
If the torsion vanishes the Dirac action is real, the 
Dirac operator is selfadjoint in Euclidean signature. 
Second remark, in the Euclidean, due to the missing
$\gamma^0$ in the scalar product, the Dirac
action for a chiral, say left-handed, fermion vanishes. 
We shall have to pay due attention to this last point
during the `Wick rotation'.

\section{The Lichn\'erowicz formula}

Dirac's first motivation for his operator was a square
root of the wave operator. Indeed, in flat Minkowski
space we have $\ddd^2=-\Box 1_4.$ Let us generalize
this formula to curved space. We suppose vanishing
torsion but allow the spinor to couple minimally
also to a Yang-Mills potential $A$ and to a Higgs scalar
$\Phi\ \in\ \hh_L^*\ot\hh_R$,
\bb\dd_{t,\rm cov}=\pp{
[\ddd\ot 1_L\,+\,i
{e^\mu}_j\gamma^j\ot\rho_L(A_\mu)]&\gamma_5
\ot\Phi
\cr \gamma_5\ot\Phi^*&
[\ddd\ot 1_R\,+\,i
{e^\mu}_j\gamma^j\ot\rho_R(A_\mu)]}.\ee
To keep notations simple we have left out the
antiparticle part. The square of this total covariant
Dirac operator is
\bb \dd^2_{t,\rm cov}=-\Box+E,\ee
  $\Box$ is the covariant wave operator
\bb \Box&=&g^{\mu\tilde\nu}\left[\left(
\frac{\partial}{\partial x^\mu} 1_4\ot 1_\hh+
{\textstyle\frac{1}{4}}\omega_{ab\mu}\gamma^a
\gamma^b
\ot 1_\hh+1_4\ot \rho(A_\mu)\right)
{\delta^\nu}_{\tilde\nu}-{\Gamma^\nu}_{\tilde
\nu\mu}1_4\ot1_{\hh}\right]
\cr &&\qq\qq\qq\times\left[
\frac{\partial}{\partial x^\nu} 1_4\ot 1_\hh+
{\textstyle\frac{1}{4}}\omega_{ab\nu}\gamma^{a}
\gamma^b
\ot 1_\hh+1_4\ot \rho(A_\nu)\right]\ee
with the internal representation
$\rho:=\rho_L\op\rho_R$ on $\hh:=\hh_L\op\hh_R$. 
$E$, for endomorphism, is a zero order operator, that is
a matrix of size
$4\dim\hh$ whose entries are functions
constructed from the bosonic fields and
their first and second derivatives,
\bb E={\textstyle\frac{1}{2}}\left[
\gamma^{\mu}\gamma^\nu
\ot1_\hh\right]\rr_{\mu\nu}\,+\,
\pp{
1_4\ot \Phi\Phi^*&-i\gamma_5\gamma^\mu\ot
\dee_\mu\Phi\cr  
-i\gamma_5\gamma^\mu\ot (\dee_\mu\Phi)^*&
1_4\ot\Phi^*\Phi}\label{E}.\ee
$\rr$ is the total curvature, a 2-form with values in
the (Lorentz $\op$ internal) Lie algebra represented
on (spinors $\ot\ \hh$). It contains the curvature
2-form
$R=\de\omega+{\textstyle\frac{1}{2}}[\omega,\omega]
$ and the field strength 2-form $F=\de
A+{\textstyle\frac{1}{2}}[A,A]$, in components
\bb\rr_{\mu\nu}={\textstyle\frac{1}{4}}
R_{ab\mu\nu}\gamma^{a}\gamma^b\ot 1_\hh+
1_4\ot\rho(F_{\mu\nu}).\ee 
An easy calculation shows that the first term in
equation (\ref{E}) produces
the curvature scalar that we also (!)
denote by $R$,
\bb {\textstyle\frac{1}{2}}\left[e^\mu_ce^\nu_d
\gamma^{c}\gamma^d\right]
{\textstyle\frac{1}{4}}R_{ab\mu\nu}\gamma^{ab}
= {\textstyle\frac{1}{4}}R1_4.\ee
 In our conventions, the curvature scalar
is positive on spheres (with signature ++).
 Finally $\dee$ is the covariant derivative appropriate
to the representation of the scalars.

The Lichn\'erowicz formula with arbitrary torsion
can be found in \cite{att}.

\section{Wick rotation}

In this section we put together the action of gravity
and of the standard model with emphasis on the
relative signs. We also indicate the changes
when passing from Minkowskian to Euclidean
signature. 

In 1983 the meter disappeared as fundamental unit of
science and technology. The conceptual revolution of
general relativity, the abandon of length in favour of
time, had made its way up to the domain of
technology. Said differently, general relativity is not
really  geo-metry, but chrono-metry. Hence our
natural choice of Minkowskian signature is $+---$. 

With this choice and the conventions,
\bb F_{\mu\nu}&=&\pa_\mu A_\nu-\pa_\nu A_\mu,\\
{\Gamma^\lambda}_{\mu\nu}&=&{\textstyle\frac{1}{2}}
g^{\lambda\kappa}\left[
\pa_\mu g_{\kappa\nu}+\pa_\nu g_{\kappa\mu}
-\pa_\kappa g_{\mu\nu}\right],\\
{R^\lambda}_{\mu\nu\kappa}&=&
\pa_\nu {\Gamma^\lambda}_{\mu\kappa}
-\pa_\kappa {\Gamma^\lambda}_{\mu\nu}
+{\Gamma^\eta}_{\mu\kappa}{\Gamma^\lambda}_
{\nu\eta}
-{\Gamma^\eta}_{\mu\nu}{\Gamma^\lambda}_
{\kappa\eta},\\
R_{\mu\kappa}&=&
{R^\lambda}_{\mu\lambda\kappa},\\
R&=&R_{\mu\nu}g^{\mu\nu},\\
\gamma^{a=0}&=&\pp{
1&0&0&0\cr 0&1&0&0\cr 0&0&-1&0\cr 0&0&0&-1},\qq
\gamma^{a=1}\ =\ \pp{
0&0&0&1\cr 0&0&1&0\cr 0&-1&0&0\cr -1&0&0&0},\\
\gamma^{a=2}&=& \pp{
0&0&0&-i\cr 0&0&i&0\cr 0&i&0&0\cr -i&0&0&0},\qq
\gamma^{a=3}\ =\ \pp{
0&0&1&0\cr 0&0&0&-1\cr -1&0&0&0\cr 0&1&0&0},\\
\gamma^\mu(x)&=&{e^\mu}_a(x)\gamma^a,\qq\qq
\gamma_5\ =\ \pp{
0&0&1&0\cr 0&0&0&1\cr 1&0&0&0\cr 0&1&0&0},\ee
the combined Einstein-Hilbert Maxwell Higgs Dirac
Lagrangian reads,
\bb 
&&\{-\,{\textstyle\frac{1}{16\pi}}m_P^{2}\,R
\,-\,
{\textstyle\frac{1}{4g^2}}\t (
F_{\mu\nu}^{*}F^{\mu\nu})\,+\,
{\textstyle\frac{1}{2g^2}}m_A^2\t (
A_{\mu}^{*}A^{\mu})\cr 
&&+\,{\textstyle\frac{1}{2}}\,(\dee_\mu\varphi)^*
\dee^\mu\varphi
\,-\,{\textstyle\frac{1}{2}}\,m_\varphi^2|\varphi|^2
\,+\,{\textstyle\frac{1}{2}}\,\mu^2|\varphi|^2
\,-\,\lambda |\varphi|^4\cr 
&&+\,
\psi^*\gamma^{a=0}\,[i\gamma^\mu\dee_\mu 
\,-\,m_\psi 1_4]\, \psi\}
\ |\!\det g_{\cdot\cdot}|^{1/2}.\ee
This Lagrangian is real if we suppose that all fields
vanish at infinity.  The relative coeffients between
kinetic terms and mass terms are chosen as to
reproduce the correct energy momentum relations
from the free field equations using Fourier transform
and the de Broglie relations as explained after
equation (\ref{deB}). With the chiral decomposition
\bb\psi_L&=&{\textstyle\frac{1-\gamma_5}{2}}\,\psi,
\qq
\psi_R\ =\ {\textstyle\frac{1+\gamma_5}{2}}\,\psi,
\label{project}\ee
the Dirac Lagrangian reads
\bb &&\psi^*\gamma^0\,[i\gamma^\mu\dee_\mu 
\,-\,m_\psi 1_4]\,
\psi\cr 
&&\qq=\psi_L^*\gamma^0\,i\gamma^\mu\dee_\mu\,
\psi_L
\,+\,
\psi_R^*\gamma^0\,i\gamma^\mu\dee_\mu\,\psi_R 
\,-\,m_\psi\psi_L^*\gamma^0\psi_R
\,-\,m_\psi\psi_R^*\gamma^0 \psi_L.\ee
The  relativistic energy momentum
relations are quadratic in the masses. Therefore the
sign of the fermion mass
$m_\psi$ is conventional and merely reflects the
choice: who is particle and who is antiparticle. We can
even adopt one choice for the left-handed fermions
and the opposite choice for the right-handed
fermions. Formally this can be seen by the change of
field variable (chiral transformation):
\bb \psi:=\exp(i\alpha\gamma_5)\,\psi'.\ee
It leaves invariant the kinetic term and the mass term
transforms as,
\bb -m_\psi{\psi'}^*\gamma^0[\cos (2\alpha)\,1_4
+i\sin (2\alpha)\,\gamma_5]{\psi'}. \ee
With
$\alpha=-\pi/4$ the Dirac Lagrangian becomes:
\bb
&&{\psi'}^*\gamma^0[\,i\gamma^\mu\dee_\mu\,+i
m_\psi\gamma_5]{\psi}'\cr
&&\qq = 
{\psi'}_L^*\gamma^0\,i\gamma^\mu\dee_\mu\,
{\psi}'_L
\,+\,
{\psi'}_R^*\gamma^0\,i\gamma^\mu\dee_\mu\,
{\psi'}_R 
\,+\,m_\psi {\psi'}_L^*\gamma^0i
\gamma_5{\psi'}_R
\,+\,m_\psi {\psi'}_R^*\gamma^0i
\gamma_5 {\psi'}_L\cr 
&&\qq = 
{\psi'}_L^*\gamma^0\,i\gamma^\mu\dee_\mu\,
{\psi}'_L
\,+\,
{\psi'}_R^*\gamma^0\,i\gamma^\mu\dee_\mu\,
{\psi'}_R 
\,+\,im_\psi {\psi'}_L^*\gamma^0{\psi'}_R
\,-\,im_\psi {\psi'}_R^*\gamma^0 {\psi'}_L.\ee
We have seen that gauge invariance forbids
massive gauge bosons, $m_A=0$, and that parity
violation forbids massive fermions, $m_\psi=0$. This
is fixed by spontaneous symmetry breaking, where we
take the scalar mass term with wrong sign,
$m_\varphi=0,\
\mu>0$. The shift of the scalar then induces masses
for the gauge bosons, the fermions and the physical
scalars. These masses are calculable in terms of the
gauge, Yukawa and Higgs couplings. 

The other relative signs in the combined Lagrangian
are fixed by the requirement that the energy density
of the non-gravitational part $\tau_{00}$ be positive
(up to a cosmological constant) and that gravity in the
Newtonian limit be attractive. In particular this
implies that the Higgs potential must be bounded from
below,
$\lambda>0$. The sign of the Einstein-Hilbert action
may also be obtained from an
asymptotically flat space of weak curvature, where we
can define gravitational energy density. Then
the requirement is that the kinetic terms
of all physical bosons, spin 0, 1 and 2, be of the same
sign. Take the metric of the form
\bb g_{\mu\nu}=\eta_{\mu\nu}+h_{\mu\nu},\ee
$h_{\mu\nu}$ small. Then the Einstein-Hilbert 
Lagrangian becomes \cite{gilles},
\bb -\,{\textstyle\frac{1}{16\pi G}}\,R\,
|\!\det g_{\cdot\cdot}|^{1/2}&=&
{\textstyle\frac{1}{16\pi G}}\{
{\textstyle\frac{1}{4}}\pa_\mu h_{\alpha\beta}
\pa^\mu h^{\alpha\beta}\,-\,
{\textstyle\frac{1}{8}}\pa_\mu{h_\alpha}^\alpha
\pa^\mu{h_\beta}^\beta\cr 
&&-\,[\pa_\nu{h_\mu}^\nu
-{\textstyle\frac{1}{2}}\pa_\mu{h_\nu}^\nu]
[\pa_{\nu'}{h^\mu}^{\nu'}
-{\textstyle\frac{1}{2}}\pa^\mu{h_{\nu'}}^{\nu'}
]\,+\,O(h^3)\}.\ee
Here indices are raised with $\eta^{\cdot\cdot}$. After
an appropriate choice of coordinates, `harmonic
coordinates', the bracket $\left[\pa_\nu{h_\mu}^\nu
-{\textstyle\frac{1}{2}}\pa_\mu{h_\nu}^\nu
\right]$ vanishes and only two independent
components of $h_{\mu\nu}$ remain,
$h_{11}=-h_{22}$ and $h_{12}$.  They represent
the two physical states of the graviton,
helicity $\pm 2$. Their kinetic terms are both positive,
e.g.:
\bb +{\textstyle\frac{1}{16\pi
G}}{\textstyle\frac{1}{4}}
\pa_\mu h_{12}\pa^\mu h_{12}.\ee 
Likewise, by an appropriate gauge transformation, we
can achieve $\pa_\mu A^\mu=0$, `Lorentz gauge',
and remain with only two, `transverse' components
$A_1,\ A_2$ of helicity $\pm 1$. They have positive
kinetic terms, e.g.:
\bb+{\textstyle\frac{1}{2g^2}}\t (\pa_\mu
A_1^*\pa^\mu A_1).\ee
Finally the kinetic term of the scalar is positive:
\bb +{\textstyle\frac{1}{2}}\pa_\mu\varphi^*
\pa^\mu\varphi.\ee

An old recipe from quantum field theory, `Wick
rotation', amounts to replace spacetime by a compact
Riemannian manifold with Euclidean signature. Then
certain calculations become feasible or easier. One of
the reasons for this is that Euclidean quantum field
theory resembles statistical mechanics, the imaginary
time playing formally the role of the inverse
temperature. Only at the end of the calculation the
result is `rotated back' to real time. In some cases, this
recipe can be justified rigorously. The precise
formulation of the recipe is that the $n$-point
functions computed from the Euclidean Lagrangian
be the analytic continuations in the complex time
plane of the Minkowskian
$n$-point functions. We shall indicate a hand
waving formulation of the recipe that for our purpose
is sufficient: In a first stroke we pass to the signature
$-+++$. In the second stroke we replace $t$ by $it$ and
replace all Minkowskian scalar products by the
corresponding Euclidean ones. 

The first stroke amounts simply to replacing the
metric by its negative. This leaves invariant the
Christoffel symbols, the Riemann and Ricci tensors,
but reverses the sign of the curvature scalar.
Likewise, in the other terms of the Lagrangian we get
a minus sign for every contraction of indices, e.g.:
$\pa_\mu\varphi^*\pa^\mu\varphi=
\pa_\mu\varphi^*\pa_{\mu'}\varphi g^{\mu\mu'}$
becomes $\pa_\mu\varphi^*\pa_{\mu'}\varphi
(-g^{\mu\mu'})=-\pa_\mu\varphi^*\pa^\mu\varphi$.
After multiplication by a conventional overall minus
sign the combined Lagrangian reads now,
\bb 
&&\{-\,{\textstyle\frac{1}{16\pi}}m_P^{2}\,R
\,+\,
{\textstyle\frac{1}{4g^2}}\t (
F_{\mu\nu}^{*}F^{\mu\nu})\,+\,
{\textstyle\frac{1}{2}}\,(\dee_\mu\varphi)^*
\dee^\mu\varphi
\,-\,{\textstyle\frac{1}{2}}\,\mu^2|\varphi|^2
\,+\,\lambda |\varphi|^4\cr 
&&\qq+\,
\psi^*\gamma^0[\,i\gamma^\mu\dee_\mu
\,+\,m_\psi 1_4\,]\psi\,\}
\ |\!\det g_{\cdot\cdot}|^{1/2}.\label{-+++}\ee

To pass to the Euclidean signature, we multiply time,
energy and mass by $i$. This amounts to
$\eta^{\mu\nu}=\delta^{\mu\nu}$ in the scalar
product. In order to have the Euclidean
anticommutation relations,
\bb   \gamma ^\mu \gamma ^\nu +\gamma ^\nu
\gamma ^\mu =  2\delta ^{\mu \nu }1,\ee  
we change the Dirac matrices to the Euclidean ones, 
(\ref{g0eucl}), (\ref{g2eucl}), (\ref{g5eucl}), that are
all self adjoint. The Minkowskian scalar product for
spinors has a
$\gamma^0$. This $\gamma^0$ is needed for the
correct physical interpretation of the energy of
antiparticles and for Lorentz invariance, $Spin(1,3)$.
In the Euclidean, there is no physical interpretation
and we can only retain the requirement of a
$Spin(4)$ invariant scalar product. This scalar
product has no $\gamma^0$. But then we have a
problem if we want to write the Dirac Lagrangian in
terms of chiral spinors as above. For instance,
$\psi_L^*\,i\gamma^\mu\dee_\mu\, \psi_L$ vanishes
identically because $\gamma_5$ anticommutes with
the four $\gamma^\mu$. The standard trick of
Euclidean field theoreticians is fermion doubling,
$\psi_L$ and $\psi_R$ are treated as two
{\it independent}, four component spinors. They are
not chiral projections of one four component spinor
as in the Minkowskian, equation (\ref{project}). The
spurious degrees of freedom in the Euclidean are kept
all the way through the calculation. They are
projected out only after the Wick rotation back to
Minkowskian, by imposing $\gamma_5\psi_L=-\psi_L,
\gamma_5\psi_R=\psi_R$.

 In noncommutative
geometry the Dirac operator must be self adjoint,
which is not the case of the Euclidean Dirac operator
$i\gamma^\mu\dee_\mu+im_\psi 1_4$ we get from
the  Lagrangian (\ref{-+++}) after multiplication of
the mass by $i$. We therefore prefer the primed
spinor variables $\psi'$ producing the self adjoint
Euclidean Dirac  operator
$i\gamma^\mu\dee_\mu+m_\psi\gamma_5$.
Dropping the prime, the combined Lagrangian in the
Euclidean then reads:
\bb 
&&\{-\,{\textstyle\frac{1}{16\pi}}m_P^{2}\,R
\,+\,
{\textstyle\frac{1}{4g^2}}\t (
F_{\mu\nu}^{*}F^{\mu\nu})\,+\,
{\textstyle\frac{1}{2}}\,(\dee_\mu\varphi)^*
\dee^\mu\varphi
\,-\,{\textstyle\frac{1}{2}}\,\mu^2|\varphi|^2
\,+\,\lambda |\varphi|^4\cr 
&&+\,
{\psi}_L^*\,i\gamma^\mu\dee_\mu\,
{\psi}_L
\,+\,
{\psi}_R^*\,i\gamma^\mu\dee_\mu\,
{\psi}_R 
\,+\,m_\psi {\psi}_L^*\gamma_5{\psi}_R
\,+\,m_\psi {\psi}_R^*\gamma_5 {\psi}_L
\}
\,(\det g_{\cdot\cdot})^{1/2}.\label{++++}\ee
In flat space, this is precisely the Yang-Mills-Higgs
Lagrangian (\ref{ymh3}) and the Dirac Lagrangian
(\ref{dirac3}) in the form obtained from Connes' first
dreisatz.

%% file: monsa7
\chapter{Connes' second dreisatz}

Again our starting point is the one--to--one
correspondence between {\it commutative} spectral
triples $(\aa,\hh,\dd)$ and compact Riemannian
manifolds $(M,g)$ with spin structure.
Noncommutative or {\it fuzzy} spaces are defined by
relaxing the condition of commutativity. In these
spaces the Dirac operator $\dd$ plays several
important roles: 
\begin{itemize}\item
It defines the differential structure in terms of the
exterior derivative $\de=[\dd,\cdot]$.
\item 
The dimension of the space can be read in the
spectrum of $\dd$, the eigenvalues $\lambda_n$
grow like $n^{1/\dim}$.
\item
The Dirac operator allows to define integration 
by regularizing the scalar product of differential
forms $\kappa$, $\varphi$: 
\bb (\kappa,\varphi) :={\textstyle\frac{1}{2}} {\rm
Re\,}\tt ([\kappa+J\kappa J^{-1}]^*
[\varphi+J\varphi J^{-1}]\,|\dd|^{-\dim}).\ee
\item 
The Dirac operator generalizes the metric. Indeed on
commutative spaces $M$, the metric $g$ can be
retrieved from the Dirac operator via the geodesic
distance between two points $x_1,x_2\in M$,
\bb d(x_1,x_2)={\rm Sup}\{|f(x_1)-f(x_2)|;\ f\in \aa,\ 
||[\dd,\rho (f)]||\le 1\},\ee
with $\aa=\ccc^\infty(M)$, 
$(\rho(f)\psi)(x)=f(x)\psi(x)$ and $\dd=\ddd$.
\end{itemize}

For gravity the last role is vital because the metric is
the dynamical variable on spacetime $M$.

\section{The spectral principle}

 Einstein used the matrix
$g_{\mu\nu}(x)$ of the metric $g$ with respect to a
holonomic frame $\pa /\pa x^\mu$ to parameterize
the set of all metrics on a fixed spacetime $M$. The
coordinate system $x^\mu$ being unphysical, Einstein
required his field equations for the metric to be
covariant under coordinate transformations, the
principle of general relativity. Following physicists'
habits we will confuse coordinate transformations and
diffeomorphisms.
 Elie Cartan used orthonormal frames, {\it
rep\`eres mobiles}, to parameterize the set of all
metrics. This parameterization allowed to generalize
the Dirac operator $\dd$ to curved space-times and also
reformulated general relativity as a gauge theory
under the Lorentz group. Connes
\cite{grav} goes one step further by relating the set
of all metrics to the set of all Dirac operators. The
Einstein-Hilbert action, from this point of view, is the
Wodzicki residue of the second inverse power of the
Dirac operator
\cite{wod} and is computed most conveniently from
the second coefficient of the heat kernel expansion of
the Dirac operator squared. 

The natural question now is: what becomes the
principle of general relativity in Connes' point of
view? Connes' answer is as natural: Invariance under
the group of automorphisms of the algebra $\aa$.
Indeed in the commutative case, $\aa=\ccc^\infty(M)$,
this group is the group of diffeomorphisms
Diff$(M)$. And what is an intrinsic property of the
Dirac operator, a property invariant under algebra
automorphisms? It is the spectrum of $\dd$ and
Connes proposes to generalize the principle of
general relativity in terms of {\it the spectral
principle:}
\begin{itemize}\item
Physics is coded in the spectrum of the Dirac
operator.
\end{itemize}
If instead of the Dirac operator we take its square, the
Laplace operator, on a flat two dimensional space, then
the spectral principle asks an old question:
\begin{itemize}\item
Can you hear the shape of a drum?
\end{itemize} 

 Let us apply the spectral principle to
almost commutative geometries,
$\aa_t=\ccc^\infty(M)\ot\aa_f$. Its group of
automorphisms is the semidirect product of the group
of diffeomorphisms with a gauge group,
\bb {\rm Diff}(M)\semi\, ^MG,\ee
where $G$ is the automorphism group of $\aa_f$.
Up to discrete symmetries, all automorphisms of the
inner space $\aa_f$ are inner automorphisms,
\bb \varphi_u(a)=uau^{-1},\qq {\rm for\ all}\
a\in\aa_f,\ee 
for a unitary element $u\in U(\aa_f)$. Consequently
(up to discrete symmetries) the automorphism group
of
$\aa_f$ is a subgroup of its group of unitaries,
$G\subset U(\aa_f)$. For instance,
$\aa_f=\hhh,\ G=U(\aa_f)=SU(2),$ and
$\aa_f=M_3(\cc),\  G=SU(3),\ U(\aa_f)=U(3).$
Therefore the spectral principle explains the
invariance group of the combined actions of gravity
with certain non-Abelian Yang-Mills theories, the
above semidirect product, in terms of almost
commutative geometries. It was precisely these
geometries, that explained the Higgs and spontaneous
symmetry breaking in Connes' first dreisatz. In other
words, as quantum mechanics is behind the (Abelian)
$U(1)$ in the gauge dreisatz, almost commutative
geometries are behind certain non-Abelian Lie
groups in the same dreisatz. 

\section{First stroke}

Let us now follow the Riemannian dreisatz in two
strokes to derive the field variables \cite{grav} and
their dynamics \cite{cc} from the spectral principle
and almost commutative geometry. 

Of course the matter equation we use in the first stroke
is the Dirac equation for a free, massive fermion
$\psi$ in inertial  coordinates (coordinates whose
holonomic frame is orthonormal) rather than
Newton's equation for a free point mass in inertial
coordinates. We have to ask how the Dirac equation
changes under an automorphism. In almost
commutative geometry an automorphism has two
parts. An outer part which is a spacetime
diffeomorphism --
$\ccc^\infty(M)$ being commutative has no inner
automorphism -- and an inner part which is a gauge
transformation. We already know how the naked Dirac
operator $\ddd$ changes under a diffeomorphism, it
becomes covariant with respect to the flat spin
connection
$\omega(e)$ induced by the diffeomorphism. This is
the gravitational coupling that the principle of
general relativity orders. The inner Dirac operator
$\dd$ or fermionic mass matrix is invariant. Let us
now see how the inner automorphism
$\varphi_u$,
$u\in U(\aa_t)$ being a  gauged unitary, modifies the
naked, total Dirac operator $\dd_t=\ddd\ot
1\,+\,\gamma_5\ot\dd_f$ . Since the spinor
transforms under this unitary as, cf. section \ref{acg},
\bb \rho_{\rm
spinor}(u)\,\psi\ =\
\rho_f(u)\,J_f\rho_f(u)J^{-1}_f\
\psi,\qq u\in U(\aa_t)=\,^M U(\aa_f),\ee
the naked, Dirac $\dd_t$ becomes:
\bb &&\left(\rho_t(u)\,J_t\rho_t(u)J_t^{-1}\right)\,
\dd_t\,
\left(\rho_t(u)\,J_t\rho_t(u)J^{-1}_t\right)^{-1}\cr 
&&=
\rho_t(u)\,J_t\rho_t(u)J_t^{-1}
\dd_t\,
\rho_t(u^{-1})\,J_t\rho_t(u^{-1})J^{-1}_t \cr 
&&=
\rho_t(u)J_t\rho_t(u)J_t^{-1}(\rho_t(u^{-1})\dd_t+
[\dd_t,\rho_t(u^{-1})])J_t\rho_t(u^{-1})J_t^{-1}\cr 
&&=
J_t\rho_t(u)J_t^{-1}\dd_t J_t\rho_t(u^{-1})J_t^{-1}
+\rho_t(u)[\dd_t,\rho_t(u^{-1})]=
J_t\rho_t(u)\dd_t \rho_t(u^{-1})J_t^{-1}
+\rho_t(u)[\dd_t,\rho_t(u^{-1})]\cr 
&&=
J_t(\rho_t(u)[\dd_t ,\rho_t(u^{-1})]+\dd_t)J_t^{-1}
+\rho_t(u)[\dd_t,\rho_t(u^{-1})]\cr 
&&=\dd_t\,-\,\pi_t(A_t)-J_t\pi_t(A_t)J^{-1}_t,
\label{fluct}\ee
with the flat connection:
\bb A_t=u\delta_tu^*=u\de u^*+u\delta_fu^*=A+H
\,\in\Omega^1(M,u(\aa_f))\,\op\,\ccc^\infty(M)
\ot\Omega^1_{\dd_f}\aa_f.\ee
In the chain (\ref{fluct}) we have used
successively the following three axioms of spectral
triples, $[\rho(u_1),J\rho(u_2)J^{-1}]=0$, the first
order condition $[[\dd,\rho(u_1)],J\rho(u_2)J^{-1}]=0$
and $[\dd,J]=0$.
The result means that the naked Dirac operator
becomes covariant with respect to the Yang-Mills
potential $A$ and with respect to the Higgs scalar $H$.
The spectral principle implies that in almost
commutative geometry, the gravitational field coded in
the metric or equivalently in the Dirac operator
 is necessarily accompanied by  the spin 1 field
$A$ and the spin 0 field $H$. 

So far the three connections $\omega(e),\ A,\ H$ have
no curvature. We now promote them to general fields.
Then we have the total, covariant Dirac operator,
\bb \dd_{t,\rm cov}=
\dd_t\,-\,\pi_t(A_t)-J_t\pi_t(A_t)J^{-1}_t,\ee
which is precisely the one of Connes' first dreisatz,
section \ref{acg}.

\section{Second stroke}

So far the gravitational, Yang-Mills and Higgs fields
are adynamical, only the fermion $\psi$ propagates
in the fixed background $((e,\omega(e)),A,H)$. In the
second stroke, Chamseddine \& Connes \cite{cc}
develop the full power of the spectral principle to
derive the dynamics of the spin 2, 1 and 0 fields from
the total, covariant Dirac operator $\dd_{t,\rm cov}$.

In even dimensions, the spectrum of the Dirac
operator is even and it is sufficient to
consider the positive part of the spectrum which in
the  Euclidean is conveniently characterized by a
distribution function
\bb S=\t f(\dd_{t,\rm cov}^2/\Lambda^2),\ee
where $\Lambda$ is an energy cutoff and 
$f:\rr_+\rightarrow\rr_+$ is a positive, smooth
function with finite `momenta',
\bb f_0&=&\int_0^\infty uf(u)\de u,\\
f_2&=&\int_0^\infty f(u)\de u,\\
f_4&=&f(0),\\
f_6&=&-f'(0),\\
f_8&=&f''(0),...\ee
Asymptotically, for large $\Lambda$, the
distribution function of the spectrum is given in
terms of the heat kernel expansion \cite{egbv}:
\bb S=\t f(\dd_{t,\rm cov}^2/\Lambda^2)=
\frac{1}{16\pi^2}\,\int_M[\Lambda^4f_0a_0+
\Lambda^2f_2a_2+f_4a_4+\Lambda^{-2}f_6a_6+...]
\sqrt{\det g}\,\de^4x, \label{master}\ee
where the $a_j$ are the coefficients of the heat kernel
expansion of the Dirac operator squared \cite{heat},
\bb a_0&=&\t (1_4\ot1_\hh),\\ 
a_2&=&{\textstyle\frac{1}{6}}R\,\t (1_4\ot1_\hh)-\t
E,\\ a_4&=&{\textstyle\frac{1}{72}}R^2\t
(1_4\ot1_\hh)-
{\textstyle\frac{1}{180}}R_{\mu\nu}R^{\mu\nu}
\t (1_4\ot1_\hh)+
{\textstyle\frac{1}{180}}R_{\mu\nu\rho\sigma}
R^{\mu\nu\rho\sigma}\t (1_4\ot1_\hh)\cr &&+
{\textstyle\frac{1}{12}}\t (\rr_{\mu\nu}
\rr^{\mu\nu}) 
-{\textstyle\frac{1}{6}}R\,\t E+
{\textstyle\frac{1}{2}}\t E^2 + {\rm surface\ terms}.\ee
We have used the Lichn\'erowicz formula for the
square of the Dirac operator, $\dd^2_{t,\rm
cov}=-\Delta+E$. Note that for large
$\Lambda$ the positive function
$f$ is universal in the sense that only the three first
momenta, $f_0,\ f_2$ and $f_4$ matter. 

 Let us first check the normalization
$16\pi^2$ of equation (\ref{master}). Again we take
$M$ to be the flat 4-torus with unit radii, $\hh_L=\cc$,
$\hh_R=0$ and
$A=\varphi=0$. Remember from section \ref{scalarp}
that for large $\Lambda$ there are 
$ 4B_4\Lambda^4$ eigenvalues (counted with their
multiplicity) whose absolute values are smaller than 
$\Lambda$. 
$ B_4=\pi^2/2$
denotes the volume of the unit ball
in $\rr^4$. On the other hand if we take for $f$ a
smooth approximation of the characteristic function
of the unit interval, then $f_0={\textstyle\frac{1}{2}}$
and
$S$ simply counts the eigenvalues of the square of the
Dirac operator less than
$\Lambda^2$:
\bb S=
4{\textstyle\frac{1}{2}}\pi^2\Lambda^4
=\frac{1}{16\pi^2}\,\Lambda^4{\textstyle\frac{1}{2}}4
(2\pi)^4.\ee 

The computation of the Chamseddine-Connes action
$S$ for the Dirac operator of the standard
model is straightforward. We give a few intermediate
steps, a full account can be found in \cite{rom}. 
\bb a_0&=&4\dim\hh,\\
\t E&=&\dim\hh\, R + 8 \t\Phi^*\Phi
=\dim\hh\,R+8 L|\varphi/v|^2,\\
L&:=&3\t (M_u^*M_u)+3\t (M_d^*M_d)+\t
(M_e^*M_e)\cr 
&=&3(m^2_t+m^2_c+m^2_u+m^2_b+m^2_s+m^2_d)+
m^2_\tau+m^2_\mu+m^2_e,\\
a_2&=&{\textstyle\frac{4}{6}}\dim\hh\, R-\dim\hh\, R
-8 L|\varphi/v|^2\cr &=&
 -{\textstyle\frac{1}{3}}\dim\hh\, R
-8 L|\varphi/v|^2,\\
\t \left( {\textstyle\frac{1}{2}}
[\gamma^{a},\gamma^b]
{\textstyle\frac{1}{2}}
[\gamma^{c},\gamma^d]\right)&=&
4\left[\eta^{ad}\eta^{bc}-\eta^{ac}\eta^{bd}\right],\\
\t \rr_{\mu\nu}\rr^{\mu\nu}&=&
-{\textstyle\frac{1}{2}}\dim\hh\,
R_{\mu\nu\rho\sigma}R^{\mu\nu\rho\sigma}
-4 \t \rho(F_{\mu\nu})^*\rho(F^{\mu\nu}),\\
\t E^2&=&{\textstyle\frac{1}{4}}\dim\hh\, R^2+
2\t \rho(F_{\mu\nu})^*\rho(F^{\mu\nu})\cr &&+
8Q|\varphi/v|^4+
8L(\dee_\mu\varphi/v)^*(\dee^\mu\varphi/v)
+4L|\varphi/v|^2R,\\ 
Q&:=&3\t \left[M_u^*M_u\right]^2+3\t
\left[M_d^*M_d\right]^2+\t
\left[M_e^*M_e\right]^2\cr 
&=&3(m^4_t+m^4_c+m^4_u+
m^4_b+m^4_s+m^4_d)+
m^4_\tau+m^4_\mu+m^4_e.
\ee 
Using the Weyl tensor,
\bb C_{\mu\nu\rho\sigma}:=
R_{\mu\nu\rho\sigma}
-{\textstyle\frac{1}{2}}(
g_{\mu\rho}R_{\nu\sigma}-
g_{\mu\sigma}R_{\nu\rho}+
g_{\nu\sigma}R_{\mu\rho}-
g_{\nu\rho}R_{\mu\sigma})
+{\textstyle\frac{1}{6}}
(g_{\mu\rho}g_{\nu\sigma}-
g_{\mu\sigma}g_{\nu\rho})R 
,\ee
we can assemble all higher derivative gravity terms 
in $a_4$ to form the square of the Weyl tensor
\bb
C_{\mu\nu\rho\sigma}C^{\mu\nu\rho\sigma}=
R_{\mu\nu\rho\sigma}R^{\mu\nu\rho\sigma}-2
R_{\mu\nu}R^{\mu\nu}+{\textstyle\frac{1}{3}}R^2 
=2R_{\mu\nu}R^{\mu\nu}
-{\textstyle\frac{2}{3}}R^2+
{\rm surface\ terms               },\ee
because
$ R_{\mu\nu\rho\sigma}R^{\mu\nu\rho\sigma}
-4R_{\mu\nu}R^{\mu\nu}+R^2$
is proportional to the Euler characteristic of $M$. Then,
up to this surface term, we have
\bb -{\textstyle\frac{1}{360}} \dim\hh\left[
7R_{\mu\nu\rho\sigma}R^{\mu\nu\rho\sigma}+
8R_{\mu\nu}R^{\mu\nu}-5R^2\right]=
-{\textstyle\frac{1}{20}}\dim \hh\ 
C_{\mu\nu\rho\sigma}C^{\mu\nu\rho\sigma}.\ee
Finally we have up to surface terms,
\bb a_4 &=& -{\textstyle\frac{1}{20}}\dim \hh\,
C_{\mu\nu\rho\sigma}C^{\mu\nu\rho\sigma}+
{\textstyle\frac{2}{3}}
\t \rho(F_{\mu\nu})^*\rho(F^{\mu\nu})\cr &&+
4Q|\varphi/v|^4+
4L(\dee_\mu\varphi/v)^*(\dee^\mu\varphi/v)
+{\textstyle\frac{2}{3}}
L|\varphi/v|^2R.\ee
 We have used a trick to compute the second and forth
power of the homogeneous scalar variable $\Phi$,
a trick proper to the noncommutative formulation of
the standard model. Remember from section
\ref{smconnes} the embedding of the scalar doublet
$\varphi=\,^t(\varphi_1,\varphi_2)$ in 
$\hh_L^*\ot\hh_R\,\op\, \hh_L\ot\hh_R^*$:
\bb \Phi=\frac{1}{v}\,\pp{
\pp{\varphi_1M_u&-\bar\varphi_2M_d\cr 
         \varphi_2M_u&\bar\varphi_1M_d}\ot 1_3&0\cr 
0&\pp{-\bar\varphi_2M_e\cr \bar\varphi_1M_e}},
\label{emb}\ee
with $v$ denoting the vacuum expectation value. This 
 embedding, which is nothing but the
Yukawa couplings, takes the form of a matrix product,
\bb \Phi=\rho_{Lw}(\phi)\mm/v,\qq 
\phi=\pp{\varphi_1&-\bar\varphi_2\cr 
\varphi_2&\bar\varphi_1}\in\hhh,\ee
and the powers of $\Phi$ follow easily from the
identity
\bb \phi^*\phi=\phi\phi^*=
(|\varphi_1|^2+|\varphi_2|^2)1_2=|\varphi|^21_2.\ee

\section{The unified action}

Chamseddine \& Connes' distribution function $S$ or
{\it spectral action} unifies the Einstein-Hilbert
action, the Yang-Mills action, the Klein-Gordon
action and the Higgs potential. 

\begin{itemize}\item
{ \qq relativity \ + \ noncommutative geometry \ \ =
\ \ Einstein-Hilbert-Yang-Mills-Higgs.  }
\end{itemize}

We still have to properly normalize the kinetic terms
of the gravitational, Yang-Mills and Higgs fields to
deduce their couplings, Newton's constant $G$, the
gauge couplings
$g_i$ and the Higgs couplings $\lambda$ and $\mu$.
We also have a cosmological constant $\Lambda_C$, the
conformal scalar gravity coupling and a higher
derivative gravity term with coefficient $a$ in the
spectral action,
\bb \t f(\dd_{t,{\rm cov}}^2/\Lambda^2)&=&\int_M[
-{\textstyle\frac{1}{16\pi}}m_P(\Lambda)^{2}\,R
\,-\,\Lambda_C(\Lambda)\cr &&\qq+\,
{\textstyle\frac{1}{2}}g_3(\Lambda)^{-2}\, \t
F_{\mu\nu}^{(3)*}F^{(3)\mu\nu}+
{\textstyle\frac{1}{2}}g_2(\Lambda)^{-2}\, \t
F_{\mu\nu}^{(2)*}F^{(2)\mu\nu}+
{\textstyle\frac{1}{4}}g_1(\Lambda)^{-2}\, 
F_{\mu\nu}^{(1)*}F^{(1)\mu\nu}\cr &&\qq
+\,{\textstyle\frac{1}{2}}\,(\dee_\mu\varphi)^*
\dee^\mu\varphi\,+\,\lambda(\Lambda) |\varphi|^4
\,-\,{\textstyle\frac{1}{2}}\,\mu(\Lambda)^2|\varphi|^2
\cr &&\qq-\,a(\Lambda)
C_{\mu\nu\rho\sigma}C^{\mu\nu\rho\sigma}\,+\,
{\textstyle\frac{1}{12}}\,|\varphi|^2
R\qq
]\,(\det
g_{\cdot\cdot})^{1/2}\,\de^4x \ +\
O(\Lambda^{-2}).\label{action}\ee   
Before identifying Newton's constant
$G=\hbar c\,m_P^{-2}$ and the cosmological
constant $\Lambda_C$, we have to shift the
Higgs field by its vacuum expectation value,
$|\varphi|=v(\Lambda)=\mu(\Lambda)/
(2\sqrt\lambda(\Lambda))$. 
With $N$
generations of quarks and leptons, $N=3$, we have:
\bb
m_P(\Lambda)^2&=&{\textstyle\frac{1}{\pi}}f_2
\left[5N -\frac{2}{3}\frac{L^2}{Q}\,\right]
\Lambda^2\,\approx\,{\textstyle\frac{1}{\pi}}f_2
[5N -2]\Lambda^2,
\\
\Lambda_C(\Lambda)&=&{\textstyle\frac{1}{4\pi^2}}
\left[\,
\frac{f_2^2}{f_4}\,\frac{L^2}{Q}\,-\,
15Nf_0\right]
\Lambda^4\,\approx\,
{\textstyle\frac{3}{4\pi^2}}
\left[
\,\frac{f_2^2}{f_4}\,-\,5Nf_0
\right]\Lambda^4,
\\
g_3(\Lambda)^{-2}&=&{\textstyle\frac{N}{3\pi^2}}f_4,
\label{stiff3}\\ 
g_2(\Lambda)^{-2}&=&{\textstyle\frac{N}{3\pi^2}}f_4,\\ 
g_1(\Lambda)^{-2}&=&{\textstyle\frac{5}{3}}
{\textstyle\frac{N}{3\pi^2}}f_4,\label{stiff1}\\ 
\lambda(\Lambda)^{-1}&=&{\textstyle\frac{1}{\pi^2}}
\;f_4\;\frac{L^2}
{Q}\,\approx\,{\textstyle\frac{3}{\pi^2}}\, f_4,\\
\mu(\Lambda)^2&=& 2\,\frac{f_2}{f_4}\,\Lambda^2
\label{stiffmu},\\
a(\Lambda)&=&{\textstyle\frac{3N}{64\pi^2}}\ f_4.\ee
The indicated approximation concerns the
dominating top mass. Comparing with the combined
Euclidean action (\ref{++++}), we see that each
relevant term comes with its correct sign! 

Identifying $f_4={\textstyle\frac{3}{2}}z$ the
constraints on the three gauge couplings from
noncommutative Yang-Mills coincide with the
constraints from noncommutative relativity.
This is not an accident. In noncommutative
Yang-Mills, we have chosen the scalar product
symmetrized with respect to charge conjugation,
\bb   
<\kappa,\varphi>& :=&{\textstyle\frac{z}{2}} {\rm
Re\,}\tt ([\kappa+J\kappa J^{-1}]^*
[\varphi+J\varphi J^{-1}]\,
|\dd|^{-\dim}),\qq\kappa, \varphi \in
\pi(\Omega^p\aa).\ee 
In the spectral action this scalar product is induced
from the symmetrized covariant Dirac operator
$\dd-\pi(A)-J\pi(A)J^{-1}$. The non-symmetrized
 covariant Dirac operator $\dd-\pi(A)$ would induce
the non-symmetrized scalar product,
\bb <\kappa,\varphi> &:=&z {\rm
Re\,}\tt (\kappa^*\varphi|\dd|^{-\dim}),\ee 
 in the spectral action.
Physics requires the use of the symmetrized Dirac
operator in the fermionic action,
$\psi^*(\dd-\pi(A)-J\pi(A)J^{-1})\psi$. In the
noncommutative Yang-Mills setting we were still
free to use either Dirac operator -- symmetrized or not
-- in the bosonic action. This is no
longer true in noncommutative relativity where the
spectral principle requires one and the same Dirac
operator in both actions, the fermionic and the
bosonic. This is why we committed to the symmetrized
scalar product already in noncommutative Yang-Mills.
Here there is no choice and we are
forced to swallow the big desert and to extrapolate
running couplings to energies
$\Lambda=10^{13}-10^{17}$ GeV where
$f_4={\textstyle\frac{3}{2}}z=(0.80-0.94)4\pi^2$.
This of course means that we have to return humblely
to flat space because, despite the higher derivative
term $a$, gravity remains unrenormalizable.
Fortunately, thanks to $f_0$ and $f_2$, the Planck
mass and the cosmological constant decouple from the
gauge couplings. Since the evolution of $\mu$
strongly depends on the regularization scheme there
is only one more unambiguous constraint from
noncommutative relativity,
\bb \lambda(\Lambda)={\textstyle\frac{N}{9}}
g_2(\Lambda)^2. \qq {\rm nc\ relat.}\ee
Remember the corresponding constraint from
noncommutative Yang-Mills, 
\bb \lambda(\Lambda)={\textstyle\frac{3N-2}{24}}
g_2(\Lambda)^2.\qq {\rm nc\ YM}\ee
They would coincide for $N=6$ generations. For $N=3$,
their mismatch is still acceptable, in terms of the
resulting Higgs mass, we have,
\bb m_H=182\pm  10 \pm  7\ {\rm GeV.\qq
nc\ relat.}\ee
The first error is from the
uncertainty in
$\Lambda=10^{13}-10^{17}$ GeV. The second is from
the present experimental uncertainty in the top mass,
$m_t=175\pm 6$ GeV. Indeed we must admit that
noncommutative relativity does not constrain the
Yukawa coupling or equivalently the top mass as was
the case in noncommutative Yang-Mills where we had
\bb 
m_H&=&197\pm 9\pm 0\  {\rm GeV},\cr 
m_t&=&187\pm 14\pm 0\ {\rm GeV}.\qq {\rm nc\ YM}
\ee

The mismatch between the two Higgs couplings or
masses from noncommutative Yang-Mills and from
noncommutative relativity is of the same order of
magnitude as the mismatch between the experimental
and theoretical values of the three gauge couplings.
We blame this mismatch on the enormous
extrapolation through the big desert. We take the
mismatch as indication that at energies
$\Lambda=10^{13}-10^{17}$ GeV almost commutative
geometry will merge into a truely noncommutative
geometry and that gravitational quantum effects will
no longer be small. In any case we find it
encouraging that noncommutative Yang-Mills and
noncommutative relativity produce comparable
results for the standard model. This is another miracle
of the standard model. Indeed applied to the
commutative example of section \ref{commym},
the two dreis\"atze produce quite different outputs,
the first has a photon the second does not. Similarly
the minimax model, \ref{mima}, with one generation
of leptons, has no spontaneous symmetry break down
in noncommutative Yang-Mills, but does enjoy
spontaneous break down in
noncommutative relativity because there junk does
not happen.

In the standard model with
$N=3$ generations, the two Higgs mass predictions
have a non-empty intersection . This intersection
is
$m_H=188-199$ GeV, an energy range experimentally
accessible to the Large Hadron Collider LHC in Geneva
within ten years.

\section{Outlook}

Connes' noncommutative geometry has impressive
unification power. Almost commutative geometry
unifies the non-Abelian gauge dreisatz with the
Riemannian dreisatz. At the same time it indicates a
sequence of dreis\"atze, the Minkowskian, Riemannian
and Connes' second dreisatz indexed by the nested
invariance groups, the Lorentz, diffeomorphism and
$\aa_t-$automorphism groups. It seems natural
to pursue this sequence to truely noncommutative
geometries. Indeed
$\aa_t=\ccc^\infty(M)\ot(\hhh\op\cc\ot M_3(\cc))$
is almost as ugly as Diff$(M)\semi\, ^M(SU(2)\times
U(1)\times SU(3))$. Noncommutative geometry grew
out of quantum mechanics. Almost commutative
geometry unifies gravity with the subnuclear forces.
We expect noncommutative geometry to reconcile
gravity with quantum field theory.

The basic variable of noncommutative geometry is the
Dirac operator acting on fermions. The fermions must
define a representation of an associative algebra and
are constrained by the axioms of noncommutative
geometry, i.e. of spectral triples. These
axioms still leave many choices, one of
which the quarks and leptons of the standard model
with their mass matrix taken from experiment. Of
course, we want an explanation for this choice. 
To define the Dirac operator in Riemannian geometry,
the spin group is essential. There is no generalization
of the spin group to noncommutative geometry yet.
According to Connes \cite{tresch}, this generalization
should be a quantum group and it should help us to get
a handle on the arbitrariness of the fermion
representation.

Minkowskian geometry explains the magnetic field,
Riemannian geometry explains gravity. Both
geometries have operated revolutions on spacetime
that today are well established  experimentally: the
loss of absolute time and the loss of universal time.
Can we observe the noncommutative nature
 of time, its uncertainty or `fuzziness', despite its
ridiculously small scale $\hbar /\Lambda=10^{-40}$ s? 

So far noncommutative geometry is developed in
Euclidean, compact spacetimes, so `Wick rotation' and
3+1 split remain to be understood \cite{kal}.
After this, we expect noncommutative geometry to
change our picture of black holes in a similar fashion
that Heisenberg's uncertainty relation has cured the
Coulomb singularity of the hydrogen atom. Also our
picture of the big bang, cosmology and the origin of
time is expected to be revised \cite{rov}.

Planatary motion has degraded circles to epicycles and
dismissed them all together in favour of ellipses.
Particle physics is about to dismiss Riemannian
geometry in favour of noncommutative geometry and
the question is, what dynamics is behind these new
ellipses?
\hfil\break
\vskip 0.1 cm 
\noindent
I am indepted to Daniel Kastler, the Emminence grise
de Marseille. It is also a pleasure to acknowledge years
of enjoyable collaboration with Lionel Carminati,
Robert Coquereaux, Gilles Esposito-Far\`ese, Meinulf
G\"ockeler, Bruno Iochum, Thomas Krajewski, Igor
Pris and Daniel Testard. Vaughan Jones and
Raymond Stora's continuous support and
friendship are behind my ellipse. Time will show if
this School was successful. If so, we owe this success to
Paulo Almeida.

%% file: monsaraz.bbl
\begin{thebibliography}{47}

\bibitem{insight}
V. M. Chernousenko, {\it Chernobyl, Insight
 from the Inside}, Springer Verlag (1991)
\bibitem{pocket}
The Particle Data Group, {\it Particle Physics Booklet},
American Institute of Physics (1996)
\bibitem{gs}
M. G\"ockeler \& T. Sch\"ucker, {\it Differential
Geometry, Gauge Theories, and Gravity}, Cambridge
University Press (1987)
\bibitem{data}
The Particle Data Group {\it Review of Particle
Properties}, Phys. Rev. D54 (1996) 1 
\bibitem{bd} J. D. Bj\o rken \& S. D. Drell, {\it
Relativistic Quantum Mechanics}, McGraw--Hill (1964)
\bibitem{book}
 A. Connes, {\it Noncommutative Geometry}, Academic 
Press (1994)
\bibitem{grav}
A. Connes, {\it Gravity coupled with matter and the
foundation of noncommutative geometry},
hep-th/9603053, Comm. Math. Phys. 155 (1996) 109
\bibitem{tresch}
A. Connes,
{\it Noncommutative geometry and reality}, 
J. Math. Phys. 36 (1995) 6194
\bibitem{joe}
J. C. V\'arilly, {\it Introduction to noncommutative
geometry}, Lectures at this School, physics/9709045
\bibitem{dix}
A. Connes, {\it The action functional in
non-commutative geometry}, Comm. Math. Phys. 117
(1988) 673
\bibitem{cl}
A. Connes, {\it Essay on physics and noncommutative
geometry}, in {\it The Interface of Mathematics and
Particle Physics}, eds.: D. G. Quillen et al., Clarendon
Press (1990)\\
 A. Connes \& J. Lott, {\it The metric 
aspect of noncommutative geometry}, in the 
proceedings of the 1991 Carg\`ese Summer Conference, 
eds.: J. Fr\"ohlich et al., Plenum Press (1992)
\bibitem{other}
$\ $\qq for other approaches along similar lines
see:\\ M. Dubois-Violette, R. Kerner \& J.
Madore, {\it Gauge bosons in a noncommutative
geometry}, Phys. Lett. 217B (1989) 485\\
J. Madore, {\it An Introduction to Noncommutative
Differential Geometry and its Physical Applications},
Cambridge University Press (1995)\\
R. Coquereaux, G. Esposito-Far\`ese \&
G. Vaillant, {\it Higgs fields as Yang-Mills fields and
discrete symmetries}, Nucl. Phys. B353 (1991) 689
\bibitem{tk}
T. Krajewski, {\it Classification of finite spectral
triples}, hep-th/9701081, J. Geom. Phys., to appear
\bibitem{sz}
T. Sch\"ucker \& J.-M. Zylinski, {\it Connes' model
building kit}, hep-th/9312186, J. Geom. Phys. 16 (1994)
1 
\bibitem{reviews}
D. Kastler, {\it A detailed account of Alain
Connes' version of the standard model in
non-commutative geometry, I and II}, Rev. Math. Phys.
5 (1993) 477 \hfil\break D. Kastler, {\it A detailed
account of Alain Connes' version of the standard
model in non-commutative geometry, III}, Rev. Math.
Phys. 8 (1996) 103  \hfil\break  
D. Kastler \& T.
Sch\"ucker, {\it A detailed account of Alain Connes'
version of the standard model in non-commutative
geometry, IV}, Rev. Math. Phys., 8 (1996) 205\\
D. Kastler \& M. Mebkhout, {\it Lectures on 
Non-Commutative Differential Geometry}, World 
Scientific, to be published\hfil\break
J. C. V\'arilly \&  J. M. Gracia-Bond\'\i a,  
{\it Connes' noncommutative differential geometry and
the standard model}, J. Geom. Phys. 12 (1993) 223 
\hfil\break
C. P. Mart\'\i n, J. M. Gracia-Bond\'\i a \& J. C. V\'arilly,
{\it The standard model as a noncommutative geometry:
the low mass regime},
 hep-th/9605001, Phys. Rep., to appear
\\
 D. Kastler
\& T. Sch\"ucker, {\it The standard model \`a la
Connes-Lott}, hep-th/9412185, J. Geom. Phys. 388
(1996) 1 
\hfil\break
L. Carminati, B. Iochum \& T. Sch\"ucker, {\it
The noncommutative constraints on the standard
model \`a la Connes}, hep-th/9604169, J. Math. Phys. 38
(1997) 1269\\
L. Carminati, B. Iochum \& T. Sch\"ucker, {\it
Noncommutative Yang-Mills and noncommutative
relativity: A bridge over troubled water},
hep-th/9706105
\bibitem{becca}
R. Asquith, {\it Non-commutative geometry and the 
strong force}, hep-th/9509163, Phys. Lett. B 366 (1996)
220
\bibitem{tkmz}
M. Paschke \& A. Sitarz, {\it Discrete spectral triples
and their symmetries}, q-alg/9612029 \\
T. Krajewski, {\it Classification of finite spectral
triples}, hep-th/9701081, J. Geom. Phys., to appear
\bibitem{anom} 
E. Alvarez, J. M. Gracia-Bond\'\i a \& C. P. Mart\'\i n,
 {\it Anomaly cancellation and the gauge group of the
Standard Model in Non-Commutative Geometry},
hep-th/9506115, Phys. Lett. B364 (1995) 33
\bibitem{versus}
B. Iochum \& T. Sch\"ucker, {\it
Yang-Mills-Higgs versus Connes-Lott}, 
hep-th/9501142, Comm. Math. Phys. 178 (1996) 1
\bibitem{lr}
B. Iochum \& T. Sch\"ucker, {\it A left-right symmetric
model \`a la Connes-Lott}, hep-th/9401048, Lett. Math.
Phys. 32 (1994)  153
\bibitem{beyond}
I. Pris \& T. Sch\"ucker, {\it Non-commutative
geometry beyond the standard model}, hep-th/9604115,
J. Math. Phys. 38 (1997) 2255\\
I. Pris \& T. Krajewski, {\it Towards a $Z'$ gauge boson
in noncommutative geometry}, hep-th/9607005, Lett.
Math. Phys. 39 (1997) 187
\bibitem{jones}
C. Ford, I. Jack \& D. R. T. Jones, {\it  The standard
model effective potential at two loops}, Nucl.
Phys. B387 (1992) 373 \\
B. Schrempp \& M. Wimmer, {\it Top quark and Higgs 
boson masses: Interplay between infrared and 
ultraviolet physics}, hep-ph/9606386,
Progress in Particle and Nuclear Physics 37 (1996) 
\bibitem{cab}
N. Cabibbo, L. Maiani, G. Parisi \& R. Petronzio, {\it
Bounds on the fermions and Higgs boson masses in
grand unified theories}, Nucl. Phys. B158 (1979) 295
\bibitem{bonse}
U. Bonse \& T. Wroblewski,
{\it Measurement of neutron quantum interference in
noninertial frames}, Phys. Rev. Lett. 1 (1983) 1401
\bibitem{Yates}
R. G. Yates, {\it Fiber bundles and supersymmetries},
Comm. Math. Phys. 76 (1980) 255
\bibitem{Cartan}
E. Cartan, {\it Le\c cons sur la
th\'eorie des spineurs}, Hermann (1938)
\bibitem{rauch}
H. Rauch et al., {\it Verification of coherent spinor
rotations of fermions}, Phys. Lett. 54A (1975) 425
\bibitem{att}
 T. Ackermann \& J. Tolksdorf, {\it A generalized
Lichnerowicz formula, the Wodzicki residue and
gravity}, hep-th/9503152, J. Geom. Phys. 19 (1996) 143
\\
 T. Ackermann \& J. Tolksdorf, {\it The generalized
Lichnerowicz formula and analysis of Dirac
operators}, hep-th/9503153, J. reine angew. Math. 471
(1996)
\bibitem{gilles} 
G. Esposito-Far\`ese, {\it Th\'eorie de
Kaluza-Klein et Gravitation Quantique}, Th\'ese de
Doctorat, Universit\'e d'Aix-Marseille II, 1989
\bibitem{wod}
A. Connes, {\it Noncommutative geometry and
physics}, in the proceedings of the 1992 Les Houches
Summer School, eds.: B. Julia, J. Zinn-Justin, North
Holland (1995)\\
D. Kastler, {\it The Dirac operator and gravitation}, 
Comm. Math. Phys. 166 (1995) 633\\
W. Kalau \& M. Walze,  {\it Gravity, non-commutative
Geometry and the Wodzicki residue}, gr-qc/9312031, 
J. Geom. Phys. 16 (1995) 327\\
\indent $\ $\qq for other approaches see:\\
J. Madore, {\it Kaluza-Klein aspects of
noncommutative geometry}, in the proceedings of the
1988 conference on {\it Differential Geometric
Methods in Theoretical Physics}, ed.: A. Solomon,
World Scientific (1989)\\
A. Chamseddine, J. Fr\"ohlich \& O. Grandjean,
{\it The gravitational sector in the Connes-Lott
formulation of the standard model}, 
hep-th/9503093, J. Math. Phys. 36 (1995) 6255\\
 T. Ackermann \& J. Tolksdorf, {\it
Unification of gravity and Yang-Mills-Higgs gauge
theories}, hep-th/9503180\\
T. Ackermann, {\it Dirac operators and Clifford
geometries - new unifying principles in particle
physics?}, hep-th/9605129\\
J. Tolksdorf, {\it The
Einstein-Hilbert-Yang-Mills-Higgs action and the
Dirac-Yukawa operator}, hep-th/9612149\\
H. Figueroa, J. M. Gracia-Bond\'\i a, F. Lizzi \& J. C.
V\'arilly, {\it A nonperturbative form of the spectral
action principle in noncommutative geometry},
hep-th/9701179
\bibitem{cc}
A. Chamseddine \& A. Connes, {\it The spectral action
principle}, hep-th/9606001, Comm. Math. Phys. 186
(1997) 731
\bibitem{egbv}
R. Estrada, J. M. Gracia-Bond\'\i a \& J. C. V\'arilly,
{\it On summability of distributions and spectral
geometry}, funct-an/9702001
\bibitem{heat}
P. B. Gilkey, {\it Invariance Theory, the Heat Equation,
and the Atiyah-Singer Index Theorem}, Publish or
Perish (1984)\\
S. A. Fulling, {\it Aspects of Quantum Field Theory in
Curved Space-Time}, Cambridge University Press
(1989)
\bibitem{rom}
B. Iochum, D. Kastler \& T. Sch\"ucker, {\it
On the Universal Chamseddine-Connes Action I:
 Details of the Action Computation}, hep-th/9607158,
J. Math. Phys., to appear\\
L. Carminati, B. Iochum, D. Kastler \& T.
Sch\"ucker, {\it On Connes' new principle of general
relativity: can spinors hear the forces of space-time?},
hep-th/9612228, Operator Algebras and Quantum Field
Theory, eds.: S. Doplicher et al., International Press,
1997
\bibitem{kal}
W. Kalau, {\it Hamiltonian formalism in
non-commutative geometry}, hep-th/9409193,
J. Geom. Phys. 18 (1996) 349\\
E. Hawkins, {\it Hamiltonian gravity and
noncommutative geometry}, gr-qc/9605068  \\
W. Kalau, work in progress
\bibitem{rov}
A. Connes \& C. Rovelli, {\it Von Neumann algebra
automorphisms and time-thermodynamics relation in
general covariant quantum theories}, gr-qc/9406019,
Class. Quant. Grav. 11 (1994) 1899


\end{thebibliography}
